\documentclass[10pt]{article} 
\usepackage{authblk}
\usepackage[utf8]{inputenc} 
\usepackage[T1]{fontenc} 
\usepackage{graphicx} 
\usepackage{longtable} 
\usepackage{hyperref}
\usepackage[inner=0.5in,outer=0.5in, top=1.0in, bottom=1.25in]{geometry}
\usepackage{sectsty}
\usepackage{gensymb}
\usepackage{textcomp}
\usepackage{alltt}
\usepackage{color}
\usepackage{longtable}
\usepackage{wrapfig}
\usepackage{rotating}
\usepackage[normalem]{ulem}
\usepackage{amsmath}
\usepackage{amssymb}
\usepackage{hyperref}
\usepackage[section]{placeins}
\usepackage{alltt}
\usepackage{color}
\usepackage{natbib}
\usepackage[lined,boxed]{algorithm2e}
\usepackage{tikz}
\usepackage{makecell}
\usepackage{siunitx}
\usepackage{xfrac}
\usepackage{enumitem}
\usepackage{listings}
\usepackage{url}
\usepackage[absolute]{textpos}

\usepackage{pgf}
\usepackage{mathrsfs}
\usetikzlibrary{arrows}

\usepackage{mathpazo} %

\usepackage[scaled]{helvet}

\DeclareUnicodeCharacter{00A0}{~}

\definecolor{NALBlueLight}{RGB}{65, 182, 230}
\definecolor{NALBlue}{RGB}{0, 76, 151}
\definecolor{NALBlueDark}{RGB}{0, 40, 85}
\definecolor{NALGreenLight}{RGB}{120, 190, 32}
\definecolor{NALGreen}{RGB}{76, 140, 43}
\definecolor{NALGreenDark}{RGB}{54, 87, 59}
\definecolor{MarlinLight}{RGB}{153, 214, 234}
\definecolor{Marlin}{RGB}{0, 181, 226}
\definecolor{MarlinDark}{RGB}{0, 133, 173}
\definecolor{MFRedLight}{RGB}{175, 39, 47}
\definecolor{MFRed}{RGB}{138, 42, 43}
\definecolor{MFRedDark}{RGB}{100, 51, 53}
\definecolor{OrangeLight}{RGB}{246, 141, 46}
\definecolor{Orange}{RGB}{203, 96, 21}
\definecolor{OrangeDark}{RGB}{185, 71, 0}
\definecolor{PrairieGoldLight}{RGB}{245, 225, 164}
\definecolor{PrairieGold}{RGB}{254, 209, 65}
\definecolor{PrairieGoldDark}{RGB}{254, 209, 65}
\definecolor{intrlred}{RGB}{169,58,63}
\definecolor{aqua}{RGB}{65,155,133}
\definecolor{prgold}{RGB}{210,140,0}
\definecolor{nalblue}{RGB}{0,51,153}
\DeclareSIUnit\deg{deg}
\DeclareSIUnit\arcsec{arcsec}
\sectionfont{\normalfont\sffamily\bfseries}
\subsectionfont{\normalfont\sffamily\itshape}
\subsubsectionfont{\normalfont\sffamily\itshape}
\newcommand\obstac{\texttt{obstac}}
\title{Dark Energy Survey's Observation Strategy, Tactics, and Exposure Scheduler}
\author[1]{Eric~H.~Neilsen~Jr.}
\author[1]{James~T.~Annis}
\author[1]{H.~Thomas~Diehl}
\author[2]{Molly~E.~C.~Swanson}
\author[3]{Chris~D'Andrea}
\author[1]{Stephen~Kent}
\author[1,4]{Alex~Drlica-Wagner}
\affil[1]{Fermi National Accelerator Laboratory, P.~O.~Box 500, Batavia, IL 60510, USA}
\affil[2]{National Center for Supercomputing Applications, 1205 West Clark St., Urbana, IL 61801, USA}
\affil[3]{Department of Physics and Astronomy, University of Pennsylvania, Philadelphia, PA 19104, USA}
\affil[4]{Kavli Institute for Cosmological Physics, University of Chicago, Chicago, IL 60637, USA}

\date{November 1, 2019}

\hypersetup{
 pdfauthor={Eric H. Neilsen, Jr.},
 pdftitle={Dark Energy Survey's Observation Strategy, Tactics, and Exposure Scheduler},
 pdfkeywords={},
 pdfsubject={},
 pdflang={English}}
\begin{document}

\maketitle

\begin{textblock*}{4in}[0, 0](.65\textwidth,0.5in)
FERMILAB-TM-2714-AE-CD-PPD
\end{textblock*}

\begin{textblock*}{0.9\textwidth}(0.75in,10.35in)
\noindent This manuscript has been authored by Fermi Research Alliance, LLC under Contract No. DE-AC02-07CH11359 with the U.S. Department of Energy, Office of Science, Office of High Energy Physics.
\end{textblock*}

\begin{abstract}
The Dark Energy Survey is a ``stage III'' dark energy experiment, performing an optical imaging survey to measure cosmological equation of state parameters using four independent methods. The scope and complexity of the survey, originally scheduled for 525 nights of observing spread over five years and combining a 5000 degree wide survey with a 10 field weekly time-domain survey, introduced complex strategic and tactical scheduling problems that needed to be addressed. We begin with an overview of the process used to develop DES strategy and tactics, from the inception of the project, to task forces that studied and developed strategy changes over the course of the survey, to the nightly pre-observing meeting in which immediate tactical issues were addressed. We then summarize the strategic choices made for each sub-survey, including metrics, scheduling considerations, choice of time domain fields and their sequences of exposures, and wide survey footprint and pointing layout choices. We go on to describe the detailed process that determined which specific exposures were taken at which specific times. We give a chronology of the strategic and tactical peculiarities of each year of observing, including the proposal and execution of a sixth year. We give an overview of \obstac, the implementation of the DES scheduler used to simulate and evaluate strategic and tactical options, and automate exposure scheduling; and describe developments in \obstac\ for use after DES. Appendices describe further details of data quality evaluation, $\tau$, and $t_{\mbox{eff}}$; airmass calculation; and modeling of the seeing and sky brightness. The significant corpus of DES data indicates that the simple scaling relations for seeing as a function of wavelength and airmass derived from the Kolmogorov turbulence model work adequately for exposure planning purposes: deviations from these relations are modest in comparison with short time-scale seeing variations.
\end{abstract}

\section{Introduction}
\label{intro}

Understanding the observed acceleration of the expansion rate of the universe is among the primary problems in contemporary cosmology. To define an experimental plan to address this challenge, DOE, NSF and NASA established the Dark Energy Task Force (DETF). The DETF defined a ``figure of merit'' (FoM) for experimental measurements of relevant cosmological parameters, and outlined a plan for a series of experiments, grouped into four stages with progressively more precise design figures of merit \citep{albrecht_report_2006}. The DETF outlined several techniques for constraining equation of state parameters. Four probes described by the DETF can be performed using optical astronomical surveys:
\begin{description}
\item[{supernova}] By measuring the light curves of type Ia supernova, the a survey can measure the redshift-distance relationship at distances sufficient to measure how the expansion rate of the universe varies as a function of the age of the universe.
\item[{weak lensing}] The gravitational field from the matter that falls along the line of sight between our observatory and a source of light (such as a galaxy) distorts the shape of the source. The parameters of the distortion are a function both of the relative distances to the source and intermediate matter and the distribution of the intermediate matter, and so can be used to constrain cosmological parameters.
\item[{galaxy clusters}] Cosmological simulations indicate that the development of small-scale density perturbations in the early universe into collapsed high-mass  structures in the distribution of galaxies is a strong function of the cosmological equation of state, so the equation of state parameters can be constrained by measuring galaxy cluster parameters at different redshifts (and therefore ages of the universe).
\item[{large scale structure}] Baryon Acoustic Oscillations (BAO) in the early universe generate a large scale cosmological structure set at the time when the plasma in the early universe combines to form neutral hydrogen. This recombination sets a characteristic scale length in the distribution of matter, which can be measured in the spatial correlations in distribution of galaxies.
\end{description}

For these measurements, two data sets are needed: a time-domain survey, and a wide-field survey. The supernova method requires the time domain survey, in which the same fields of sky are observed in multiple bands on a regular cadence over the course of several months. With this set of images, the light curves (rise and fall of the brightness) of the supernovae in the observed fields can be measured, and calibrated for use as standard candles to measure their distances. The remaining methods require statistics on large numbers of galaxies, but do not require tracking changes over time: they require a survey over a large volume of sky, but these methods place few constraints on exposure timing. The different methods that use wide-field data place different (although compatible) constraints on other aspects of how the wide-field survey data is collected.

The Dark Energy Survey (DES) \citep{dark_energy_survey_collaboration_dark_2016} is an astronomical observing program to measure cosmological equation of state parameters with a time domain survey consisting of ten 3.1 sq. deg. fields, combined with a wide field survey of 5000 square degrees in the southern Galactic cap. To perform these surveys, DES used 577 nights spread over 6 years on the DECam camera \citep{flaugher_status_2012} and the Victor Blanco 4-m telescope at the Cerro Tololo Inter-American Observatory (CTIO) in Chile.\footnote{The original plan was for DES to use 525 nights spread over five years, but this was supplemented with additional year with 52 nights to make up for poor weather in the third year of operations; see section~\ref{year6} for more details.} The expected constraints on the DETF FoM qualify it as a DETF ``stage III'' dark energy experiment, improving the DETF FoM by a factor of 3 to 5 over stage II experiments. 

This note begins with an overview of the process the DES collaboration used to design and update DES strategy and tactics. Section~\ref{timedomainstrat} summarizes the choice of time domain fields and their sequences of exposures. Section~\ref{widestrat} discusses the wide survey observing strategy, including motivations, observing metrics, scheduling considerations, and footprint and pointing layout choices. Section~\ref{tactics} describes observing tactics: the detailed process that determined which specific exposures were taken and which specific times. Section~\ref{chronology} discusses the observing strategy, tactics, and results of each year of observing. Section~\ref{obstac} gives an overview of the implementation of the DES scheduler used to simulate and evaluate strategic and tactical options, and automate exposure scheduling. Section~\ref{afterdes} describes developments in \obstac\ for use after DES, and section~\ref{conclusion} gives some concluding thoughts. Appendices describing further details of data quality evaluation, $\tau$, and $t_{\mbox{eff}}$ (appendix~\ref{dataquality}), airmass calculation (appendix~\ref{airmass}), modeling of the seeing (appendix~\ref{seeing}) and sky brightness (appendix~\ref{skybrightness}) follow, and the note ends with a reference table for notation used (appendix~\ref{notation}) and references.

\section{The survey strategy development and tactics process}
Decisions that affect which exposure are taken at what times are made at a variety of time scales, specificity, and levels of abstraction, ranging from the foundational choices of scientific objectives and instrument, to strategic choices such as overall depth and footprint area and location, to the immediate tactical choice of which specific exposures to take at any specific times. Higher level, more abstract, and strategic decisions set the constraints and objectives for more tactical, concrete, and specific choices. 

The strategic decisions with the widest impact were made at the project inception, proposal, and funding stages. These initial choices, which set the constraints under which all other strategic decisions are made, resulted in the following basic survey parameters:
\begin{itemize}
    \item The cosmological equation of state would be measured using supernova, weak lensing, galaxy clusters, and large scale structure.
    \item The survey data would be collected using a new imaging camera, DECam, on the Victor Blanco 4.0 meter telescope at the Cerro Tololo Inter-American observatory (CTIO) in Chile.
    \item The survey would be completed using 525 nights of observing spread over five years.
    \item The 525 nights would include a mix of dark time (time when the moon is below the horizon) and gray or bright time (time when it is above it).
\end{itemize}
Motivation for these decisions is outside the scope of this document. Interested readers should consult \cite{lahav_dark_2019}.

The choice of instrument set a variety of constraints on all choices that followed. The location of the telescope at a latitude of $-30\degree$ set the limits on which declinations are accessible at what airmasses, over what period of time; the field of view, readout and slew times, sensor sensitivity, and aperture of the telescope set the parameters for the trade-off between area, depth, and mean number of exposures per unit area of sky. The equatorial mount of the Blanco telescope and the lack of rotator on the DECam establish a constant orientation of the camera relative to celestial equatorial coordinates.

Within the parameters set by these foundational parameters, a survey strategy was developed by the DES survey scientist and the DES Survey Strategy Task Force (SSTF). This group included the DES project scientist and representatives of each of the different science working groups.   
Initial designs for the survey were set to achieve a complete cluster catalog to $z=1$ over 4000 sq-degrees; thus no u-band for example, as it was unnecessary to locate clusters at very low $z$. The large scale structure science was met automatically. Initial simulations made it clear 5000 sq-degrees was possible and preferred. Incorporation of the weak lensing program started placing requirements on the PSF and pushed the idea of the maximal number of tilings rather than increasing the exposure time as the survey went along. Finally, the incorporation of the SN program added a time domain component. As a side effect, this component allowed us to do non-wide survey science during non-optimal observing conditions. 
The most dramatic change to come from the weak lensing group was not to aim to get past $i=24$; the thought being that even if we did, there wouldn't be photometric redshift training samples to calibrate at that depth.
The outcome of the DES Survey Strategy Task Force was a survey strategy that could be summed up simply: a {\it g, r, i,} and {\it z} survey consisting of two tilings of the whole survey area per year per band, using 90 second exposures.\footnote{These were supplemented by an additional set of exposures of 45 seconds each in an additional {\it Y} filter (see item~\ref{req:photocalib} of the list in section~\ref{wideobjectives}).}

Over the course of the survey, the SSTF met regularly to discuss progress, updated survey simulations, and potential problems or improvements. Questions and concerns of general interest were presented at biannual DES collaboration meetings. In addition to these formal mechanisms, the survey scientists and other members of the SSTF informally answered questions and collected suggestions from members of the collaboration at large. Ultimately, high level strategy decisions were made by the DES project director, operations scientist, and executive committee, informed by input from the SSTF and survey scientist.

Before each year, the survey scientist implemented the latest strategic plans in \obstac, the DES observing simulator and automated scheduler (see section~\ref{obstac}), and ran suites of simulations with different combinations of tactics and possible schedules for the year to follow. The results were used by the survey scientist, operation scientist, and project director to inform the construction of a schedule request to NOAO for use in allocating nights of observing. NOAO shared candidate schedules with DES for evaluation and feedback, so that high priority community programs could be scheduled in ways that accommodated the needs both of DES and the respective community programs.

After each night of observing, the DES data management team (DESDM) processed the data, calculated quality metrics, and determined which exposures met scientific requirements. These evaluations were then fed back into the observing database, so that bad exposures could be repeated as soon as possible. Before each night of observing, an automated process used \obstac\ to simulate the following night using the latest quality evaluations available, under a variety of seeing conditions. At 4:00pm Chilean local time before most nights of observing, the observers, operations scientist, and survey scientist reviewed the latest data quality evaluations, expectations of \obstac's behavior based on the latest simulations, expected weather for the upcoming nights, and tactical decisions the observers might need to make during the night.\footnote{This meeting was sometimes skipped if there were no tactical changes from the previous night, no new observers, and nothing else to be discussed.} On most nights of observing, the tactical instruction to the observing staff was to turn on \obstac\ and let it run the whole night. Under these conditions, the tactics were set in the design and implementation of \obstac. If there were unusual observing programs to be run, or if there were hand-designed tactical improvements that could be implemented, the observers could be provided with observing scripts (files with exposure specifications that could be read and executed by SISPI, the readout and control system for DECam) and instructions for when and under what conditions to run them.

See \cite{diehl_dark_2019} for a more detailed description of nightly observing procedures.

\section{Time-domain survey strategy}
\label{timedomainstrat}
The time domain survey (also referred to as the ``supernova survey'') was designed to detect type Ia supernova and measure their light curves so that they may be used to measure the dark energy equation of state. These objects brighten and fade over the course of a couple of months. When sequences of exposures on the same field are taken on a regular cadence over a given time window, the light curves of any supernova that both fall within the field and brighten and fade within that time window can be measured. To characterize the light curve well, exposures must be taken at a cadence that is short compared to the timescale over which the brightness of the supernova changes. This cadence must be maintained in multiple wavelength bands in order both to distinguish supernova from other variable objects, and to measure the intrinsic properties of each supernova necessary to estimate the intrinsic brightness (and therefore be used as a standard candle for distance measurement).  The DES time domain survey collected light curves on type Ia supernovae following such a strategy. \cite{bernstein_supernova_2012} and \cite{kessler_difference_2015} describe the time-domain survey strategy in detail; only a brief summary is given here.

One factor that needed to be optimized was the total exposure time accumulated in each sequence, which determined magnitude limits and uncertainties on individual points on the light curves. More exposure time for each sequence improves photon statistics and increases the redshift out to which supernovae may be measured, while shorter sequences take less time and therefor allow the monitoring of more fields, increasing the total number of supernovae. Simulations described in \cite{bernstein_supernova_2012} indicated that the Dark Energy Task-Force (DETF) figure of merit \citep{albrecht_report_2006} is optimized by a hybrid strategy including both a few fields observed with long sequences of exposures and additional fields observed with shorter exposures. DES selected two ``deep'' and eight ``shallow'' fields in total.

When observing shallow fields, exposures in all bands were combined into a single sequences, described in table~\ref{tab:snshallowseqexptimes}. For deep fields, exposures in each band were taken in separate sequences; table~\ref{tab:sndeepseqexptimes} describes these. In total, there were 16 sequence of exposures targeted for observing on a one week or better cadence: one on each of eight ``shallow'' fields, and four (using $g$, $r$, $i$, and $z$ filters) on each of two ``deep'' fields. 

In addition to the exposures listed in each table, a short (10 second) ``pilot'' exposure was taken at the start of each sequence, and used to correct the pointing before the start of the science exposures. (If sequences at different epochs are taken at slightly different pointings, then many objects near the edges of the field of view will be missed at some epochs: small misalignment between the pointings in different epochs can results in a significant reduction in the effective area over which light curves can be collected at an acceptable cadence.)

\begin{table}
\begin{center}
\begin{tabular}{lllll}
Filter & \# & time/exposure (s)& total time (s)& $m_{\mbox{lim}}$ \\
\hline
g & 1 & 175 & 175 & 24.0  \\
r & 1 & 150 & 150 & 23.7 \\
i & 1 & 200 & 200 & 23.2 \\
z & 2 & 200 & 400 & 22.9 \\
\end{tabular}
\end{center}
\caption{\label{tab:snshallowseqexptimes}
Parameters for different types of time-domain survey fields. \(m_{\mbox{lim}}\) is a roughly $10\sigma$ point source magnitude limit for a sequence with a mean $\tau=1.0$ (dark sky, seeing=0.9''). The full shallow sequence takes roughly 20 minutes of wall clock time (including overhead between exposures). See appendix~\ref{dataquality} for a definition of $\tau$.
}
\end{table}

\begin{table}
\begin{center}
\begin{tabular}{llllll}
Filter & \# & time/exposure (s)& total time (s)& $m_{\mbox{lim}}$ & sequence duration\\
\hline
g & 3 & 200 & 600 & 24.7 & 14 min.\\
r & 3 & 400 & 1200 & 24.8 & 23 min.\\
i & 5 & 360 & 1800 & 24.4 & 34 min.\\
z & 11 & 330 & 3630 & 24.1 & 67 min.\\
\end{tabular}
\end{center}
\caption{\label{tab:sndeepseqexptimes}
Parameters for different types of time-domain survey fields. \(m_{\mbox{lim}}\) is a roughly $10\sigma$ point source magnitude limit for a sequence with a mean $\tau=1.0$ (dark sky, seeing=0.9''). The durations include typical overhead between exposures (slew and readout time).
}
\end{table}

In sequences within which multiple exposures are taken of the same filter, the exposures are ``dithered'' into up to three pointings, offset from each other by a few arcseconds. This prevents objects from falling on the same pixels in every exposure. Tables~\ref{tab:sndeepseqexps} and~\ref{tab:snshallowseqexps} show the pointings and numbers of exposures at each dither.

\begin{figure*}
\centering
\includegraphics[width=\linewidth]{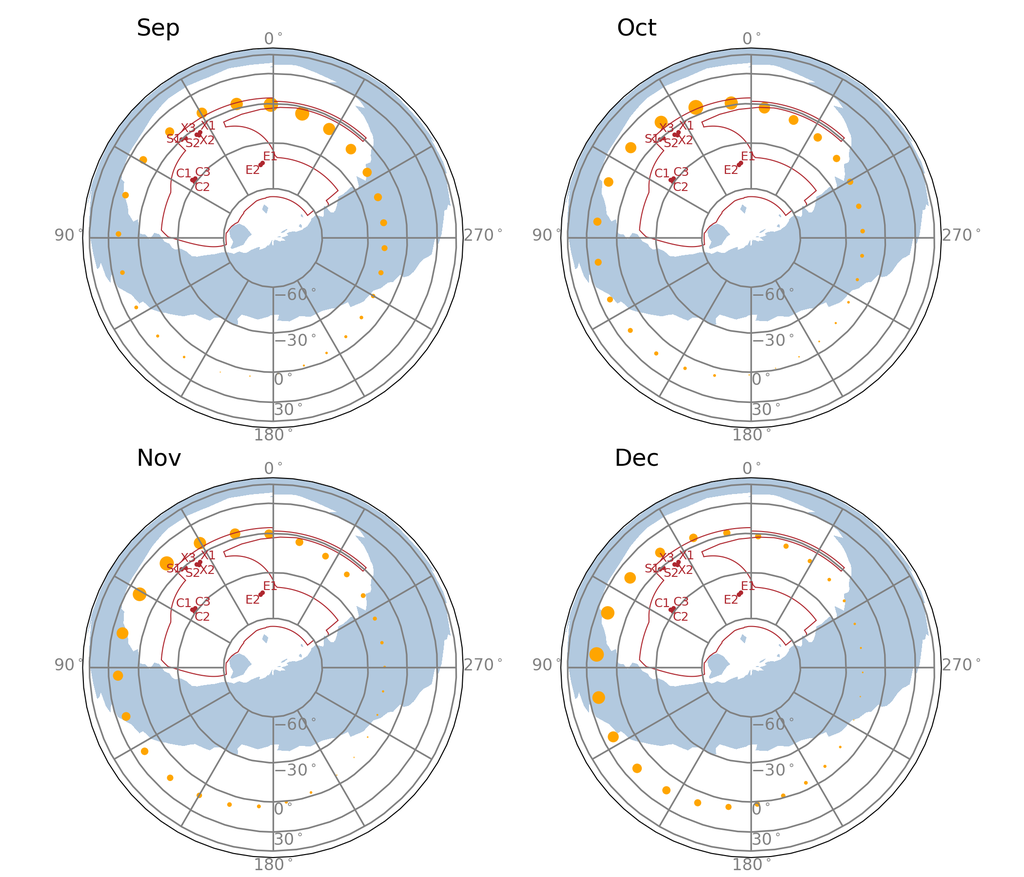}
\caption{\label{fig:moonpositionmap}
Positions and phase of the moon on Lambert azimuthal equal area projection, for four months in 2016. Orange points designate the positions of the moon for the lunations on these months. The sizes of the points indicate the phase, with the largest point representing a full moon. The blue shaded area shows stellar density (stars in 2MASS \citep{skrutskie_two_2006} with $m_J < 16$, mostly from the Milky Way) for context. The red line shows the ultimate DES wide-survey footprint, and red dots show the time-domain (SN survey) pointings. The spacing of the orange points indicates the motion of the moon from one night to the next, with an {\sc r.a.} which increases by $\sim13\degree$ per day (counterclockwise in this projection). 
}
\end{figure*}

The collaboration selected fields for monitoring based on several criteria:
\begin{enumerate}
 \item the field must be observable from CTIO during the same observing season used for the DES wide survey (August to February);
 \item fields visible from the northern hemisphere for spectroscopic followup were preferred;
 \item galaxies within the fields must be included in available historical surveys;
 \item fields that fall within the footprint of the VISTA survey \citep{emerson_visible_2004} were preferred;
 \item fields with bright stars were avoided; and
 \item fields with low Galactic extinction were preferred.
\end{enumerate}
Table~\ref{tab:snfields} and figure~\ref{fig:moonpositionmap} show the fields selected for monitoring, and tables~\ref{tab:snshallowseqexptimes} and~\ref{tab:sndeepseqexptimes} describe the details of the exposures for each sequence on each field.

\begin{table}
    \centering
    \begin{tabular}{lllll}
         DES fields &  R.A. & decl. & overlap field & reference \\
         \hline
         E1, E2 & $0.5\degree$ & $-43.0\degree$ & ELAIS S1 & \cite{oliver_european_2000}\\
         X1, X2, X3 & $34.5\degree$ & $-5.5\degree$ & XMM-LSS & \cite{pierre_xmm-lss_2004}\\
         C1, C2, C3 & $52.5\degree$ & $-27.5\degree$ & Chandra Deep Field-S & \cite{giacconi_first_2001}\\
         S1, S2 & $42\degree$ & $0\degree$ & SDSS Stripe 82 & \cite{stoughton_sloan_2002}
    \end{tabular}
    \caption{ \label{tab:snfields} The fields selected for DES time-domain survey observing.}
\end{table}

\begin{table}
\begin{center}
\begin{tabular}{llllllll}
sequence & filter & R.A. & decl. & exptime (s) & \# exposures & rise {\sc lst} & set {\sc lst}\\
\hline
X3 g & g & \(36.4495\degree\) & \(-4.6013\degree\) & 200 & 1 & \(-7\degree\) & \(80\degree\)\\
 &  & \(36.4500\degree\) & \(-4.6000\degree\) & 200 & 1 & \(-7\degree\) & \(80\degree\)\\
 &  & \(36.4487\degree\) & \(-4.6005\degree\) & 200 & 1 & \(-7\degree\) & \(80\degree\)\\
\hline
X3 r & r & \(36.4500\degree\) & \(-4.6000\degree\) & 400 & 1 & \(-7\degree\) & \(80\degree\)\\
 &  & \(36.4487\degree\) & \(-4.6005\degree\) & 400 & 1 & \(-7\degree\) & \(80\degree\)\\
 &  & \(36.4495\degree\) & \(-4.6013\degree\) & 400 & 1 & \(-7\degree\) & \(80\degree\)\\
\hline
X3 i & i & \(36.4495\degree\) & \(-4.6013\degree\) & 360 & 2 & \(-7\degree\) & \(80\degree\)\\
 &  & \(36.4500\degree\) & \(-4.6000\degree\) & 360 & 2 & \(-7\degree\) & \(80\degree\)\\
 &  & \(36.4487\degree\) & \(-4.6005\degree\) & 360 & 1 & \(-7\degree\) & \(80\degree\)\\
\hline
X3 z & z & \(36.4487\degree\) & \(-4.6005\degree\) & 330 & 3 & \(-7\degree\) & \(80\degree\)\\
 &  & \(36.4500\degree\) & \(-4.6000\degree\) & 330 & 4 & \(-7\degree\) & \(80\degree\)\\
 &  & \(36.4495\degree\) & \(-4.6013\degree\) & 330 & 4 & \(-7\degree\) & \(80\degree\)\\
\hline
C3 g & g & \(52.6479\degree\) & \(-28.1013\degree\) & 200 & 1 & \(-3\degree\) & \(108\degree\)\\
 &  & \(52.6469\degree\) & \(-28.1005\degree\) & 200 & 1 & \(-3\degree\) & \(108\degree\)\\
 &  & \(52.6484\degree\) & \(-28.1000\degree\) & 200 & 1 & \(-3\degree\) & \(108\degree\)\\
\hline
C3 r & r & \(52.6479\degree\) & \(-28.1013\degree\) & 400 & 1 & \(-3\degree\) & \(108\degree\)\\
 &  & \(52.6469\degree\) & \(-28.1005\degree\) & 400 & 1 & \(-3\degree\) & \(108\degree\)\\
 &  & \(52.6484\degree\) & \(-28.1000\degree\) & 400 & 1 & \(-3\degree\) & \(108\degree\)\\
\hline
C3 i & i & \(52.6469\degree\) & \(-28.1005\degree\) & 360 & 1 & \(-3\degree\) & \(108\degree\)\\
 &  & \(52.6484\degree\) & \(-28.1000\degree\) & 360 & 2 & \(-3\degree\) & \(108\degree\)\\
 &  & \(52.6479\degree\) & \(-28.1013\degree\) & 360 & 2 & \(-3\degree\) & \(108\degree\)\\
\hline
C3 z & z & \(52.6469\degree\) & \(-28.1005\degree\) & 330 & 3 & \(-3\degree\) & \(108\degree\)\\
 &  & \(52.6479\degree\) & \(-28.1013\degree\) & 330 & 4 & \(-3\degree\) & \(108\degree\)\\
 &  & \(52.6484\degree\) & \(-28.1000\degree\) & 330 & 4 & \(-3\degree\) & \(108\degree\)\\
\end{tabular}
\end{center}
\caption{\label{tab:sndeepseqexps}
Exposure parameters for deep sequences. Rise and set sidereal times ({\sc lst}s) are for rising above and falling below 1.5 airmasses, and show when each sequence may be observed; see figure~\ref{fig:nightsvslst}. Each sequence defines a set of exposures taken as a single block, and processed together. The trio of pointings prevent the same object from appearing on the same pixels in all exposures. 
}
\end{table}

\begin{table}
\begin{center}
\begin{tabular}{llllllll}
sequence & filter & R.A. & decl. & exptime (s) & rise {\sc lst} & set {\sc lst}\\
\hline
SN E1 & g & \(7.8744\degree\) & \(-43.0096\degree\) & 175 & \(-51\degree\) & \(67\degree\)\\
 & r & \(7.8744\degree\) & \(-43.0096\degree\) & 150 & \(-51\degree\) & \(67\degree\)\\
 & i & \(7.8744\degree\) & \(-43.0096\degree\) & 200 & \(-51\degree\) & \(67\degree\)\\
 & z & \(7.8744\degree\) & \(-43.0096\degree\) & 200 & \(-51\degree\) & \(67\degree\)\\
 & z & \(7.8738\degree\) & \(-43.0109\degree\) & 200 & \(-51\degree\) & \(67\degree\)\\
\hline
SN E2 & g & \(9.5000\degree\) & \(-43.9980\degree\) & 175 & \(-50\degree\) & \(69\degree\)\\
 & r & \(9.5000\degree\) & \(-43.9980\degree\) & 150 & \(-50\degree\) & \(69\degree\)\\
 & i & \(9.5000\degree\) & \(-43.9980\degree\) & 200  & \(-50\degree\) & \(69\degree\)\\
 & z & \(9.5000\degree\) & \(-43.9980\degree\) & 200  & \(-50\degree\) & \(69\degree\)\\
 & z & \(9.4993\degree\) & \(-43.9993\degree\) & 200  & \(-50\degree\) & \(69\degree\)\\
\hline
SN X1 & g & \(34.4757\degree\) & \(-4.9295\degree\) & 175 & \(-9\degree\) & \(78\degree\)\\
 & r & \(34.4757\degree\) & \(-4.9295\degree\) & 150 & \(-9\degree\) & \(78\degree\)\\
 & i & \(34.4757\degree\) & \(-4.9295\degree\) & 200  & \(-9\degree\) & \(78\degree\)\\
 & z & \(34.4757\degree\) & \(-4.9295\degree\) & 200  & \(-9\degree\) & \(78\degree\)\\
 & z & \(34.4752\degree\) & \(-4.9308\degree\) & 200  & \(-9\degree\) & \(78\degree\)\\
\hline
SN X2 & g & \(35.6645\degree\) & \(-6.4121\degree\) & 175  & \(-9\degree\) & \(80\degree\)\\
 & r & \(35.6645\degree\) & \(-6.4121\degree\) & 150  & \(-9\degree\) & \(80\degree\)\\
 & i & \(35.6645\degree\) & \(-6.4121\degree\) & 200  & \(-9\degree\) & \(80\degree\)\\
 & z & \(35.6640\degree\) & \(-6.4134\degree\) & 200  & \(-9\degree\) & \(80\degree\)\\
 & z & \(35.6645\degree\) & \(-6.4121\degree\) & 200  & \(-9\degree\) & \(80\degree\)\\
\hline
SN C1 & g & \(54.2743\degree\) & \(-27.1116\degree\) & 175  & \(-1\degree\) & \(110\degree\)\\
 & r & \(54.2743\degree\) & \(-27.1116\degree\) & 150  & \(-1\degree\) & \(110\degree\)\\
 & i & \(54.2743\degree\) & \(-27.1116\degree\) & 200  & \(-1\degree\) & \(110\degree\)\\
 & z & \(54.2738\degree\) & \(-27.1129\degree\) & 200  & \(-1\degree\) & \(110\degree\)\\
 & z & \(54.2743\degree\) & \(-27.1116\degree\) & 200  & \(-1\degree\) & \(110\degree\)\\
\hline
SN C2 & g & \(54.2743\degree\) & \(-29.0884\degree\) & 175  & \(-2\degree\) & \(110\degree\)\\
 & r & \(54.2743\degree\) & \(-29.0884\degree\) & 150 & \(-2\degree\) & \(110\degree\)\\
 & i & \(54.2743\degree\) & \(-29.0884\degree\) & 200  & \(-2\degree\) & \(110\degree\)\\
 & z & \(54.2738\degree\) & \(-29.0897\degree\) & 200  & \(-2\degree\) & \(110\degree\)\\
 & z & \(54.2743\degree\) & \(-29.0884\degree\) & 200 & \(-2\degree\) & \(110\degree\)\\
\hline
SN S1 & g & \(42.8200\degree\) & \(0.0000\degree\) & 175 & \(3\degree\) & \(82\degree\)\\
 & r & \(42.8200\degree\) & \(0.0000\degree\) & 150  & \(3\degree\) & \(82\degree\)\\
 & i & \(42.8200\degree\) & \(0.0000\degree\) & 200  & \(3\degree\) & \(82\degree\)\\
 & z & \(42.8195\degree\) & \(-0.0013\degree\) & 200  & \(3\degree\) & \(82\degree\)\\
 & z & \(42.8200\degree\) & \(0.0000\degree\) & 200  & \(3\degree\) & \(82\degree\)\\
\hline
SN S2 & g & \(41.1944\degree\) & \(-0.9884\degree\) & 175 & \(1\degree\) & \(82\degree\)\\
 & r & \(41.1944\degree\) & \(-0.9884\degree\) & 150  & \(1\degree\) & \(82\degree\)\\
 & i & \(41.1944\degree\) & \(-0.9884\degree\) & 200  & \(1\degree\) & \(82\degree\)\\
 & z & \(41.1944\degree\) & \(-0.9884\degree\) & 200  & \(1\degree\) & \(82\degree\)\\
 & z & \(41.1939\degree\) & \(-0.9897\degree\) & 200  & \(1\degree\) & \(82\degree\)\\
\end{tabular}
\end{center}
\caption{\label{tab:snshallowseqexps}
Exposure parameters for shallow sequences. Rise and set local sidereal times ({\sc lst}s) are for falling below and rising above 1.5 airmasses.
}
\end{table}

The placement of the S and X fields near the celestial equator presented special challenges to observing these fields on a regular cadence.\footnote{In retrospect, stronger guidance on field placement should have been given to the SN working group.} The ecliptic crosses the equator within the DES footprint, and the moon's orbit is only inclined $\sim5\degree$ from that of the ecliptic, so fields near the celestial equator are close to the path of the moon on the sky. In a window of five nights of centered on the date of the full moon, Rayleigh scattering of moonlight fills the sky for a significant fraction of the night, making observing inefficient or even futile (see appendix~\ref{skybrightness}). Such conditions affect both the time-domain and wide survey, and the collaboration requested schedules that avoid this time.\footnote{This is good time for the observatory to schedule engineering and maintenance.} This avoidance of time severely degraded by the moon results in a five day gap each month. Even on nights other than these severely affected five, observing a field when it is close to the moon on the sky is often inefficient or futile, and avoiding observing a field when it is close to the moon can introduce a gap of two or three days as well. During October and November, the full moon occurs when the moon is near the S and X fields, so these two gaps overlap. Provided the weather is reasonable for one of the few days on either side of the five day gap around full moon, the target seven day cadence can be maintained. In September and December, on the other hand, these two gaps can be adjacent, which can make it difficult to maintain the cadence even when the weather is good. Furthermore, the S and X fields are only at an observable airmass for half of nights in September and October, further limiting opportunities to maintain the cadence. So, while detailed scheduling of time-domain sequences was generally performed automatically by {\tt obstac}, careful hand-tuned intervention near the full moons of September and December was sometimes required, and even then long gaps in the cadence were sometimes created.

\section{Wide survey strategy}
\label{widestrat}
\subsection{Objectives and requirements}
\label{wideobjectives}

Three of the four DES dark energy probes are being measured from data from the wide survey. Each of these three probes place requirements on the wide survey data \citep{annis_dark_2010}: 

\begin{enumerate}
    \item \label{req:photozphot} All three wide survey dark energy probes rely on photometry in $g$, $r$, $i$, and $z$ bands to calculate redshift estimates (photo-z's). Photometric precision at a given depth (or, equivalently, limiting magnitude at a given photometric precision) depends on the total exposure time, reddening by dust in the Milky Way, and observing conditions such as sky brightness and point spread function (PSF) width. Both sky brightness and the width of the PSF depend on the zenith distance (which maps to airmass -- see appendix~\ref{airmass}) and (statistically) the date (time of year and phase and coordinates of the moon).
    \item \label{req:photoztrain} Photo-z redshift estimates rely on training using large samples of galaxies with both spectroscopic redshift measurements and photometry in DES bands, at similar depths to DES images.
    \item \label{req:numgals} All three probes improve with a greater numbers of detected galaxies. Although both deeper images (more time spent on any given area of the sky) and greater footprint increase the number of galaxies detected, the number of galaxies detected increases faster with footprint area than depth, such that a larger footprint area improves this factor even at the expense of depth. Confusion by stars also affects the number of galaxies that can be measured: where there is a greater density of local stars (from the Milky Way and Magellanic Clouds), a greater fraction of galaxies will fall behind such stars and therefore appear as ``blended'' objects in the images. Neither the photometery nor the shape of such blended objects can be measured at the precision possible with otherwise similar unblended ones.
    \item \label{req:cosmicvar} All three probes are limited by cosmic variance, the variation between different parts of the universe due to large scale structure and statistics. The effect of cosmic variance is inversely proportional to survey area, motivating a large footprint.
    \item \label{req:angscale} Measurement of cosmological parameters using large scale structure relies on measuring correlations across different angular scales.
    \item \label{req:psfstat} Weak lensing relies on measurements of ellipticities of galaxies. The PSF of images strongly affects the precision of such ellipticity measurements, and the weak lensing probe therefore relies on images with sharp PSFs. This drives a strategy that takes images at low airmass and in good seeing conditions (even beyond such preferences introduced by the need for good photometry).
    \item \label{req:psfsyst} Weak lensing also relies on precise modeling of the PSF. Errors in PSF models can be correlated across different exposures in a night due to slow telescope and optics variation, and different exposures on the same location in the focal plane. To average PSF model errors across many independent instances, different detections of the same object in the same filter should be made on different nights and on different locations in the focal plane.
    \item \label{req:photocalib} The ``ubercal'' method for photometric calibration relies on having a contiguous footprint, and is more effective when there are many overlapping images, thus many independent observations of points but on differing positions on the focal plane (\cite{padmanabhan_improved_2008, tucker_photometric_2007}; see also \cite{burke_forward_2017}).%
\end{enumerate}

In addition to the factors driven directly by science requirements, there are several requirements arising from agreements with other institutions \citep{annis_dark_2010}:
\begin{enumerate}[resume]
    \item \label{req:noaoagreement} The Memorandum of Understanding (MOU) between Fermilab, NCSA, and NOAO describing the Dark Energy Survey allocated it 525 nights of observing using DECam on the Blanco telescope, spread over five years, with an even mixture of bright or gray (moon above the horizon) and dark (moon below the horizon) observing time.
    \item \label{req:sptoverlap} The DES collaboration and the South Pole Telescope (SPT) survey agreed to share cluster data on overlapping footprint, providing DES with cluster mass estimates using the Sunyaev-Zeldovich effect measured in SPT data, and providing the SPT collaboration with DES cluster photo-z and weak lensing mass estimates. This drove the DES collaboration to design a footprint with as much overlap as possible with the SPT footprint.
    \item \label{req:vhs} Combining photometry from DES $g$, $r$, $i$, and $z$ band images with deep photometry in $J$, $H$, and $K$ bands can significantly improve photo-z measurements. The DES collaboration therefore entered into an agreement with the VISTA Hemisphere Survey (VHS), under which VISTA takes deeper exposures in $J$, $H$, and $K$ in exchange for DES providing data in $Y$ band, introducing the requirement that DES include the $Y$ band in the survey.
\end{enumerate}

\subsection{Survey area and depth}
\label{areaanddepth}

Estimating the DETF or other direct dark energy related figures of merit from survey design parameters was impractical. The quality of cosmological measurement clearly improves with improvements in several parameters directly related to survey design and quality. Statistical limitations in these measurements improve with the number of galaxies measured, which scales with the survey area and depth: these are the primary factors which limits the statistical uncertainty in the cosmology.\footnote{There are other less prominent factors as well. Sharper images (smaller PSFs), for example, improve the precision of ellipticity measurements, affecting the measurement of weak lensing more strongly than would be indicated by its affect on limiting magnitude alone.}\textsuperscript{,}\footnote{Measurement of cosmological parameters by most probes begins by selecting a subset of the area covered by the survey that meets some minimal quality threshold, the primary element of which is the depth of the image. Over the course of the survey, the DES collaboration developed a rough mapping between the depth threshold used, the area that above that threshold, and the DETF figure of merit. %
}
Given 525 nights of allocated observing time, a balance must be made between these two features: a larger area means less time spent per unit area and therefore reduced depth, leading to less precise photometric redshifts. Simulations \citep{cuhna_survey_2010, cuhna_survey_2011, cuhna_survey_2011-1} indicated that the DETF figure of merit for the survey as a whole would be maximized by a total footprint of $\sim5000$ sq. deg. and exposure times evenly divided among the different filters. Although a larger area or more emphasis on specific filters is useful for increasing the total number of objects detected or area covered, this benefit was counterbalanced by uncertainties in photometric redshifts introduced by such changes.

After the establishment of the survey footprint area, the science of the survey could be maximized by optimizing the depth over that footprint. Astronomers traditionally quantify the ``depth'' of an image as the limiting magnitude: the magnitude at which the flux of point source can be measured at better than a given signal to noise ratio (or, alternately, detected with some reference completeness and contamination). When the photometric uncertainty is dominated by photon statistics from the sky background,\footnote{foreground, really} the limiting magnitude varies as
\begin{equation}
m_{\mbox{\scriptsize lim}} = m_{0,\mbox{circ}} + 1.25  \log{(\mbox{\sc exptime})}
\end{equation}
where $m_{0,\mbox{circ}}$ depends on observing conditions. Instead of using the limiting magnitude, the DES collaboration often uses an alternate quantity, $\tau$,\footnote{Within the DES collaboration, $\tau$ is colloquially referred to as ``$t_{\mbox{\scriptsize eff}}$'' after ``effective exposure time,'' although conceptually the effective exposure time is really $\tau \times \mbox{\sc exptime}$.} defined such that:
\begin{eqnarray}
\tau & = & 10^{\frac{4}{5}(m_{0} - m_{0,\mbox{\scriptsize circ}})} \\
     & = & \eta^2 \left( \frac{0.9"}{\mbox{\sc fwhm}} \right)^2 \left( \frac{b_{\mbox{\scriptsize dark}}}{b} \right)
\end{eqnarray}
in which $m_{0}$ is constant for all observing conditions, {\sc fwhm} is the PSF full width at half maximum in arcseconds, $\eta$ is the atmospheric transmission, $b$ is the sky brightness in the image, and $b_{\mbox{\scriptsize dark}}$ is a reference sky brightness (defined as a value typical for zenith under dark conditions). The magnitude limit of a coadded image is then related to the $\tau$ of each of the contributing images as
\begin{equation}
m_{\mbox{\scriptsize lim}} = m_{0} + 1.25 \log\left(\sum_{i} \tau_{i} \times \mbox{\sc exptime}_{i}\right),
\end{equation}
or, when the exposure times of all contributing images are the some (which is always the case for DES wide survey {\it g, r, i} and {\it z} images),
\begin{equation}
m_{\mbox{\scriptsize lim}} = m_{0, \mbox{\scriptsize\sc exptime}} + 1.25 \log\left(\sum_{i} \tau_{i} \right).
\end{equation}
Appendix~\ref{dataquality} provides more detail on $\tau$, including estimation of its variation with airmass, hour angle, moon phase and position, and time in twilight.

\subsection{Scheduling and the observable sky}
\label{observablesky}
\begin{figure*}
\centering
\includegraphics[width=0.99\linewidth]{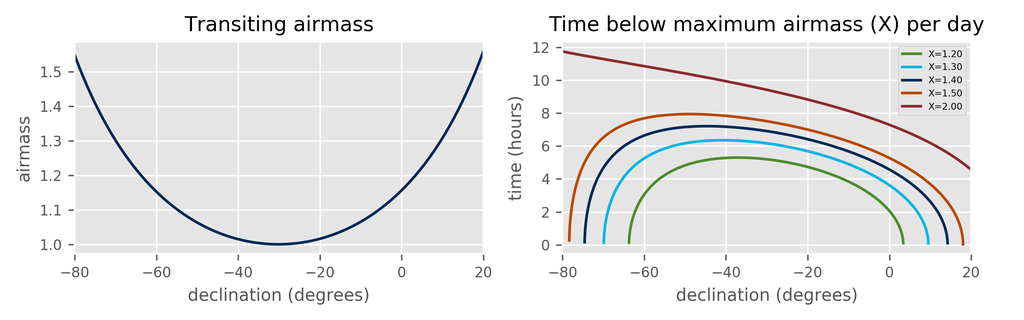}
\caption{\label{fig:txvsdecl}
The left plot shows the minimum airmass a point at a given airmass ever takes at CTIO (when it transits).  The right plot shows the time during which a pointing at a given declination has an airmass below a given limit at CTIO, as a function of declination. }
\end{figure*}

The quality of exposures depends critically on the zenith distance (see appendixes~\ref{dataquality} and~\ref{airmass}). The angle between zenith and either celestial equatorial pole is determined by the latitude of the observatory: the angle between the axis of the Earths rotation (which defines the celestial equatorial poles) and the zenith from a given point on the Earth is $90\degree - \phi$, where $\phi$ is the latitude of the observatory. The area of the sky limited by a given zenith distance (and therefore airmass) is a spherical cap centered on the zenith. With each rotation of the Earth (sidereal day), the zenith completes a circuit in a small circle around the south celestial pole with an angular radius of of $90\degree + \phi$. 

The latitude of the observatory therefore sets limits on the northern and southern extents of candidate survey footprints. The left-hand plot of figure~\ref{fig:txvsdecl} shows the minimum airmass reached by pointings as a function of their declination. Note that the minimum airmass is symmetric about the the point at which the declination $\delta$ of the pointing equals the latitude $\phi$ of the observatory. Furthermore, the time over which any given pointing remains below a given airmass limit (if it falls below that limit at all) varies with declination as well, and is neither symmetric, nor reaches a peak at $\delta=\phi$. See the right plot of figure~\ref{fig:txvsdecl}. Pointings not on the celestial equator move along small circles on the celestial sphere, not great circles, and so move along curved paths (relative to great circles) with angular velocities that vary with declination. So, pointings near the south pole move through the spherical cap defined by the airmass limit along more curved path and with a lower angular velocity than pointings near the celestial equator.

These considerations place natural limits on survey footprint area for a given site. Footprint area which never reaches low airmass should be avoided, because reaching an acceptable depth in this area will either require disproportionate observing time, or simply be impossible. Footprint area which remains at an acceptable airmass for only limited amounts of time should not be automatically excluded. Such area needs to be approached with care, however, because it must be observed in specific time windows, and so imposes observing and scheduling constraints and increases vulnerability to variations in weather.

\begin{figure*}
\centering
\includegraphics[height=0.4\textheight]{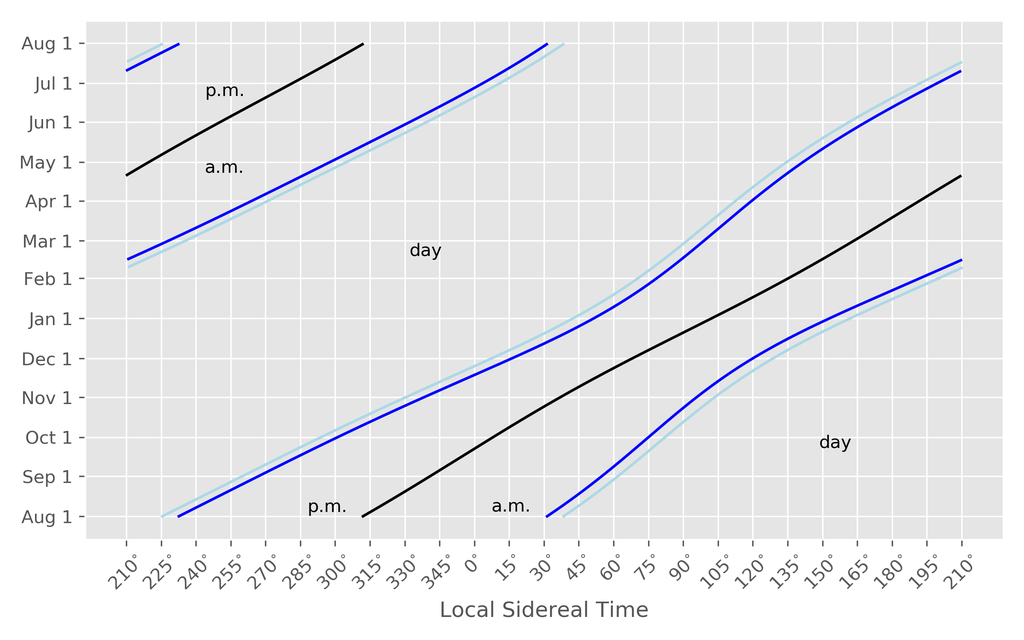}
\caption{\label{fig:nightsvslst}
The sidereal ``hour glass'' plot for Cerro Tololo. The black line shows the local sidereal time (LST) of midnight over the course of a year, the dark blue line astronomical ($18\degree$) twilight, the light blue line nautical ($12\degree$) twilight. So, the areas marked ``p.m.'' show the LSTs covered by first half-nights over the course of a year, and ``a.m.'' by second half nights. Accessible sky corresponding to each sidereal time can be read from figure~\ref{fig:airmasslimits}.
}
\end{figure*}

\begin{figure*}
\centering
\includegraphics[height=0.80\textheight]{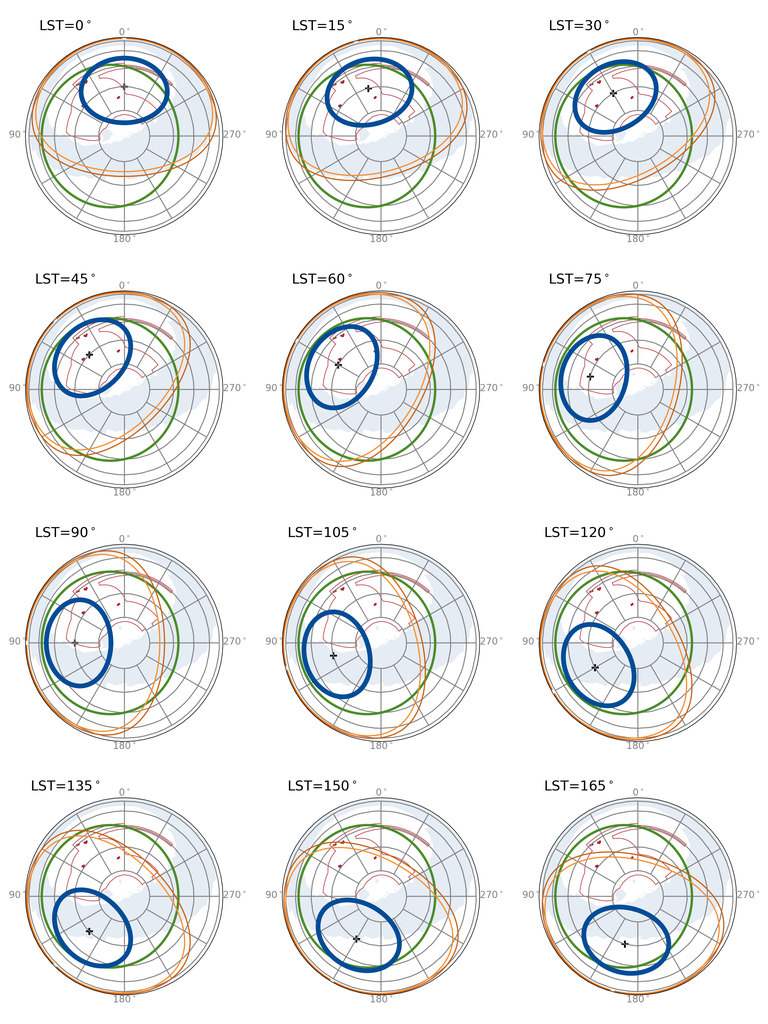}
\caption{\label{fig:airmasslimits}
The blue oval marks area accessible within a 1.4 airmass limit on a Lambert azimuthal equal area projection, for different local sidereal times (at one hour intervals). The blue shaded area shows regions of high stellar density (stars in 2MASS \citep{skrutskie_two_2006} with $m_J < 16$, mostly from the Milky Way) for context. 
The green circle shows the ecliptic: the path of the apparent location of the sun, which it traverses counter-clockwise once a year. The orange lines show the limits of nautical and astronomical twilight: when the sun falls within the inner orange circle, it is day, and optical observing is not possible during that LST. When it is outside the outer orange circle, it is fully dark. The red line shows the ultimate DES wide-survey footprint, and red dots show the time-domain (SN survey) pointings. The sidereal time at the start and end of a given half night can be read from figure~\ref{fig:nightsvslst}. A planisphere, a mechanical cardboard version of this plot, was routinely used as an aid for human understanding of survey strategy.
}
\end{figure*}

\begin{figure*}
\centering
\includegraphics[height=0.80\textheight]{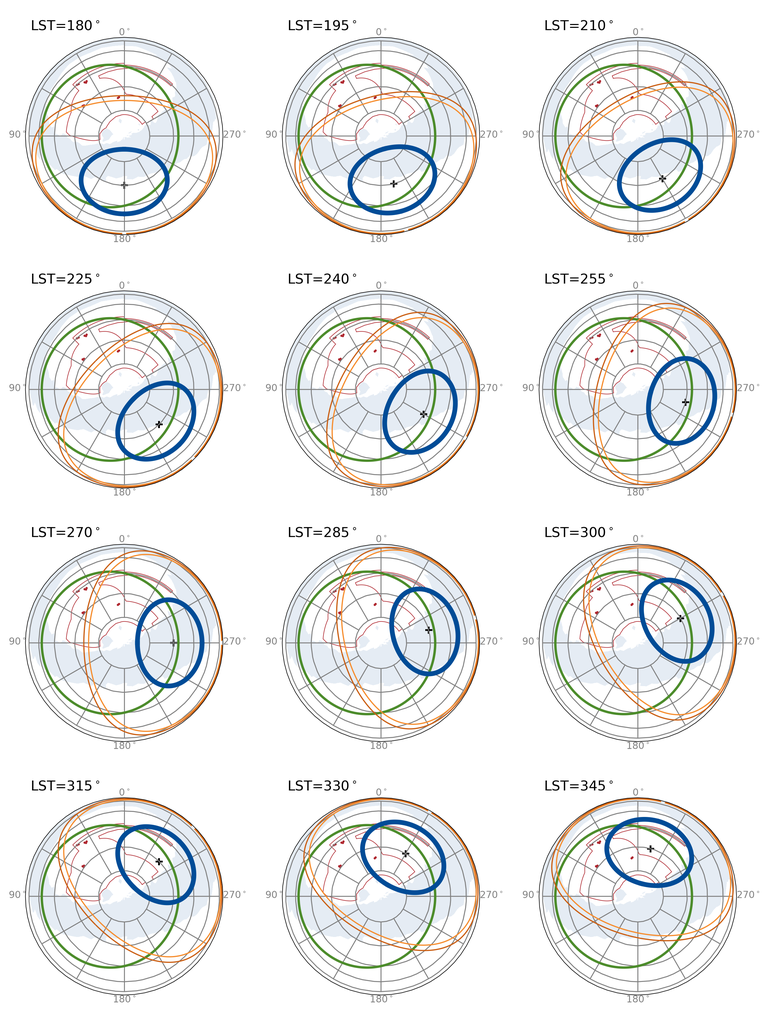}
\caption{\label{fig:airmasslimitscontd} A continuation of figure~\ref{fig:airmasslimits}, for later sidereal times.
}
\end{figure*}

While the declination of zenith remains constant, constraining it to remain along a constant small circle, the {\sc r.a.} position of zenith within that cone depends on the rotation angle of the Earth and the longitude of the observatory. The zenith sweeps through the entire range of {\sc r.a.} over the course of one rotation of the Earth (approximately\footnote{"Approximately" because a day is defined by the rotation angle of the Earth with respect to the Sun, the apparent position of which moves as the Earth orbits the Sun, adding one complete rotation per year. So, a sidereal day (the period of rotation of the earth) is $\frac{365.24}{1+365.24}=0.99727$ solar days, about 4 minutes short of one mean solar day.} one day). The Greenwich sidereal time (GST) is defined to be the {\sc r.a.} of the zenith at a longitude of $0\degree$, and represents the position angle of the Earth. The local sidereal time (LST), the {\sc r.a.} of the current zenith, is offset from the GST by the longitude of the observatory. The blue ovals in figure~\ref{fig:airmasslimits} show the 1.4 airmass limits for a sequence of sidereal times, with the high stellar density area of the sky (from Galactic plane) shown for reference.

Although the zenith passes through all sidereal times over the course of one day, only a fraction of that time is useful for optical observing: the rest is during the day. Proper handling of the position of the Sun is involved and best left to standard libraries, but very rough approximations adequate for determining which parts of the sky are practically observable on which nights are straightforward. The day time is determined by the apparent position of the Sun, which has an {\sc r.a.} of $0\degree$ at the vernal equinox (about March 20). It makes one complete rotation through {\sc r.a.} per year, so its {\sc r.a.} increases by roughly $\frac{360\degree}{365.24 \mbox{~days}} \sim 1\degree/\mbox{day}$.\footnote{The actual orbit of the Earth about the Sun is elliptical and in a plane at an angle with the celestial equator, so this is only an approximation, but it is adequate for estimating which areas of the footprint can be observed by a given observing schedule.}\textsuperscript{,}\footnote{This angular variation corresponds to a temporal variation of $4 \mbox{~minutes} / \mbox{day}$: events that occur at a constant sidereal time each day (such as a given pointing rising or setting) happen 4 minutes earlier each day in solar time.} The {\sc r.a.} of the Sun is the LST of noon, so solar midnight (the middle of a night of observing) is $12 \mbox{~hours}=180\degree$ away: the {\sc lst} of the middle of a night of observing is $180\degree$ at about March 20, $270\degree$ on the summer solstice (about June 21), $0\degree$ on the autumnal equinox (about September 21), and $90\degree$ on the winter solstice (about December 21). The black line of figure~\ref{fig:nightsvslst} shows the solar midnight, calculated to much greater precision, as a function of the date of the year. 

The sidereal times of night, day, and twilight can also be estimated using figure~\ref{fig:airmasslimits}. The green circle marks the ecliptic: the path the Sun takes over the course of a year. The ecliptic is offset from the celestial equator, such that the center of the green circle in the plots is slightly to the right of the south pole, because the ecliptic is at an angle with the celestial equator, crossing at $0\degree$ and $180\degree$ (by construction). On the vernal equinox (about March 20), the Sun is at $\mbox{\sc r.a.}=0\degree$, and moves counterclockwise along the ecliptic, completing one full circuit per year, or about one radial ({\sc r.a.}) graticule on the plot per month. The orange lines in figure~\ref{fig:airmasslimits} mark the boundaries of nautical and astronomical twilight: when the Sun falls within the inner orange circle, it is day, and optical observing is not possible during that {\sc lst}. When it is outside the outer orange circle, it is fully dark.\footnote{Note that the seasonal variation of the duration of the night can be read from these maps. Because the center of the circle representing the ecliptic is offset from the pole, some points along the ecliptic fall within the orange (twilight limit) lines for a higher fraction of sidereal times than others. Therefore, when the Sun is on these locations on the ecliptic, the day time is longer, and the date is closer to the summer solstice.} The sidereal times of sunset and sunrise for a given date can also be read form figure~\ref{fig:nightsvslst}.

\begin{figure*}
\centering
\includegraphics[width=0.9\linewidth]{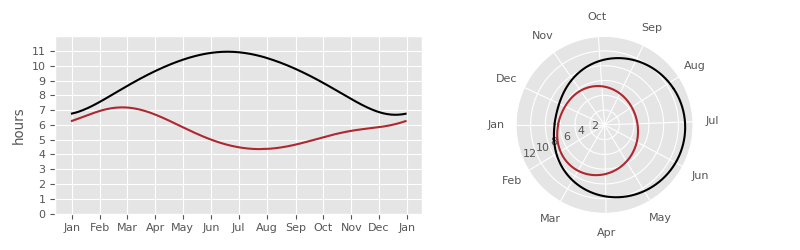}
\caption{\label{fig:tausvsdoy}
These figures show the total duration (black) and total effective duration (red) per night as a function of the time of year, where the effective duration is the product of the duration of the night and the fits to the cloud and seeing contributions to $\tau$. The angular axis in the right-hand (polar) plot corresponds to the transiting R.A. (LST) of midnight of each night on the maps given in figures~\ref{fig:galaxymap},~\ref{fig:moonpositionmap}, and~\ref{fig:nightsvslst}.
}
\end{figure*}

\begin{figure*}
\centering
\includegraphics[width=0.9\linewidth]{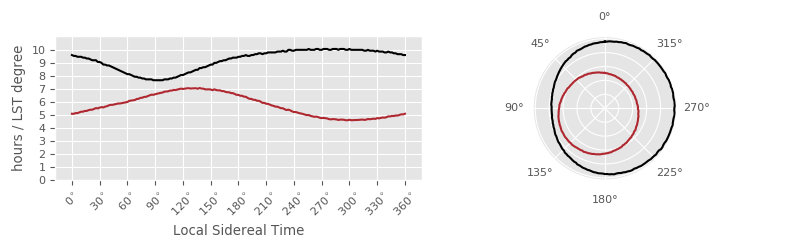}
\caption{\label{fig:teffvslst}
Each night covers a range of sidereal times (shown in figure~\ref{fig:nightsvslst}), so weather affecting one night affects a range of sidereal times, not just the sidereal time of midnight (shown in figure~\ref{fig:tausvsdoy}). The black lines above show the total time at each sidereal degree. Assuming one observes every night in a year, night durations and weather conditions combine to give a total effective time available for observing at any given sidereal time, shown in red. Again, angles in the right-hand (polar) plot correspond to transiting R.A. in figures~\ref{fig:galaxymap},~\ref{fig:moonpositionmap}, and~\ref{fig:nightsvslst}. Survey footprints centered on R.A. values corresponding to sidereal times with more effective hours per LST degree are easier to schedule and complete.
}
\end{figure*}

Because the accessible sky (the local sidereal time during the night) varies by time of year, a survey's schedule must correspond to its footprint. The {\sc r.a.} distribution of pointings in the survey footprint should approximate the effective {\sc lst} distribution provided by the schedule. The effective {\sc lst} distribution is not quite equal to the true calendar distribution, however: the mean accumulated $\sum \tau \times \mbox{\sc exptime}$ for a given night varies because weather conditions also vary by time of year. The seasonal variation of weather conditions therefore places practical constraints on the {\sc r.a.} distribution of a survey footprint.\footnote{While it would be intuitive for the seasonal variation in the duration of the night to also be a major factor, it is much less of a factor than it naively appears. The local sidereal time ``lost'' in the spring and summer is not taken from the center of the night ($\sim0\degree < \mbox{\sc lst} < \sim180\degree$), but rather near the beginning and ending of these nights: $\sim-75\degree < \mbox{\sc lst} < \sim105\degree$ and $\sim105\degree < \mbox{\sc lst} < \sim165\degree$; while in the autumn and winter, the time ``gained'' is not in the center of the night ($\sim180\degree < \mbox{\sc lst} < \sim360\degree$), but near the beginning and ending of {\em these} nights: $\sim105\degree < \mbox{\sc lst} < \sim180\degree$ and $\sim-100\degree < \mbox{\sc lst} < \sim75\degree$. In total, much of the time at any given {\sc lst} ``lost'' in the spring and summer is counterbalanced by time ``gained'' in the autumn and winter. This can be seen by comparing the black line figure~\ref{fig:teffvslst} to that in figure~\ref{fig:tausvsdoy}, in which the former is much more uniform than the later.
}

\begin{figure*}
\centering
\includegraphics[width=0.9\linewidth]{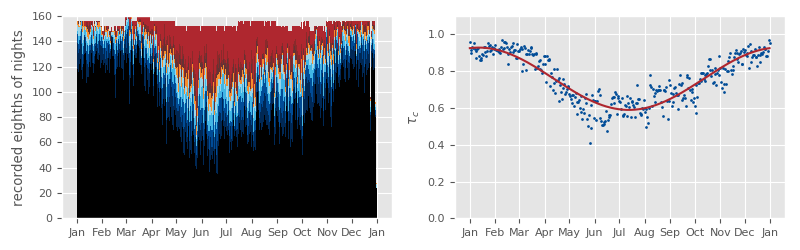}
\caption{\label{fig:cloudsvsdoy}
The left hand figure above shows the distribution of cloud levels for each day of the year, mapped from the eighths recorded by CTIO to estimates of usability: black corresponds to no clouds, shades of blue to \sfrac{1}{2} cloud cover (mostly to sometimes useful), orange and red from \sfrac{5}{8} to fully overcast (mostly useless). Not all nights reach the same overall height due to missing data. The right hand plot maps these values to extinction factors in $\tau$ following the mapping in \cite{neilsen_clouds_2015}, and shows a simple sinusoidal fit in red.
}
\end{figure*}

\begin{figure*}
\centering
\includegraphics[width=0.9\linewidth]{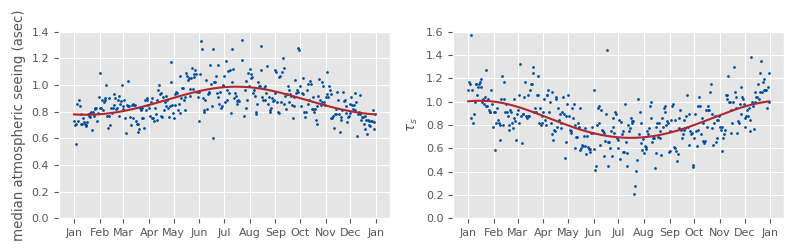}
\caption{\label{fig:seeingsvsdoy}
The left-hand plot above shows the median seeing on different nights as estimated by the DIMM from April, 2004 to March, 2010. The right-hand plot transforms these measurements into seeing factors in $\tau$ following equation~\ref{taudef}, assuming an instrumental contribution to the seeing of $0.45"$. The red line shows a simple sinusoidal fit.
}
\end{figure*}

Cloud cover and seeing both show strong seasonal variations at CTIO, and play a major role in determining available time. Cerro Tololo has collected cloud-cover records for quarter-nights since 1975 \citep{cerro_tololo_inter-american_observatory_ctio_2014}, as recorded by human observers at different telescopes at the site. \cite{neilsen_clouds_2015} describes an approximate mapping between this cloud cover and $\tau$ during the corresponding quarter night, using early DES data for which both these human recorded cloud levels and measured values of $\tau$ are available. Figure~\ref{fig:cloudsvsdoy} shows the distribution of cloud levels by the day of the year, and the corresponding variation in $\tau$ (with a simple sinusoidal fit).

A DIMM \citep{els_four_2009, els_results_2011} monitors the seeing at Cerro Tololo. Figure~\ref{fig:seeingsvsdoy} shows the median seeing (measured by the DIMM) as a function of the day of year, and the corresponding $\tau$ and sinusoidal fits. Figure~\ref{fig:tausvsdoy} combines the effects of seeing and clouds on $\tau$, showing the overall observing efficiency due to weather as a function of the day of the year. Each {\sc lst} can be observed on many different nights, and the overall efficiency of observing depends on the combination of the nights on which that {\sc lst} occurs at night and the efficiency of observing on those nights.  Figure~\ref{fig:teffvslst} combines these considerations, and shows the overall efficiency of observing by {\sc lst}, combining all nights.

In contrast with the accessibility restrictions on declination, these constraints on {\sc r.a.} do not place rigid constraints on the survey footprint: no area on the sky is strictly ruled out. However, they do have a significant effect on what depth and uniformity the survey can expect to attain for a given area and number of allocated calendar nights, and how sensitive the science of the survey is to deviations from strictly optimal allocations of nights: a sub-optimal distribution in footprint {\sc r.a.} has consequences for overall science quality and flexibility in the placement of the nights of observing allocated to DES. A survey footprint with more area observable at $\mbox{\sc lst}\sim135\degree$ and less at $\mbox{\sc lst}\sim315\degree$ will be higher quality and easier to schedule than the converse. The position of the Milky Way, however, prevents our selecting a footprint on this basis alone.

\subsection{Obstructing astronomical sources}
\label{obstructing}
\begin{figure*}
\centering
\includegraphics[width=0.9\linewidth]{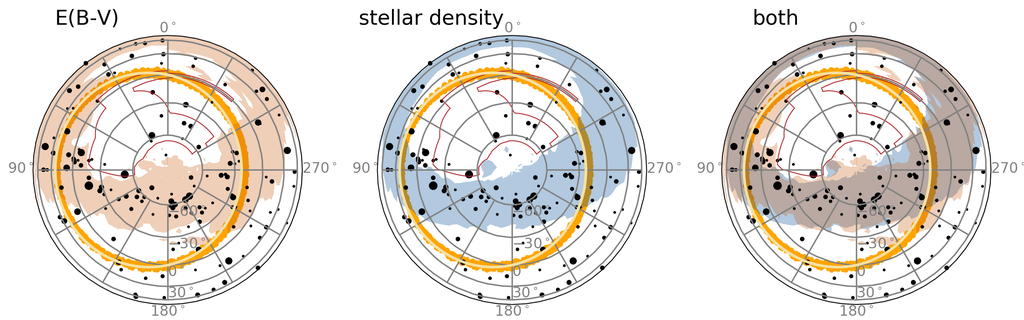}
\caption{\label{fig:galaxymap}
Astronomical obstacles to cosmological surveys of galaxies in Lambert azimuthal equal area projection centered on the south celestial pole. In all three sub-figures, black dots mark bright stars, orange dots mark the positions of the moon over the course of the survey, and a yellow line shows the ecliptic. In the left and right sub-figures, brown shading marks areas of high extinction from dust in the Milky Way (as measured by Planck \citep{planck_collaboration_planck_2014}). In the center and right sub-figures, blue shading marks area of high stellar crowding (stars in 2MASS \citep{skrutskie_two_2006} with $m_J < 16$). Note that the extinction and stellar density from the Milky Way are highly correlated, but do not match perfectly. The red line shows the ultimate DES footprint. This footprint avoids high extinction, stellar crowding, and proximity to the path of the moon, to the extent possible given the requirement that it include area along the celestial equator for overlap with other surveys.
}
\end{figure*}

The Earth (and therefore the observatory) sits in the plane of the Milky Way galaxy, which creates a band of stars and dust that divide the sky roughly in half, creating two separated areas with minimal obscuration by such material: the northern and southern Galactic caps. The red and blue shaded areas of figure~\ref{fig:galaxymap} mark the areas of high extinction due to dust (where flux from extra-Galactic sources, and therefore signal to noise, is reduced) and high stellar density, respectively. These areas largely coincide, but not perfectly.\footnote{For example, the area at an R.A. between $30\degree$ and $80\degree$ near the celestial equator has high extinction, but a reasonable stellar density, while the area near an R.A. of $280\degree$ and a declination between $-60\degree$ and $-40\degree$ has high stellar density but lower extinction.}

The northern Galactic cap, centered at roughly {\mbox{R.A}=193\degree, \mbox{Decl.}=+27\degree}, is mostly too far north to be well observed from Cerro Tololo, while the southern Galactic cap, centered at {\mbox{R.A}=13\degree, \mbox{Decl.}=-27\degree}, is well positioned. The importance of a large, contiguous survey area (items~\ref{req:numgals},~\ref{req:angscale} and~\ref{req:photocalib} in section~\ref{wideobjectives}) restricts the survey footprint to the southern Galactic cap, and the need to avoid areas of strong Galactic reddening and high stellar density (item~\ref{req:numgals}) set eastern and western limits on the wide survey footprint. The region of high Galactic extinction near the celestial equator between $\mbox{\sc R.A.}=40\degree$ and $90\degree$ sets an additional limit in the north for these values of R.A.

\begin{figure*}
\centering
\includegraphics[width=0.9\linewidth]{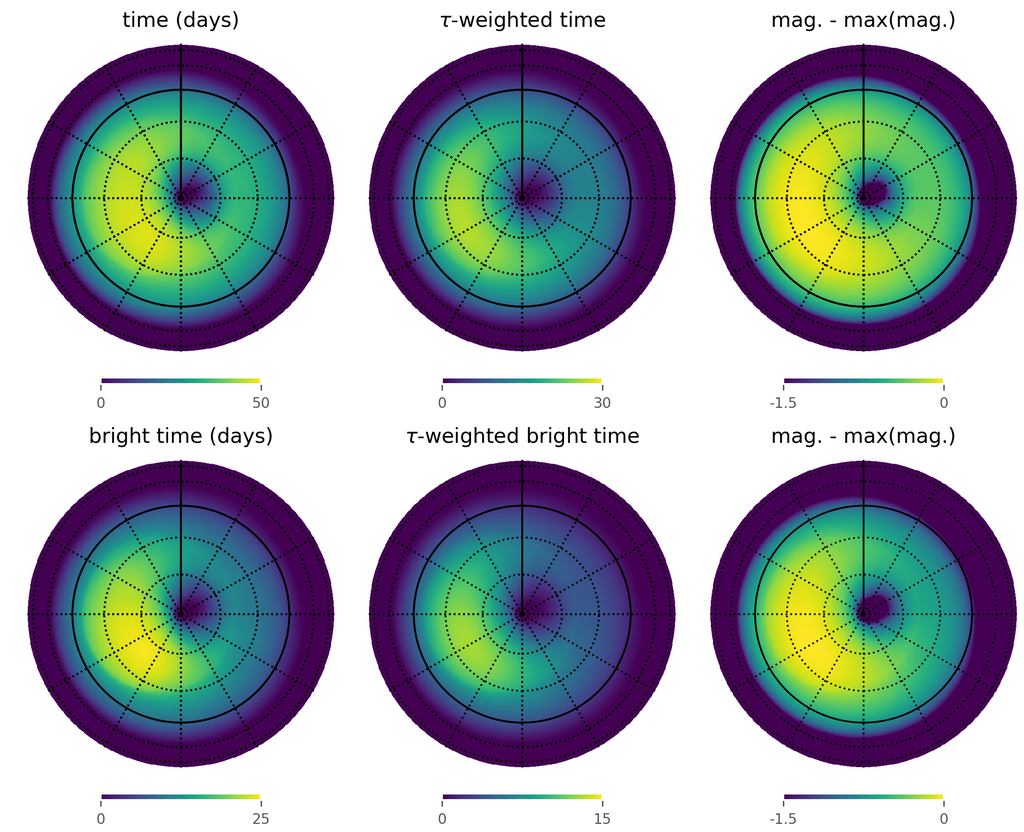}
\caption{\label{fig:taumap}
These maps show the integrated effects of night duration (see figure~\ref{fig:nightsvslst}), moon position (see figure~\ref{fig:galaxymap}), seasonal variation in seeing and clouds (see figures~\ref{fig:cloudsvsdoy},~\ref{fig:seeingsvsdoy},~\ref{fig:tausvsdoy}, and~\ref{fig:teffvslst}), and telescope pointing limits on the time available for observing over a single year
in {\it z} band.
The maps on the left show the total time (in days) during which that pointing could be observed (within the telescope limits, more than $30\degree$ from the moon, solar zenith distance of at least $102\degree$, and $\tau>0.3$) during the year. The central maps weight this time by $\tau$, and so represent the ``effective exposure time'' of observing that pointing at every possible moment during the year. The maps on the right convert this $\tau$ to differences in limiting magnitude. The top row includes all time during the year. To allow sufficient time for observing in {\it g}, {\it r}, and {\it i} in dark time in a schedule evenly split between moonlit and dark time, the {\it z} band must be observed almost exclusively when the moon is above the horizon. The lower plot therefore includes only ``bright'' time, excluding the dark time that was used for {\it g}, {\it r}, and {\it r}, and so better represents the available time for observing in {\it z}.
}
\end{figure*}

The moon, the positions of which are shown by the orange band in figure~\ref{fig:galaxymap}, presents an additional challenge. Proximity of the moon significantly increases the sky brightness (see appendix~\ref{scatteredmoonlight}). As described in the agreement with NOAO (item~\ref{req:noaoagreement}), roughly half of DES observing time is scheduled when the moon is up. Observing in the {\it g} and {\it r} filters is generally futile when the moon is up, so most of the dark time was be used to observe in these filters, and the redder filters ({\it i}, {\it z}, and {\it Y}) were observed mostly in bright time. Even in {\it i} and {\it z}, however, the sky brightness from the moon prevents efficient observing when observing within $30\degree$ of the moon (see figure~\ref{fig:skytvsmoonangle}). Footprint area within $30\degree$ of the moon's path is therefore challenging to observe. This effect can be seen in figure~\ref{fig:taumap}: both the total time and the integrated effective time when {\sc R.A.} is between $0\degree$ and $180\degree$ (where the moon is south of the celestial equator) is more limited than when it is between $180\degree$ and $360\degree$ (where it passes north of the equator).

\subsection{Overlapping surveys}
\label{overlapping}
The five-band optical photometric survey to be produced by DES is not a stand-alone dark energy measurement data set. In same cases, it relies on other data sets for calibration and validation, and in others, the precision of DES dark energy measurements can be greatly enhanced by complementary data from other surveys. In particular,
\begin{enumerate}
    \item Most DES dark energy probes rely on photometric redshifts of galaxies catalogued by the survey. Calibration of photometric redshift methods depends critically on a large spectroscopic data set of objects within the DES survey footprint \citep{annis_report_2012, newman_calibrating_2008}.
    \item Photometric redshifts can be better estimated with the addition of near infrared data, so overlap with deep near-IR surveys provides areas of improved photo-z's.
    \item Overlap with other optical surveys allows photometric calibration against those surveys.
    \item Overlap with deeper optical surveys allows studies of the completeness of DES catalogs.
    \item Overlap with microwave surveys that generate catalogs of galaxy cluster masses measured using the Sunyaev–Zeldovich provide an independent mass measurement for the cluster cosmology probe. Furthermore, overlap with such surveys enables calculation of constraints on cosmological parameters using the cross-correlation between the gravitational lensing of the cosmic microwave background (measured in the microwave surveys) and wide-survey observables such as galaxy density and cosmic sheer \citep{des_and_spt_collaborations_dark_2019, des_&_spt_collaborations_dark_2019}.
\end{enumerate}

\begin{figure*}
\centering
\includegraphics[height=0.8\textwidth]{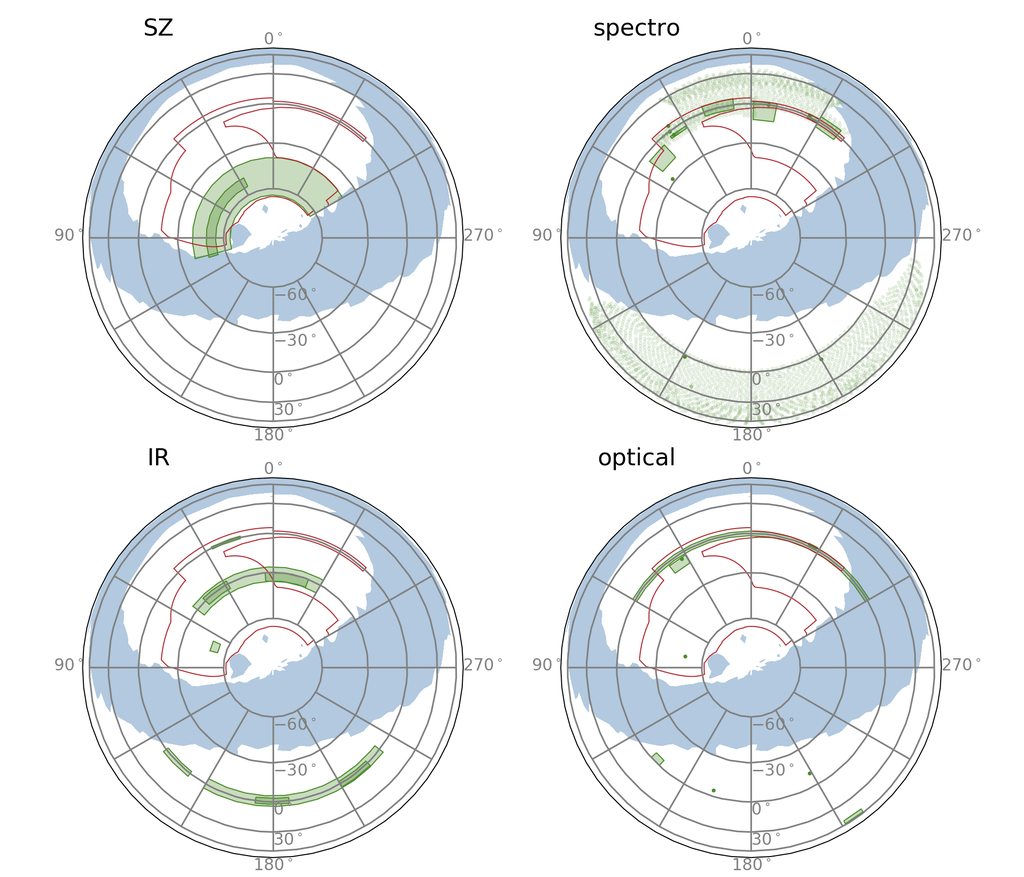}
\caption{\label{fig:overlapmap}
These maps show the footprints of other surveys with which overlaps with DES would be scientifically useful. The ultimate DES footprint is outlined in red. The blue shaded area shows areas of high stellar density (from the Milky Way, the LMC, and the SMC) from 2MASS. Filled green boxes show areas of surveys well described by equatorial spherical quadrangles, and green points show surveys areas with small footprints. Points in the spectroscopy subplot show locations of SDSS BOSS and eBOSS plates. Other areas are listed in table~\ref{tab:overlappingsurveys}.
}
\end{figure*}

Table~\ref{tab:overlappingsurveys} lists different data sets proposed as valuable for these purposes. Figure~\ref{fig:overlapmap} shows the footprints of these data sets, separated by motivation for inclusion. Not all data sets were of equal priority: overlap with the SPT survey, the SDSS imaging survey, and the BOSS and eBOSS spectroscopic surveys was vital. Inclusion of other surveys was beneficial, but not a driving consideration of the final footprint selection.

\begin{table}
\begin{tabular}{lllllll}
Survey & \multicolumn{2}{c}{R.A.} & \multicolumn{2}{c}{declination} & Reason & Reference \\
& min. & max. & min. & max & &\\
\hline
SPT & \(-60\degree\) & \(105\degree\) & \(-64\degree\) & \(-45\degree\) & SZ & \cite{the_spt_collaboration_overview_2013}\\
ACT & \(26\degree\) & \(107\degree\) & \(-55\degree\) & \(-49\degree\) & SZ & \cite{marriage_atacama_2011} \\
DEEPLens F3 & \multicolumn{2}{c}{\(80.0\degree\)}  & \multicolumn{2}{c}{\(-49\degree\)}  & Optical imaging & \cite{wittman_deep_2002}\\
DEEPLens F4 & \multicolumn{2}{c}{\(163.0\degree\)}  & \multicolumn{2}{c}{\(-5\degree\)} & Optical imaging & \cite{wittman_deep_2002}\\
DEEPLens F5 & \multicolumn{2}{c}{\(208.833\degree\)}  & \multicolumn{2}{c}{\(-11.05\degree\)} & Optical imaging & \cite{wittman_deep_2002}\\
DEEPLens F6 & \multicolumn{2}{c}{\(32.5\degree\)}  & \multicolumn{2}{c}{\(-4.5\degree\)} & Optical imaging & \cite{wittman_deep_2002}\\
CFHTLS W1 & \(30.25\degree\) & \(38.75\degree\) & \(-10.75\degree\) & \(-3.25\degree\) & Optical imaging & \cite{hudelot_t0007_2012}\\
CFHTLS W2 & \(132\degree\) & \(136.85\degree\) & \(-5.03\degree\) & \(-0.33\degree\) & Optical imaging & \cite{hudelot_t0007_2012}\\
CFHTLS W4 & \(331\degree\) & \(335\degree\) & \(1\degree\) & \(2.5\degree\) & Optical imaging & \cite{hudelot_t0007_2012}\\
SDSS Stripe82 & \(-60\degree\) & \(60\degree\) & \(-2\degree\) & \(2\degree\) & Optical imaging + spectro. & \cite{abazajian_seventh_2009} \\
WiggleZ 1 hr & \(7.5\degree\) & \(20.6\degree\) & \(-3.7\degree\) & \(5.3\degree\) & Spectroscopy & \cite{drinkwater_wigglez_2010}\\
WiggleZ 22 hr & \(-40\degree\) & \(-30\degree\) & \(-5\degree\) & \(5\degree\) & Spectroscopy & \cite{drinkwater_wigglez_2010}\\
WiggleZ 0 hr & \(-11\degree\) & \(-1\degree\) & \(-13\degree\) & \(2\degree\) & Spectroscopy & \cite{drinkwater_wigglez_2010}\\
WiggleZ 3 hr & \(43\degree\) & \(53\degree\) & \(-19\degree\) & \(-6\degree\) & Spectroscopy & \cite{drinkwater_wigglez_2010}\\
VIPERS W1 & \(30.5\degree\) & \(38.5\degree\) & \(-6\degree\) & \(-4\degree\) & Spectroscopy & \cite{scodeggio_vimos_2018}\\
VIPERS W4 & \(330.2\degree\) & \(335\degree\) & \(-1.2\degree\) & \(2.4\degree\) & Spectroscopy & \cite{scodeggio_vimos_2018}\\
DEEP2 field 3 & \multicolumn{2}{c}{\(352.5\degree\)} & \multicolumn{2}{c}{\(0\degree\)}  & Spectroscopy & \cite{newman_deep2_2013} \\
DEEP2 field 4 & \multicolumn{2}{c}{\(37.5\degree\)} & \multicolumn{2}{c}{\(0\degree\)}  & Spectroscopy & \cite{newman_deep2_2013} \\
VVDS Wide 1003+01 & \multicolumn{2}{c}{\(150.8\degree\)} & \multicolumn{2}{c}{\(1.5\degree\)} & Spectroscopy & \cite{le_fevre_vimos_2013}\\
VVDS Wide 1400+05 & \multicolumn{2}{c}{\(210\degree\)} & \multicolumn{2}{c}{\(5\degree\)} & Spectroscopy & \cite{le_fevre_vimos_2013}\\
VVDS Wide 2217+00 & \multicolumn{2}{c}{\(334.5\degree\)} & \multicolumn{2}{c}{\(0.4\degree\)} & Spectroscopy & \cite{le_fevre_vimos_2013}\\
VVDS Deep 0226-04 & \multicolumn{2}{c}{\(36.5\degree\)} & \multicolumn{2}{c}{\(4.5\degree\)} & Spectroscopy & \cite{le_fevre_vimos_2013}\\
VVDS Deep ECDFS & \multicolumn{2}{c}{\(53.1\degree\)} & \multicolumn{2}{c}{\(-27.8\degree\)} & Spectroscopy & \cite{le_fevre_vimos_2013}\\
VVDS Ultra-Deep & \multicolumn{2}{c}{\(36.6\degree\)} & \multicolumn{2}{c}{\(-4.4\degree\)} & Spectroscopy & \cite{le_fevre_vimos_2013}\\
VIKING SGP & \(330\degree\) & \(52.5\degree\) & \(-36\degree\) & \(-26\degree\) & IR imaging & \cite{banerji_combining_2015}\\
VIKING NGP & \(150\degree\) & \(232.5\degree\) & \(-5\degree\) & \(4\degree\) & IR imaging & \cite{banerji_combining_2015}\\
VIKING GAMA09 & \(129\degree\) & \(141\degree\) & \(-2\degree\) & \(3\degree\) & IR imaging & \cite{banerji_combining_2015}\\
Herschel Atlas SGP W & \(-22\degree\) & \(4.7\degree\) & \(-29.8\degree\) & \(-35.8\degree\) & IR imaging & \cite{smith_herschel-atlas_2017}\\
Herschel Atlas SGP E & \(28.2\degree\) & \(45.2\degree\) & \(-27.7\degree\) & \(-33.7\degree\) & IR imaging & \cite{smith_herschel-atlas_2017}\\
Herschel Atlas GAMA09 & \(127.5\degree\) & \(142\degree\) & \(-2\degree\) & \(3.1\degree\) & IR imaging & \cite{smith_herschel-atlas_2017}\\
Herschel Atlas GAMA12 & \(172.5\degree\) & \(186.9\degree\) & \(-3\degree\) & \(2\degree\) & IR imaging & \cite{smith_herschel-atlas_2017}\\
Herschel Atlas GAMA15 & \(210.2\degree\) & \(224.8\degree\) & \(-2.2\degree\) & \(3.1\degree\) & IR Imaging & \cite{smith_herschel-atlas_2017}\\
SHELA & \(14\degree\) & \(27\degree\) & \(-1\degree\) & \(1\degree\) & IR imaging & \cite{papovich_spitzer-hetdex_2016} \\
ADF-S & \(66\degree\) & \(75\degree\) & \(-55\degree\) & \(-50\degree\) & IR imaging & \cite{matsuhara_deep_2006} \\
\end{tabular}
\caption{\label{tab:overlappingsurveys}
Other sky surveys with which overlap with the DES footprint would be useful.
}
\end{table}

\subsection{The wide survey footprint}
\label{footprint}

Summarizing the factors considered when designing the footprint for the DES survey:

\begin{enumerate}
    \item \label{fp:decaccess} The declination of pointings within the footprint should be between about $-65\degree$ and $5\degree$, because pointings outside of this areas can only be observed at high airmass (if at all), and only in limited times. See section~\ref{observablesky}, particularly figure~\ref{fig:txvsdecl}.
    \item \label{fp:contig} The footprint should encompass $\sim 5000$ sq. deg. of contiguous area.
    \item \label{fp:galaxy} The footprint should avoid dust and stars from the Milky Way. See section~\ref{obstructing}, particularly figure~\ref{fig:galaxymap}.
    \item \label{fp:overlap} The footprint should overlap SZ, spectroscopic, optical, and IR imaging surveys where possible, with the SPT survey area, the SDSS stripe 82 imaging data, and BOSS and eBOSS spectroscopic footprint given particular importance. See section~\ref{overlapping}, particularly figure~\ref{fig:overlapmap}.
    \item \label{fp:powersample} The footprint should sample a wide range of spatial frequencies well, including large distances and small spatial frequencies.
    \item \label{fp:moonpath} The footprint should avoid area significantly obstructed by the moon during bright time. See section~\ref{obstructing}, particularly figure~\ref{fig:taumap} (which also incorporates preferences in {\sc r.a.} due to weather patterns).
\end{enumerate}

Considerations~\ref{fp:decaccess},~\ref{fp:contig},~\ref{fp:galaxy}, and~\ref{fp:overlap} combine to restrict the DES footprint area to the southern Galactic cap. Consideration~\ref{fp:overlap} constrains the footprint to include large fractions of three equatorial quadrangles:
\begin{itemize}
    \item $-60\degree <= \mbox{\sc r.a.} <= 105\degree$ and  $-65\degree <= \mbox{Dec.} <= -40\degree$, the SPT survey area.
    \item $-43\degree <= \mbox{\sc r.a.} <= 0\degree$ and  $-2\degree <= \mbox{Dec.} <= 2\degree$, the southern edge of SDSS imaging and the BOSS and eBOSS spectroscopic surveys in the {\sc r.a.} range where the spectroscopic footprint does not extend much south of the celestial equator. (Consideration~\ref{fp:decaccess} sets the northern limit of this quadrangle.)
    \item $0\degree <= \mbox{\sc r.a.} <= 45\degree$ and  $-7\degree <= \mbox{Dec.} <= 5\degree$, the southern edge of the BOSS and eBOSS surveys where they extend a little farther south of the equator. (Consideration~\ref{fp:decaccess} sets the northern limit of this quadrangle.)
\end{itemize}

\begin{figure*}
\centering
\includegraphics[width=0.9\linewidth]{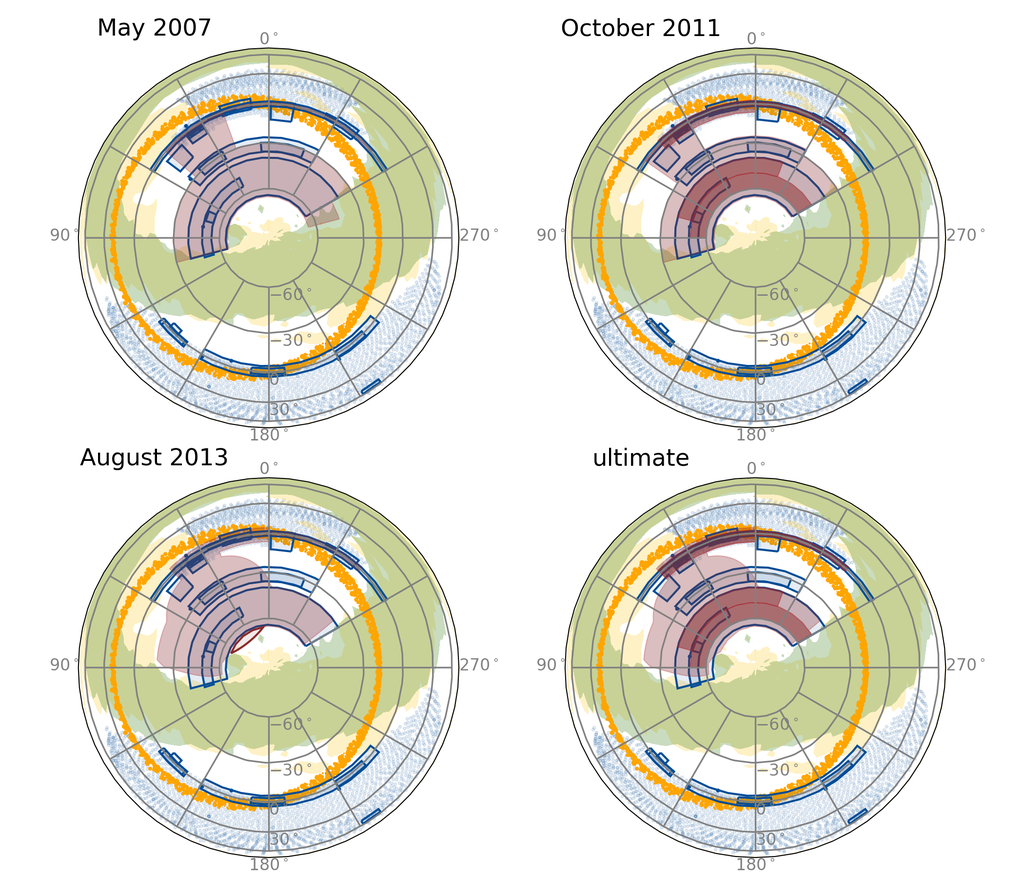}
\caption{\label{fig:footprints}
The DES footprint evolved in the years leading up to the start of the survey; each subplot represents a stage in that evolution. The yellow and green mark areas of high extinction and stellar density, mostly due to the Milky Way. Orange points show the position of the moon for each night in the years of DES observing. Blue points and outlines show areas of other surveys for which overlap with the DES would be useful. Light red marks the DES footprint itself. The darker red regions show a subset of the footprint schedule for observing in the first year of DES. The lighter, outlined area in the August 2013 (lower left) map show an area declared to be low priority, because it is unnecessary to make the goal of 5000 square degrees, is only ever available at high airmass, suffers from proximity to the Large Magellanic Cloud, and competes with higher priority parts of the footprint for local sidereal time.
}
\end{figure*}

The earliest footprints proposed for DES, some examples of which are shown in the top subplots of figure~\ref{fig:footprints}, were designed to maximize overlap with other surveys as well, particularly the VIKING SGP footprint (and the area required to connect it to the SPT quadrangle).

Survey strategy simulations during the summer of 2013 clarified the importance of other considerations, and the footprint was modified in August of 2013. The lower left hand subplot of figure~\ref{fig:footprints} shows the August 2013 footprint. Several considerations motivated the change in footprint:
\begin{itemize}
    \item High stellar density near the plane of the Milky Way makes the usefulness of this area questionable, so a strictly imposed limit of seven times the minimal stellar density in 2MASS was applied to the footprint, reducing the extension of the footprint into the Galactic plane. This change slightly reduced the coverage of the SPT area. %
    \item A large, circular area allows improved sampling of large spatial correlations for the large scale structure (LSS) dark energy probe.
    \item The path of the moon and seasonal variability in weather conditions result in {\sc r.a.} dependent footprint accessibility (see section~\ref{observablesky} and figure~\ref{fig:taumap}). Therefore, within the limits imposed by extinction, stellar density, and overlap with SDSS, BOSS, eBOSS, and SPT, the footprint was moved as far east as practical, resulting in the placement of the center of the LSS circular area at $\mbox{\sc r.a.}=38.3\degree$ and $\mbox{decl.}=-39.5\degree$, and the extension of the footprint to the east beyond it to the high stellar density limit. This shift came at the expense of $\sim40\%$ of the overlap with the VIKING footprint.
\end{itemize}

The total area of this footprint is 5122 sq. deg., slightly larger than the nominal 5000 sq. deg. footprint. The LSS circle extends slightly south of the southern edge of the SPT overlap which ends at decl.=$-65\degree$, and exclusion of this region reduces the footprint to be 5027 sq. deg., achieving our goal of 5000 sq. deg. This area is also challenging to observe at high quality due to its low declination (see figure~\ref{fig:txvsdecl}) and contaminated by stars from the Magellenic clouds, and so was designated as lower-priority, optional area.

The initial plan for performing the survey was to complete the full footprint in two tilings (so \sfrac{1}{5} depth in the entire footprint each year) in all filters in each year. With two or fewer exposures over the footprint, however, the state of the survey after the first year following this plan introduces processing and calibration challenges, and results in a shallow survey. Instead, for the first year of observing the project collected four tilings on a smaller area of the footprint, described in table~\ref{tab:y1footprint} and shown in dark red in the two plots on the right of figure~\ref{fig:footprints}. Note that these quadrangles were defined before the adaption of the August 2013 footprint, and so the southernmost quadrangle extends slightly farther into the Galactic plane than is included in that footprint.

\begin{table}
\begin{center}
\begin{tabular}{llll}
\multicolumn{2}{c}{R.A.} & \multicolumn{2}{c}{declination} \\
min. & max. & min. & max\\
\hline
$-43\degree$ & $0\degree$ & $-2\degree$ & $2\degree$ \\
$0\degree$ & $45\degree$ & $-7\degree$ & $3\degree$ \\
$-20\degree$ & $75\degree$ & $-50\degree$ & $-40\degree$ \\
$-60\degree$ & $90\degree$ & $-60\degree$ & $-50\degree$
\end{tabular}
\caption{\label{tab:y1footprint}
Equatorial quadrangles comprising the year 1 footprint.
}
\end{center}
\end{table}

The footprint in the lower right subplot of figure~\ref{fig:footprints} shows the final footprint as actually observed. It is similar to the August 2013 footprint, except that it includes the western edge of the year 1 footprint that extends a little farther into the Galactic plane, and completes only a portion of the optional area where the LSS circle extends south of the overlap with the SPT quadrangle.

\begin{figure*}
\centering
\includegraphics[width=0.9\linewidth]{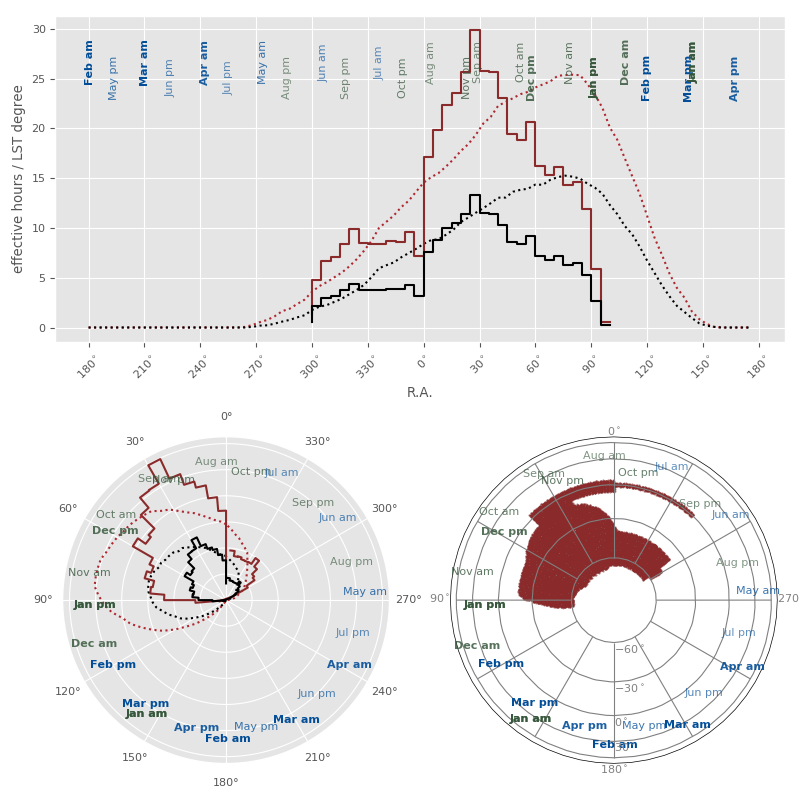}
\caption{\label{fig:wideexptimevsra}
Solid lines show the R.A. distribution of planned exposures in the DES wide-survey footprint. The black lines show exposures in $g$ and $r$, only practical in ``dark'' (moonless) time, and red show total exposures. Dotted lines show the distribution of scheduled time over the six years of observing, scaled by the expected contributions to $\tau$ from seeing and clouds using simple sinusoidal fits, black and red showing moon-free and total effective time, respectively. Text labels show the central LSTs of half-nights in each month. The weight of the text indicates the expected effective duration of these half nights (duration times the fit $\tau$ contribution from clouds and seeing). Text in blue indicates half nights in the NOAO ``A'' semester, green in the ``B'' semester.
}
\end{figure*}

\subsection{Tiling the sky}
With the decision to observe for equal times in {\it g, r, i} and {\it z} (see section~\ref{areaanddepth}), the agreement to observe for half that in {\it Y} (section~\ref{wideobjectives}), and establishment of a survey footprint (section~\ref{footprint}), it remained to determine the total numbers of exposures and their arrangement within the footprint.

The DECam camera contains 62 science CCDs of $2048 \times 4096$ pixels, separated by gaps with the widths of 153 and 201 pixels. The width and height of each pixel is 0.263''. Allowing for a half-gap surrounding each CCD such that neighboring pairs result in a full gap, each CCD has live area of $2048 \times 0.263'' \times 4096 \times 0.263'' = 0.0448~\mbox{sq. deg.}$, for $0.0448'' \times 62 = 2.78~\mbox{sq. deg.}$ of focal plane area covered by pixels. Allowing for the gaps between CCDs, each CCD occupies an area of $(2048+201)\times(4096+153)\times62\times(0.263'')^{2}=0.0510~\mbox{sq. deg.}$ on the focal plane, for $0.0510'' \times 62 = 3.16~\mbox{sq. deg.}$. This area, however, allows for chip gaps along the outside of the footprint, such that CCDs between neighboring pointings are separated by the chip gap as well. The focal plane is 12 CCDs high and 7 CCDs wide, so the padded height is $12\times(2048+201)\times0.263''=1.972\degree$ and width is $7\times(4096+153)\times0.263=2.173\degree''$. Without the between-pointing padding, the height is $1.972\degree - 201\times0.263''=1.957\degree$ and width, $2.173\degree - 163\times0.263''=2.176\degree$, resulting in $1.972\degree \times 2.173\degree - 1.957\degree \times 2.173\degree = 0.054~\mbox{sq. deg.}$ of recoverable area, for a total efficiency of tightly packed pointings of 
$
\frac{2.78~\mbox{sq. deg.}}
{3.16~\mbox{sq. deg.} - 0.05~\mbox{sq. deg.}}
= 0.89
$.
Therefore, because of the gaps between CCDs, the efficiency of coverage for a set of pointings with minimal overlap is at best 89\%.\footnote{In production, up to 2\sfrac{1}{2} CCDs could be in an unusable state, and a border of 15 pixels at the edges of CCDs was masked due to strong distortion. Combining these factors, the good area could be as poor as $(62-2.5)\times( ((2048 - 2\times15) * 0.263'') \times ((4096 - 2\times15) * 0.263'') = 2.61~\mbox{sq. deg.}$, for a fill factor as poor as 0.84. This was not a consideration in the layout of pointings for DES. The unusable CCDs are along the top and bottom edges of the camera, however, so future surveys may wish to consider a pointing layout that is more tightly packed in declination than the DES layout.}

In the context of survey strategy,\footnote{Confusingly, the term ``tile'' was used to mean something completely different in the context of DES data management.} a ``tiling'' is a collection of pointings that cover the entire footprint with minimal overlap and a pixel coverage of about 89\%, such that missing area is dominated by gaps between CCDs. To design a list of specific exposures (exposures with specific filers, pointings, and exposure times) based on the basic parameters of survey area and distribution of exposure time, several questions needed to be answered:
\begin{enumerate}
    \item How many tilings should there be per filter? Another way of phrasing the same question is to ask how many different exposures we want of a given object in the footprint: the mean number of exposures on a given set of coordinates in the sky will be 89\% of the number of tilings.
    \item How are the different tilings to be dithered? That is, should one tiling have the same set of pointings as another, or a different one, and if different, what should that pattern of differences be?
\end{enumerate}
Several factors need to be considered in answering these questions:
\begin{enumerate}
    \item More tilings require more exposures, which in turn results in greater overhead and reduced observing efficiency. The shortest overhead between DECam exposures is about 27 seconds, so given a constant interval of ``wall clock'' time spent observing, each additional tiling reduces the final total exposure time by at least 27 seconds. If this were the only consideration, a single tiling would be optimal.
    \item Weather conditions vary from one exposure to another. If the survey observes many tilings, then each area of the footprint can be observed under a variety of observing conditions, while if there are few tilings, the variations in weather will result in variations in imaging quality on the footprint.
    \item Systematic errors (including errors in the photometric calibration, PSF model, and astrometric calibration) vary by exposure, night, and/or placement on focal plane. If there are many tilings, then different exposures of the same patch of sky can be spread across nights and (if there are large dithers between tilings) placement on the focal plane, averaging these errors over instances.
    \item DES achieves uniform photometric calibration by taking advantage of partially overlapping exposures. A dither pattern in which the relative photometric calibration of neighboring (non-overlapping) exposures can be derived using many other exposures which overlap both images is therefore important. 
\end{enumerate}

The layout of the CCDs on the focal plane results in an approximately hexagonal camera footprint. Note that a plane is efficiently tiled with a hex  pattern. The Blanco telescope has an equatorial mount, and DECam does not have a rotator, so the orientation of the focal plane on the sky is constant with respect to declination. Near the celestial equator, the sphere of the sky is well approximated by a plane, and the density of pointings required to cover 89\% of the sky with pixels is close to the nominal planar density. With a camera footprint area of 3.11 sq. deg., the density of pointings required would be 0.32 pointings/sq. deg. of footprint, or roughly 1600 pointings for the nominal footprint area of 5000 sq. deg. Simulations indicated that the survey could perform roughly 80,000 exposures in the allocated 525 nights, which is a good match for ten tilings in each of five filters at 1600 pointings in each tiling and filter.

\begin{figure*}
\centering
\includegraphics[width=\linewidth]{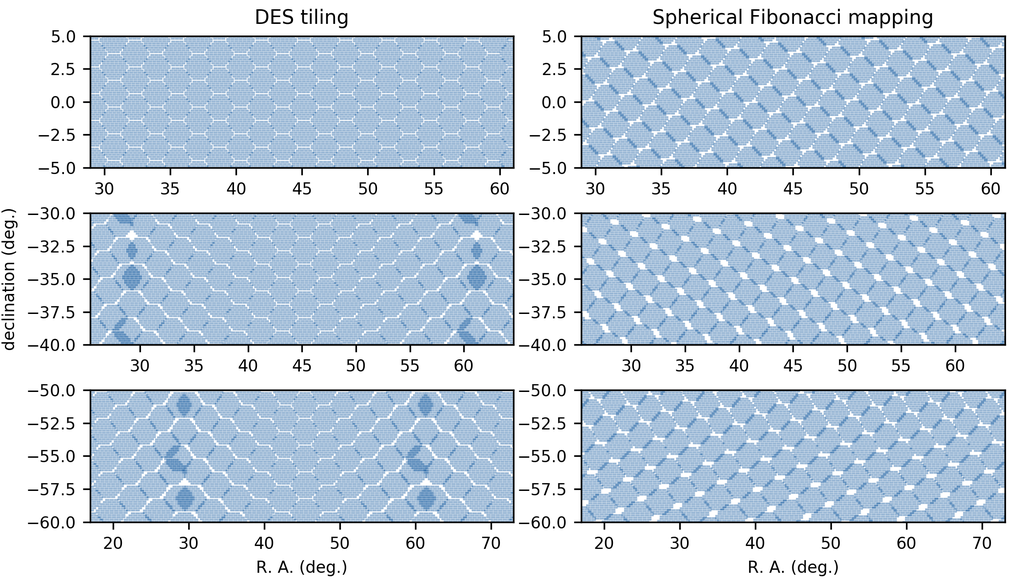}
\caption{\label{fig:hexlayout}
These plots show the pointing layouts both for the ultimately selected DES pointing scheme (left) and spherical Fibonacci mapping \citep{keinert_spherical_2015} (right), a typical example of a layout method that results in a more uniform distribution of pointings (which does not take advantage of the near-hexagonal shape of the DECam footprint). The top row shows the layout near northern edge of the DES footprint (near the celestial equator), the center row shows the center of the DES footprint, and the bottom row shows an area near the southern end. The DES tiling scheme does well in areas of the sky near the equator (top) or centers of DES tiling neighborhoods ($\mbox{\sc r.a.}=45\degree$), and poorly near neighborhood boundaries far from the equator; while the more uniform distributions leave overlaps and uncovered islands throughout.
}
\end{figure*}

The sky is spherical rather than planar, so this hex tiling approximation breaks down as the footprint moves farther from the equator. If the (planar) hexagonal coordinates are naively matched to {\sc r.a.} and declination, then the spacing of pointings becomes compressed in the {\sc r.a.} direction as one moves farther from the celestial equator. It is tempting to take advantage of the significant body of literature in mathematics that addresses the problem of packing points on a sphere. The Tammes problem \citep{aste_pursuit_2008}, the problem of deriving the distribution of a given number of points on a sphere such that the minimum distance between any two points is maximized, is one example of such a problem. However, optimization according to these metrics does not map directly onto the scientific problem faced by DES for several reasons. First, these metrics do not take into account that the shape of the camera footprint is not circular, but rather roughly hexagonal, and in a fixed orientation relative to the equatorial coordinate system. Second, the behaviour of the tiling scheme outside the survey footprint is not a concern: a solution that works well within the DES footprint but does poorly at the equatorial poles, for example, would be acceptable. Finally, overall uniformity is less of a concern than total coverage at a minimum threshold depth.

When evaluating pointing layout schemes, therefore, three criteria more directly related to survey science needs were used instead: the footprint area covered by a single tiling, the distribution across CCDs of pairs of exposures on the same point of the sky, and the footprint area covered more than 8 times in a complete set of ten tilings. In practice, these were estimated by sampling points and calculating the following statistics:
\begin{eqnarray}
T_{1} & \equiv & \frac{N^{\mbox{1 tiling}}_{\mbox{pt}}(\ge {\mbox{1 hex}})A_{\mbox{pt}}}{N_{\mbox{hex}}A_{\mbox{hex}}} \\
T_{2} & \equiv & \frac{\left(\prod^{62}_{i=1,j \ge i,n_{ij} \ge 0} n_{ij}\right)^{\frac{1}{M_{nz}}}}{\frac{1}{M_{nz}} \sum^{62}_{i=1,j \ge i,n_{ij} \ge 0}  n_{ij}} \frac{M_{nz}}{M} \\
T_{3} & \equiv & \frac{N^{\mbox{10 tilings}}_{\mbox{pt}}(\ge {\mbox{8 hexes}})A_{\mbox{pt}}}{A_{\mbox{DES}}} \frac{1500}{N_{\mbox{hex}}} 
\end{eqnarray}
where $N_{\mbox{pt}}$ is the number of sampling pixels (healpix \citep{gorski_healpix:_2005} with nside=2048), $N_{\mbox{hex}}$ is the number of hexes in one tiling, $n_{ij}$ is the number of sampling pixels observed in CCDs $i$ and $j$, $M$ is the number of pairs of CCDs ($(62^{2} - 62)/2$), $M_{nz}$ is the number of pairs of CCDs for which $n_{ij}$ is non-zero, $A_{\mbox{pt}}$ is the area of one healpix pixel, $A_{\mbox{hex}}$ is the area of one hex, and $A_{\mbox{DES}}$ is the area of the DES footprint\footnote{The tiling schemes were evaluated before the final footprint was established, so the footprint considered in this optimization is slightly different than that actually used in the survey.}.

The $T_{2}$ and $T_{3}$ metrics both depend not only on the pointing layout of a single tiling, but the pointings used in all tilings. We adopted a scheme in which the pointings in different tilings used the same reference pointing layout, but dithered by offsets of order the radius of the camera footprint. A variety of different dithering schemes were studied, and evaluated in combination with different single tiling layout patters.\footnote{The same dither schemes did not always perform equally well across different tiling layout schemes: the pairs of such needed to be evaluated in combination, rather than independently.} Although we found uniform pointing layouts more aesthetically appealing, when evaluating based on criteria of direct scientific relevance (metrics $T_{2}$ and $T_{3}$), we did not find such a pointing and dither layout combination that outperformed the simple process of laying the pointings in a hex pattern in lunes (spherical segments bounded by lines of constant {\sc r.a.}), so this was what was ultimately used.

\begin{table}
\begin{center}
\begin{tabular}{lll}
tiling & $\Delta$ {\sc r.a.} & $\Delta$ decl. \\
\hline
1 &  0.0000\degree & 0.0000\degree \\
2 & -0.76668\degree & 0.473424\degree \\
3 & -0.543065\degree &  -0.828492\degree \\
4 &  0.0479175\degree &  0.777768\degree \\
5 &  0.06389\degree &  0.287436\degree \\
6 & -0.4632025\degree &  0.490332\degree \\
7 &  0.9423775\degree &  0.405792\degree \\
8 & -0.2395875\degree &  -0.135264\degree \\
9 &  0.76668\degree &  0.4227\degree \\
10 &-0.0479175\degree &  0.388884\degree \\
\end{tabular}
\caption{\label{tab:wideoffsets}
Offsets between wide survey tilings.
}
\end{center}
\end{table}

\section{Observing tactics}
\label{tactics}
\subsection{Tactics as a Markov Decision Process}
\label{mdp}

\begin{figure*}
\centering
\includegraphics[width=0.8\linewidth]{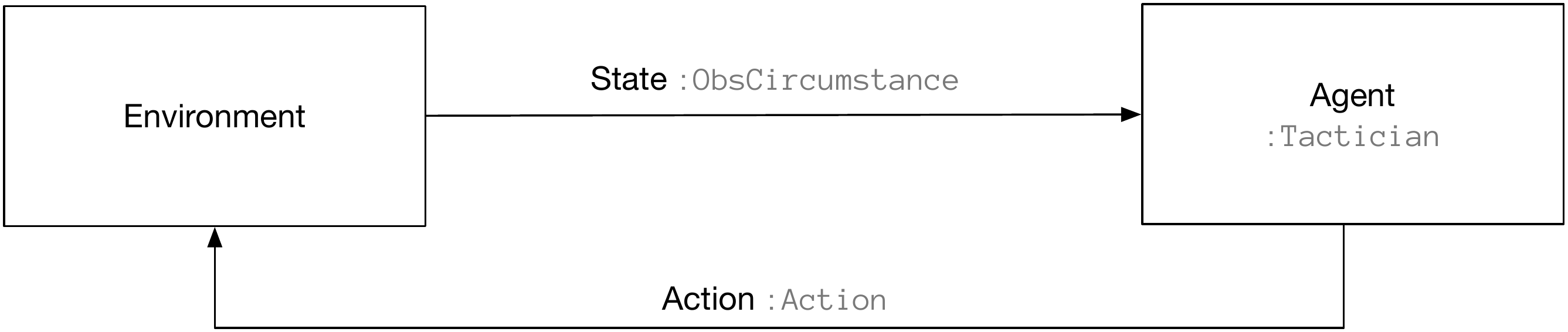}
\caption{\label{fig:obstacmdp}
DES modeled tactical decision making as a Markov decision process (MDP). An {\it agent} (the core of the scheduler) makes a sequence of decisions through interactions with the environment: for each decision, it reads the {\it state} of the {\it environment}, and responds with an {\it action}. The environment (perhaps non-deterministically) changes state as a result of the the action, and the process begins for the next decision. The text in light gray indicates the {\tt python} classes used to implement these elements in \obstac. 
}
\end{figure*}

DES implemented observing tactics following the architecture of a Markov decisions process (MDP) \citep{sutton_reinforcement_2018}, represented visually by figure~\ref{fig:obstacmdp}. \obstac, the DES scheduler, reads the state of the survey, instrument, and environment; selects an action (either one wide survey exposure or one sequence of supernova exposures) and returns it to the environment; and the environment responds by moving to a different state. When another exposures is required, the cycle begins again. \obstac\ implements action selection through a decision tree. Nodes in the tree correspond observing program selection (time-domain vs. wide survey), cuts on various parameters, database queries that sort candidate exposures based on a variety of conditions, and other factors.

\subsection{Observable exposures}
\label{observable-exposures}

In a general MDP, not all actions are available from all states, and this is the case for DES tactics as well. Several factors render an exposure or supernova sequences unavailable at a given time.
\begin{enumerate}
    \item Wide survey exposures that are either completed (and not declared bad), already in the observing queue, or currently in progress are considered unavailable for scheduling.
    \item When a pointing is too far from the zenith (at too high an airmass), the data quality is likely to be severely degraded. \obstac\ imposed a (configurable) hard limit on the predicted airmass of the exposures it will select from. For the wide survey, this limit was 1.4 throughout the survey. The supernova survey usually had an airmass limit of 1.5, although this was raised to 2.0 during a few brief periods in order to extend the season over which a field could be observed. (The Blanco telescope also imposes limits, but these are looser than those adopted for data quality reasons, and therefore never relevant for DES tactics.)
    \item If the sky brightness is too high, the resultant $\tau$ will be severely degraded, so \obstac\ imposed hard limits on the maximum sky brightness. This hard limit was 1 mag. per square arcsecond brighter than full dark for a wide survey exposure, 3 mag. per square arcsecond for $g$ and $r$ supernova exposures, and 2 mag. per square arcsecond for $i$ and $z$ supernova exposures.
    \item There were seeing limits of 1.8'' {\sc fwhm} for shallow SN sequences, and 1.3'' for deep SN sequences. There was no upper limit on the seeing for wide exposures to be attempted, although exposures with a delivered {\sc fwhm} of worse than 1.6'' were declared bad and redone. Ultimately, it made little difference whether data taken in such poor conditions were wide survey exposures or supernova sequences, because such exposures are not useful in either case. However, if there were brief deviations from the expected seeing, it is more likely that an occasional wide exposure might be good than a (much longer) supernova sequence.
    \item Although the sky brightness limit usually prevents observing close to the moon indirectly, an additional requirement that the field be at least $30\degree$ from the moon was imposed because, when a field is that close to the moon, the gradients in sky brightness can make exposures difficult to process even if the overall level is acceptable.
    \item Wide survey exposures whose fields overlapped that of other wide survey exposures in the same filter on the same night were considered unavailable. The primary motivation for observing ten separate tilings, rather than fewer tilings with longer exposure times, was to ensure that each object was observed under a variety of conditions. Observing objects in the same filter multiple times on the same night compromises this objective.
    \item To promote uniformity in the covered area of the footprint at the end of each observing season, the scheduler did not attempt to work on all ten tilings simultaneously: available exposures were added in batches each year. In the first year, the first four tilings in a reduced area footprint were scheduled (see section~\ref{year1}). In year two, the rest of the footprint was scheduled in these same four filters. In each succeeding year, an additional two tilings were added over the whole footprint. Tilings and areas not yet scheduled were considered unavailable (except by the desperation tactician; see section~\ref{desperate-tactician}).
    \item A human can explicitly declare to \obstac\ that it should not do a specific exposure by adding it to a table in the SISPI database. This will prevent \obstac\ from scheduling the exposure as long as it remains in the table. Exposures could be added to this table for any reason, the most common of which was to prevent \obstac\ from attempting (or re-attempting) exposures where there were stars bright enough that scattered light will always contaminate the field enough that exposures on this field are not useful.
\end{enumerate}

\subsection{Program selection}
\label{program-selection}

\begin{figure*}
\centering
\includegraphics[width=\linewidth]{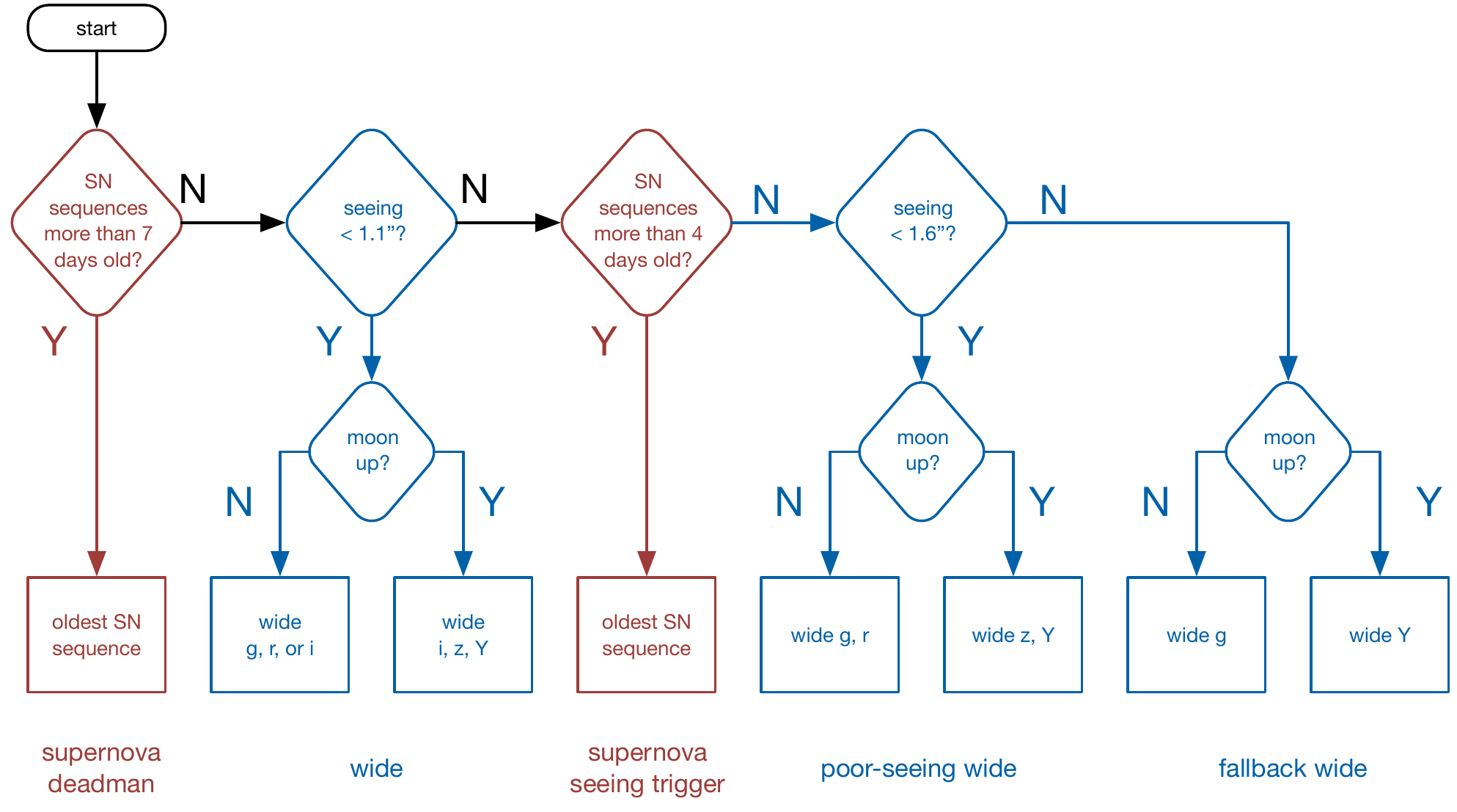}
\caption{\label{fig:obstac}
In years 1 through 5, \obstac\ followed this basic decision tree for deciding whether to select supernova sequences or wide survey exposures.
}
\end{figure*}

The initial nodes in \obstac's decision tree concern the selection of observing program: whether to observe a supernova (time-domain) sequence or wide survey exposure. Figure~\ref{fig:obstac} shows this decision process graphically. The most time critical element in the survey is the maintenance of cadence for each supernova sequence. So, the first node in the decision tree is to check whether there are any such sequences with an age of more than 7 days (that is, where there are any that have not been observed in the last seven days). If there are any such sequences observable in current conditions, the oldest such observable sequence is returned by \obstac. This selection is colloquially called the ``supernova dead-man,'' because the supernova sequence is selected without being triggered by observing conditions.

There is little benefit to observing supernova sequences with a cadence shorter than four days, so, if all observable sequences are four days old or younger, then a wide survey exposure is returned by \obstac.

Supernova light curves do, however, benefit from cadences of 5 or 6 days (compared to 7), even at the expense of image depth ($\tau$). In contrast, delivered image {\sc fwhm} is vitally important to the wide survey science. Shape measurements for weak lensing depend on a small image {\sc fwhm} even beyond its impact on survey depth, because a poor {\sc fwhm} prevents shape measurement even in deeper images. Therefore, if the seeing (expected {\sc fwhm} of an exposure in $i$ band, at zenith) is worse than 1.1'', and there are observable supernova sequences older than 4 days, then \obstac\ will choose the oldest supernova sequence. This simultaneously improves the cadence of the time-domain survey and improves the likelihood that any given wide survey image have a good delivered {\sc fwhm}.

This process for program selection had several interesting features. First, because no information about gaps in the observing schedule is used, it can sometime result in gaps in the supernova cadence of much greater than the desired 7 days. The DES schedule typically had gaps of five nights of no DES observing each month, and some gaps were as long as eight days. To minimize the damage to the supernova cadence due to these gaps, as many supernova sequences need to be completed as possible on the nights immediately prior to the gap. This gap anticipation was implemented in \obstac\ by specifying a configurable set of dates on which to trigger all supernova sequences, even if they have been completed recently. These nights were then entered by hand in the \obstac\ configuration file once the schedule for each year was finalized.

Another interesting behaviour of this decision tree is its reaction to long runs of poor weather. If the distribution of overcast or poor-seeing nights were distributed evenly across scheduled nights of DES observing, the decision tree would work particularly well, and would result in a balanced set of exposures in which the time-domain survey maintains its cadence at the same time the wide survey makes steady progress. Weather patterns are not so evenly distributed, however; weather is consistently better in the late months of each DES observing season (which runs from August into February) than the early, and there are long time-scale variations such that some years are much better than others. When there are long sequences of overcast nights, such that there is no good data collected for many consecutive nights, on those nights when productive observing {\em is} possible, all of the supernova sequences are ``on dead-man,'' and the wide survey gets shut out. In weather patterns with many cloudy nights and poor seeing as well (common in August through October), not only are the supernova sequences on dead-man more often, but they are triggered by poor seeing more often as well.

The seeing-based choice of program was removed from the decision tree before year 6, because only the wide survey was scheduled for year 6.

\subsection{Selection of wide survey exposures}

\subsubsection{General approach}

Once the scheduler determines that it is to schedule a wide-survey exposure, it must select from among the potential exposures that pass the criteria in section~\ref{observable-exposures}. Note that figure~\ref{fig:obstac} shows three paths by which wide survey exposures may be chosen; each path corresponds to a different execution of the same python code (three callable objects belonging to the same class), configured with different parameters which set seeing and $\tau$ limits, and preferred filters for each path. This code constructs an SQL query to the SISPI database which returns prioritized table of candidate exposures, executes it, performs some additional filtering to ensure that the selected exposure is indeed observable, and returns the top exposure. The details of this process were modified and adjusted many times over the course of the survey, but maintained a behavior that was consistent in its overall approach. An example of the sorting criteria, implemented by this combination of python and SQL code, prioritized exposures by the following criteria:
\begin{enumerate}
    \item Calculate a revised airmass limit based on the current estimate of the seeing, the configured seeing limit, and the Kolmogorov relation between seeing and airmass (see appendix~\ref{seeing}); and select exposures with pointings within that airmass limit (or 1.4, whichever is more restrictive).
    \item Select exposures with a predicted $\tau$ greater than a configurable limit. The value is independently configurable by filter and path through the code (``wide'', ``poor-seeing wide'', or ``fallback wide'' in figure~\ref{fig:obstac}).
    \item In dark time, group available exposures by where they are in a set of preferred filters based on the conditions, and select exposures with filters in this group. 
    \item Group remaining selected exposures by footprint priority, and select exposures in the highest priority group. 
    \item At early\footnote{The local sidereal time considered ``early'' was a tunable parameter.} sidereal times, group remaining selected exposures by {\sc h.a.} into one hour ($15\degree$) bins, and select exposures in the populated bin closest to transiting; at later sidereal times, group remaining selected exposures by the sidereal time at which they reach an airmass of 1.4 into one hour ($15\degree$) bins, and select exposures in the earliest populated bin.
    \item Group remaining selected exposures by whether they can be reached with a slew $4\degree$ or less, and select those that are (if there are any), otherwise preserve all remaining exposures.
    \item Group remaining selected exposures by tiling number, and select exposures with the lowest tiling.
    \item Group remaining selected exposures by whether they can be reached with a slew of $35\degree$ or less, and select exposures that can be, if there are any.
    \item Select the northernmost (if the prior pointing has a declination of greater than $-40\degree$) or southernmost (otherwise) remaining selected exposure. So, if a long slew was absolutely necessary, \obstac\ would work from the northern or southern edge inward, thereby working on the harder to observe areas of the footprint first.
\end{enumerate}

\subsubsection{Airmass limit}

Each execution of the wide-survey exposure selection code is configured with a seeing limit. The delivered image {\sc fwhm}, however, is dependent not only on the atmospheric seeing, but also on the airmass. \obstac\ approximates the variation in seeing with airmass using the Kolmogorov model: $\mbox{\sc fwhm} \propto X^{\frac{5}{3}}$, where $X$ is the airmass. (If the modified airmass limit is less than 1.1, then an airmass limit of 1.1 is used.) Expressed more precisely, \obstac\ uses an airmass limit, $X_{\mbox{max}}$, of
\begin{equation}
X_{\mbox{max}} = 
    \begin{cases}
    1.1 &  X_{\mbox{seeing}} \le 1.1, \\
    X_{\mbox{seeing}} & 1.1 < X_{\mbox{seeing}} < 1.4, \\
    1.4 & X_{\mbox{seeing}} > 1.4
    \end{cases}
\end{equation}
where 
\begin{equation}
    X_{\mbox{seeing}} = \left( \frac{\mbox{\sc fwhm}_{\mbox{max}}}{\mbox{\sc fwhm}_{\mbox{pred}}}\right) ^{\frac{5}{3}},
\end{equation}
$\mbox{\sc fwhm}_{\mbox{pred}}$ is the predicted seeing at zenith, and $\mbox{\sc fwhm}_{\mbox{max}}$ is the configured {\sc fwhm} limit for this wide survey exposure selection path.

This variable airmass limit causes \obstac\ to select exposures near zenith during marginal conditions, maximizing the likelihood of their being useful, while still allowing it flexibility to select exposures at higher airmasses when conditions permit.

\subsubsection{\texorpdfstring{$\tau$}{t}}

When DES data management determines that exposures fail data quality cuts, exposures are declared bad and \obstac\ must re-observe them. One of the data quality cuts applied is to $\tau$, and \obstac\ prioritized exposures whose predicted $\tau$ (based on its seeing and sky brightness model) are above this cut value by a configurable margin. The precise level of the cut depends on the band and the path by which the wide survey exposure was selected. In the standard case (seeing < 1.1 in figure~\ref{fig:obstac}), the $\tau$ limits were 0.5 in {\it g, r, i} and {\it z}, and 0.3 in {\it Y}. In the poor seeing path, the limits were 0.3 in {\it g, r, i} and {\it z}, and 0.2 in {\it Y}. In the fallback path, the $\tau$ limits were removed entirely.

\subsubsection{Filter selection}

The initial decision points for selection of specific wide survey exposures are for the filter to be used. If the seeing is very poor, only {\it g} and {\it Y} exposures were selected, because these filters were not used for weak lensing shape measurements\footnote{The exact value of ``very poor'' was a configuration parameter, adjusted a few times over the course of the survey in response to the relative progress in {\it g} and {\it Y} and the other filters. A typical value was 1.4''.}. If the moon is down, then exposures in {\it g, r} or {\it i} were selected if possible, because the sky brightness makes these fields difficult and inefficient (low $\tau$) when the moon is up: unless most dark (moonless) time is used observing in these filters, the survey would not be completed in these filters. See appendices~\ref{skybrightness} and~\ref{dataquality} for more, particularly figures~\ref{fig:skytvsmoonangle} and~\ref{fig:skytvsmoonphase}. There was no explicit selection of preferred filters in good seeing and bright time, although in practice sky brightness and $\tau$ limits prevented the selection of exposures in {\it g, r} and often {\it i} in bight time.

\subsubsection{Footprint priority}

Not all parts of the DES footprint were of equal priority. A table in the SISPI database contained rows for all DES wide survey pointings, with priorities that could be set independently. These priorities were primarily used to reduce the priority of the ``low priority'' area described in section~\ref{footprint}, and outlined in red in the lower left map of figure~\ref{fig:footprints}. At the end of the survey, this feature was also sometimes used to force \obstac\ to observe specific areas of the sky at specific nights, fine-tuning the ``end game.'' 
\subsubsection{{\sc h.a.} and airmass}

The exposure depth depends critically on airmass, which in turn is optimized when each field is observed when it transits (see appendix~\ref{dataquality}, particularly figure~\ref{fig:tauvsha} and equation~\ref{eq:tauandx}). With a schedule in which the allocated distribution of local sidereal time exactly matches the distribution of {\sc h.a.} of the footprint, the scheduler should always schedule exposures as close to transiting as possible (where $\mbox{\sc h.a.} \simeq \mbox{\sc lst}$). Such an exact match was not possible, however, given the narrow distribution in {\sc r.a.} of the wide survey footprint (see the solid histogram in figure~\ref{fig:wideexptimevsra}). The dotted lines in figure~\ref{fig:wideexptimevsra} show the distribution of {\sc lst} from the schedules actually granted to DES. This distribution was as narrow as could be managed given the scheduling constraints imposed by the number of nights in each month available for observing. As shown in this figure, the distribution of time in {\sc lst} roughly matches that of the footprint {\sc r.a.} at early sidereal times ({\sc r.a.} in the western part of the footprint), but has less time than necessary for a perfect match near the center of the footprint, and additional time at later sidereal times. 

Use of these later sidereal time, much but not all of which occurs at the end of each DES observing season, therefore requires careful planning. If observing tactics were to observe transiting fields at all times, then significant progress would be made on the eastern end of the footprint in the middle of the season (a.m. half nights in October and November). When the survey reaches the end of the season, and there is significant time during which the eastern edge of the footprint is all that is observable, most of this area may be complete, while area farther west (in the center of the footprint) will have been missed entirely, and no longer accessible -- we will have ``painted ourselves into a corner.'' \obstac\ must, therefore, observe west of the meridian in order both to complete the center of the footprint, and also allow use of the scheduled time at the end of the observing season.

Early in the season, the {\sc lst} and {\sc r.a.} distributions agree moderately well, and there are dangers associated with working in the west at early sidereal times. Not only will observing too far west early in the the season result in unnecessarily high airmass exposures, it can paint us into a corner in the early part of the year. In the early parts of nights early in the year, the limited area in the west of the footprint is all that is visible. If we observe this area in later parts of those nights and the weather is consistently good, it may be completed before the time during which this part of the footprint is all that is observable is exhausted, again leaving us time during which nothing is observable.\footnote{It may be wondered why we did not request a schedule in which the distribution of {\sc lst} was better centered on the footprint {\sc r.a.}. There are several reasons for this. First, the height of the {\sc lst} peak just east of the footprint peak is partially enabled by the good weather at the site in December and January: the distribution is not only a function of selected nights. Second, if the mismatched time is mostly at the end of the observing season rather than the beginning, then the degree to which the observing tactics depart from observing at transit by going farther west can be tuned to the weather and corresponding progress actually accomplished during the year. If the extra time were at the start of the observing season, we would need to guess at the weather later in the year, and potentially run into more severe issues with either not going far enough east (in a good year) or to far east (in a poor one). Finally, the weather is consistently better at CTIO in December and January.} 

\obstac\ therefore used a hybrid strategy. At early {\sc lst}, it observed in a band $15\degree$ (one hour) wide in {\sc r.a.} centered on the exposure {\sc lst} (transiting). After the {\sc lst} passes a (configurable) boundary, the prioritization switches to observe exposures that reach the 1.4 airmass limit as early as possible, in a band of oblique ascension again $15\degree$ (one hour) wide. In other words, at a certain sidereal time, \obstac\ switches from observing fields as they transit to observing them according to how soon in the year the become inaccessible.

\obstac\ selected $15\degree$ wide bands in each of these cases, rather than simply taking the most extreme pointings, in order to enable other factors to play a significant role. In particular, if the earliest setting or closest to transit exposure is always chosen, then \obstac\ would bounce wildly in declination, resulting in many long slews and wasting telescope time.

The ultimate result of this priority by {\sc h.a.}/airmass was a distribution of {\sc h.a.} values dominated by exposures between 0 and 3 hours, with sharp peaks at both edges of this band.

\subsubsection{Short slews}

The time it takes for the Blanco telescope to slew between pointings can be significant: long slews can take several minutes. A single wide survey exposure takes roughly two minutes, including readout time, so a single unnecessary long slews can cost the survey more than the time it would take for an entire exposure. Minimizing the time spent slewing is therefore a critical factor in observing efficiency. Camera readout and short slews can, however, be completed simultaneously: while longer slews come at a considerable cost in time, slews of less that $\sim 4 \degree$ are usually ``free'', adding no overhead beyond what is required for readout.\footnote{In practice, there is scatter in the times for non-zero slews which can result in an additional overhead of a few seconds, particularly for slews longer than $2\degree$.} \obstac\ therefore prefers exposures it can reach with slews of less than $4\degree$.

Wide survey exposures in {\it g, r, i,} and {\it z} take roughly 2 minutes to complete, and the telescope tracks the field during each exposures, so the {\sc h.a.} of the telescope pointing shifts by $\sim 0.5\degree$ west during each exposure (in horizon coordinates). A pure short slew time requirement therefore pushes the telescope pointing near the edge of the window imposed by airmass limit (or binned tracking window, described above). Rather than preferring short slews exactly, \obstac\ introduces a slight preference for $0.5\degree$ slews to the east, smoothly accommodating the tracking during each exposure using the slew following it.

The combination of the short slew prioritization and working in hour wide bands in {\sc r.a.} or setting {\sc lst} results in sequences of exposures that move in trails across the footprint, moving from west to east in celestial coordinates and maintaining a roughly consistent pointing in horizon coordinates.

\subsubsection{Tilings}

When there are either multiple appropriate exposures reachable with short slews, or no such exposures at all, \obstac\ preferred to select exposures with earlier tiling numbers, encouraging \obstac\ to complete early tilings before making progress on later ones. This had the overall effect of encouraging survey uniformity, and simplified the human task of tracking which exposures had been completed: when there are only a few tilings ``in progress,'' and all others are either complete or unstarted, then only the partially completed tilings need to be mapped and tracked.

\subsubsection{Long slews}

When there are no short slews available, \obstac\ chose among exposures with the same tiling by minimizing the slew distance if any exposures were available less than $35\degree$ away. If it had to slew more than $35\degree$, it selected an exposure as far north (if the starting pointing was north of $-40\degree$ in declination) or south (otherwise) as possible. Exposures near the center of the footprint, within the same band of either {\sc r.a.} or setting {\sc lst}, are observable at a good airmass for a longer period of time, and therefore easier to complete later. When there were unavoidable holes in the footprint of completed exposures, this prioritization tended to place those holes near the central declination of the footprint, where they were easiest to fill in in later years.

\subsection{The desperate tactician}
\label{desperate-tactician}

Although rare, it was possible for \obstac\ to reach the end of the sequence in figure~\ref{fig:obstac} without finding any observable exposures, and there was a short sequence of additional steps it would take. First, it would search for available exposures in tilings not scheduled until future years. Next, it would attempt to schedule a supernova sequence (even if it had been observed recently). Finally, it would select an exposure near zenith, even if it had already been observed, or even fell outside of the DES footprint. The only time it would not schedule an exposure at all is during the day or twilight, when the sky was too bright for observing.

\section{The chronology of the survey}
\label{chronology}
\label{yearly-chronology}

\subsection{Year 1}
\label{year1}

During the first year of observing, which ran from the last day of August, 2013 into the second week of February, 2014, the survey planned to observe a 2000 sq. deg. subset of the full footprint in the first four tilings, in all filters.  This approach was used instead of a steady full footprint, two tilings per year strategy in order to maximize the science possible after the first year: if only two tilings were obtained, it would limit the effectiveness of relative photometric calibration, the gaps between CCDs result in holes in coverage, and the simplest cosmic ray rejection algorithms cannot be applied. With four tilings, these are not issues, and the greater scientific usefulness the area obtained outweighs loss in area. The darker red area in the lower right map of figure~\ref{fig:footprints} shows the area attempted in year 1, which consisted of two sections: one in the south, covering much of the overlap with the SPT footprint, and one in the north, covering much of the equatorial region. The central area and the remainder of the SPT overlap were left for the second year.

The schedule for DES nights in year one was late relative to the footprint, and \obstac\ used an aggressive ``go west'' at all times, always selecting wide survey pointing that were approaching 1.4 airmasses in the west.

\begin{figure*}
\centering

Time usage in year 1

\includegraphics[height=0.4\textheight]{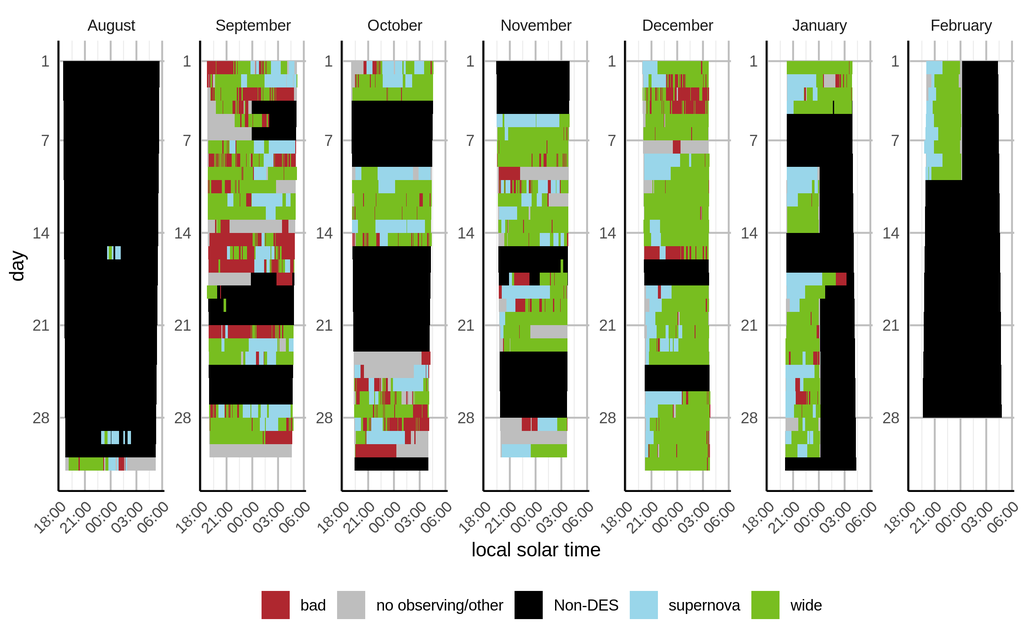}
\caption{\label{fig:desschedy1}
A graphical representation of the first year of DES observing. Blue areas mark time during which useful supernova sequences were completed, and green time during which useful wide survey exposures were taken. Red indicates time during which DES exposures were attempted, but declared bad (for example due to clouds or very poor seeing). Gray indicates time allocated to DES, but during which no science exposures were taken (typically due to the dome being closed due to weather). Black indicates time not allocated to DES.
}
\end{figure*}

Figure~\ref{fig:desschedy1} shows the DES time usage and schedule for year 1. The time lost to clouds during year 1 was typical for the site, with substantial time lost in September, but improving over the the course of the observing season so that very little time was lost after the start of December. The seeing, however, was unexpectedly poor, a problem exacerbated by problems with the dome floor and primary mirror cooling systems, which were repaired in November. Exposures corresponding to the red points in the upper left corner of the left-hand plot of figure~\ref{fig:fwhmvsdimmkolmogorov} were mostly taken before this repair. An unexpectedly high overhead between exposures also reduced the number of good exposures collected. The result of these factors was incomplete coverage in the west of the footprint and the northern (equatorial) area, but complete coverage farther east in the year one area, and even extending beyond it. The generally poor delivered seeing, particularly in the beginning of the year, also resulted in an imbalance in the completing in different filters, with exposures in {\it Y} bands outnumbering those in other bands.

The effects of the supernova vs. wide survey decisions tactics can be seen clearly in figure~\ref{fig:desschedy1}. Note the blue bands (nights of mostly supernova exposures) that follow each of the longer sequences of non-DES nights (indicated by black bands). Over the course of the first year, roughly 30\% of useful observing time was used for supernova sequences, which resulted in a mean cadence of 6.8 nights.

For a more details on year one of DES observing, consult \cite{diehl_dark_2016}.

\subsection{Year 2}
\label{year2}

\begin{figure*}
\centering

Time usage in year 2

\includegraphics[height=0.4\textheight]{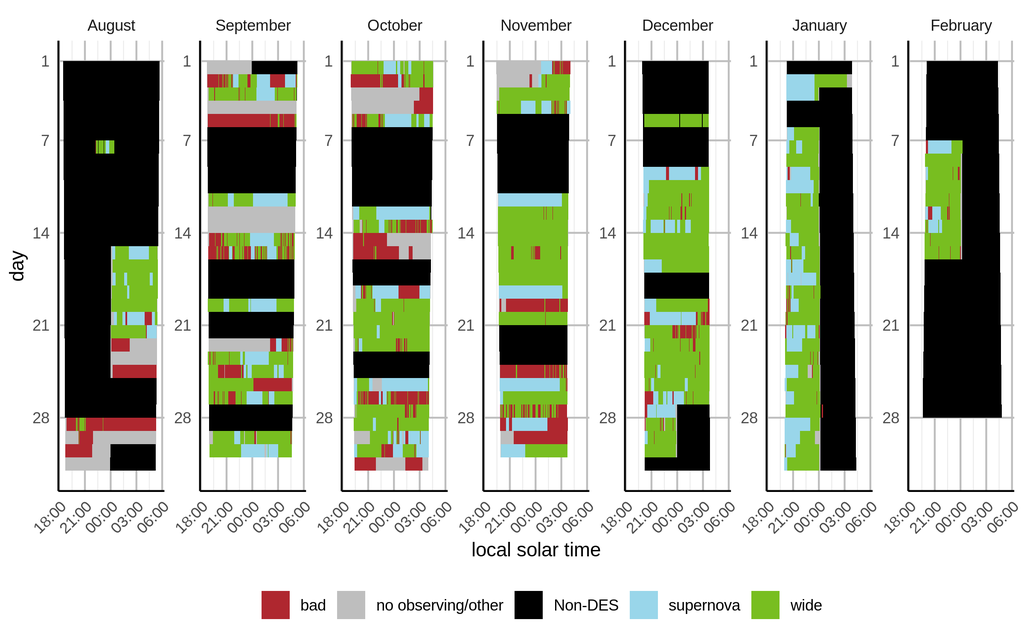}
\caption{\label{fig:desschedy2}
A graphical representation of the second year of DES observing. Blue areas mark time during which useful supernova sequences were completed, and green time during which useful wide survey exposures were taken. Red indicates time during which DES exposures were attempted, but declared bad (for example due to clouds or very poor seeing). Gray indicates time allocated to DES, but during which no science exposures were taken (typically due to the dome being closed due to weather). Black indicates time not allocated to DES.
}
\end{figure*}

The survey worked in the first four tilings again in year two, this time on the complement of the area covered in year one, with the goal of completing the first four tilings over the whole footprint. The overall weather was poor compared to historical averages, but not exceptionally so, again following the typical trend for the site of starting the season with many cloudy nights, which diminish over the course of the observing season (see figure~\ref{fig:desschedy2}). The delivered seeing was better than in year one, resulting in a roughly even distribution of exposures in the different bands at the end of the year. The generally poor weather did lead to the survey being unable to complete the full four tilings over the whole footprint by the end of the year. The combination of the sharp peak in the number of pointings in the center of the footprint and \obstac's tactic of observing in the north or south concentrated the incomplete area in the center of the footprint, creating a hole. Although this concentration in {\sc r.a.} exacerbates the problem of required pointings being concentrated in future years, the concentration in declination allows the completing of these exposure at reasonable airmass even at large hour angles, making it easier to fill in the hole. 

Over the course of the second year, roughly 30\% of useful observing time was used for supernova sequences, which resulted in a mean cadence of 7.1 nights.

For a more details on year two of DES observing, consult \cite{diehl_dark_2016}.

\subsection{Year 3}
\label{year3}

Year three began with the goal of ending the year with six complete tilings over the entire footprint, completing those exposures in the first four tilings not completed in years one and two, and adding two more. In addition to the normal DES wide and supernova observing, the DES shared scheduled nights with a ``target of opportunity'' (ToO) program, such that the two programs were allocated a single block of 108 nights, during which it was expected that the non-DES program use a combined 3 nights. These three nights were to be selected based on an external trigger, independent of predicted weather conditions: the weather ``risk'' was shared equally between the two.

The weather this year began with 10 consecutive nights completely lost to weather, including a blizzard. The overall weather continued to be poor for the rest of the season, resulting in the worst weather recorded for the history of the site. The first four tilings were completed over most of the footprint, but only about half of the footprint reached 5 tilings in {\it g, r, i} or {\it z}, with very little progress in tiling 6 anywhere. (Slightly more progress was made in the {\it Y} band.) Over the course of year three, the wide survey fell about half a year behind the pace needed to finish in 5 years.

Due to the preponderance of poor weather, the 7 night supernova ``dead-man'' trigger resulted in 35\% of useful observing time being spent on supernova sequences, compared to 30\% during the first two years of observing. (See section~\ref{program-selection}.) In spite of this shift in emphasis, the poor weather degraded the time-domain survey as well, resulting in a mean cadence of 7.2 days, a shorter season, and more long gaps.

For a more details on year three of DES observing, consult \cite{diehl_dark_2016}.

\begin{figure*}
\centering

Time usage in year 3

\includegraphics[height=0.4\textheight]{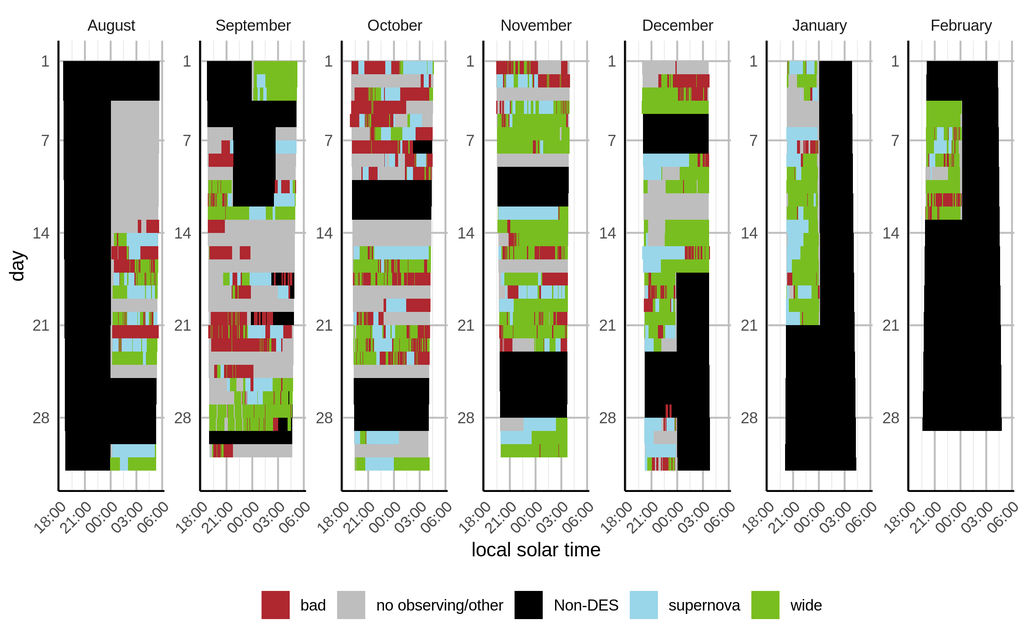}
\caption{\label{fig:desschedy3}
A graphical representation of the third year of DES observing. Blue areas mark time during which useful supernova sequences were completed, and green time during which useful wide survey exposures were taken. Red indicates time during which DES exposures were attempted, but declared bad (for example due to clouds or very poor seeing). Gray indicates time allocated to DES, but during which no science exposures were taken (typically due to the dome being closed due to weather). Black indicates time not allocated to DES.
}
\end{figure*}

\subsection{Year 4}
\label{year4}

Before the start of year 4, two optimizations were made that resulted in improved observing efficiency. First, improvements in the Blanco slew and dome controls significantly reduced the overhead between exposures. Second, instead of completing two 45 second exposures in {\it Y} band tilings 7 through 10, tilings 8 and 10 were dropped and the exposures in tilings 7 and 9 in {\it Y} band were raised to match that of the other filters: 90 seconds instead of 45, thus reducing the total number of exposures (and therefore overhead) without reducing the final photometric signal to noise. This merger was only possible because {\it Y} band exposures are not being used for weak lensing, and therefore do not need exposures spread across many nights.

The sharing of nights with a ToO program continued into year 4.

The overall weather in year 4 was a significant improvement over the previous three, matching the long term average for the site. During this year, all of tiling 6 and most of tiling 7 was completed. For more details on year four of DES observing, consult \cite{diehl_dark_2018}.

\begin{figure*}
\centering

Time usage in year 4

\includegraphics[height=0.4\textheight]{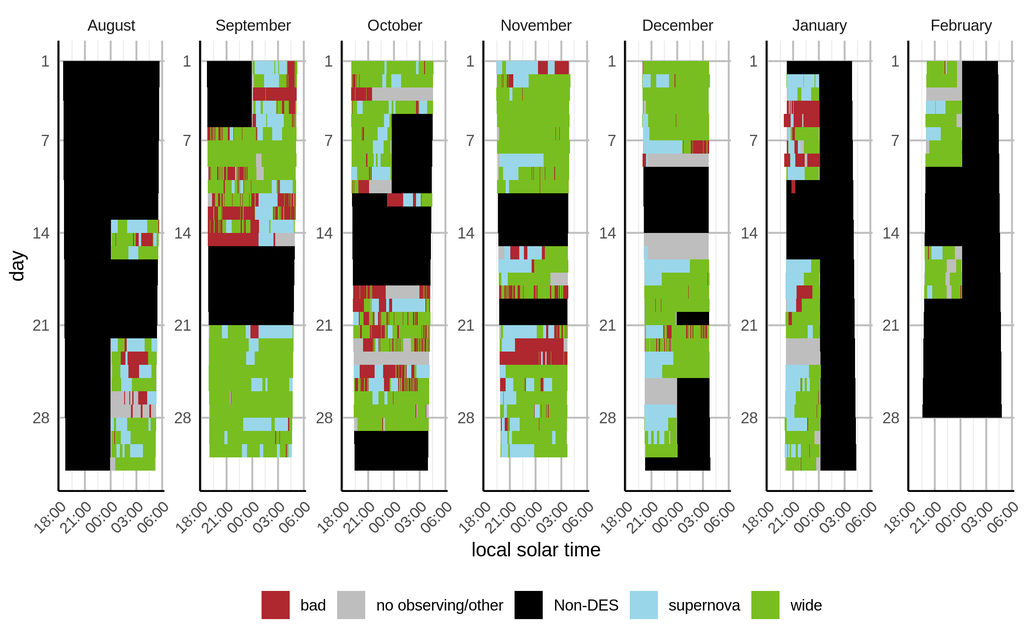}
\caption{\label{fig:desschedy4}
A graphical representation of the fourth year of DES observing. Blue areas mark time during which useful supernova sequences were completed, and green time during which useful wide survey exposures were taken. Red indicates time during which DES exposures were attempted, but declared bad (for example due to clouds or very poor seeing). Gray indicates time allocated to DES, but during which no science exposures were taken (typically due to the dome being closed due to weather). Black indicates time not allocated to DES.
}
\end{figure*}

\subsection{Year 5}
\label{year5}

\begin{figure*}
\centering

Time usage in year 5

\includegraphics[height=0.4\textheight]{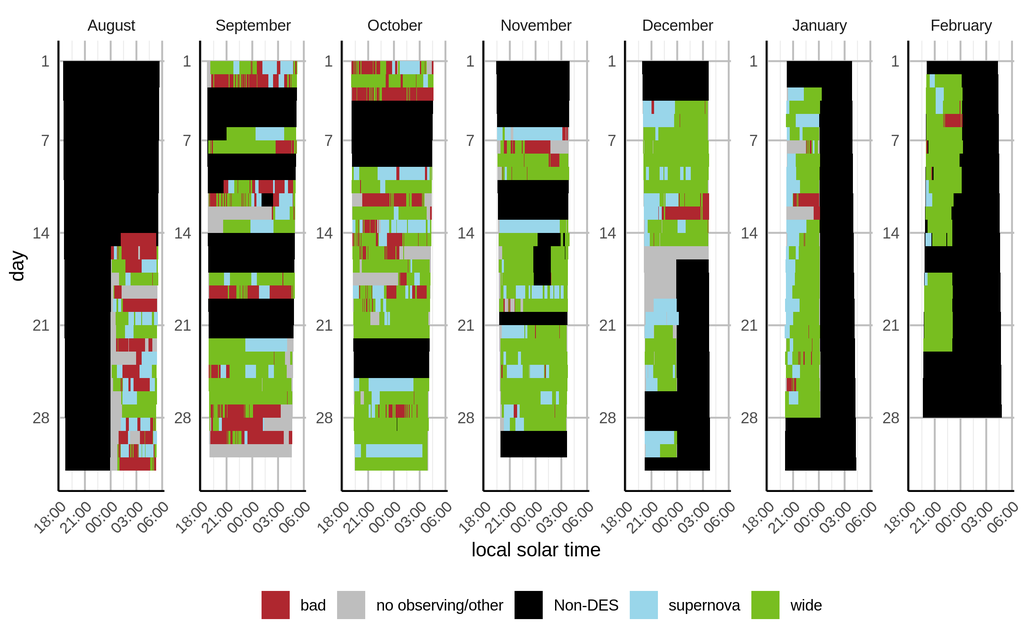}
\caption{\label{fig:desschedy5}
A graphical representation of the fifth year of DES observing. Blue areas mark time during which useful supernova sequences were completed, and green time during which useful wide survey exposures were taken. Red indicates time during which DES exposures were attempted, but declared bad (for example due to clouds or very poor seeing). Gray indicates time allocated to DES, but during which no science exposures were taken (typically due to the dome being closed due to weather). Black indicates time not allocated to DES.
}
\end{figure*}

The fifth year of DES observing was largely a continuation of year four in survey strategy, and a slight improvement in weather conditions (see figure~\ref{fig:desschedy5}). At the end of this year, 9 tilings were complete over most of the footprint, except for a band along $\mbox{\sc r.a.} = 30\degree$, where the peak in the {\sc r.a.} distribution is strongest (see figure~\ref{fig:wideexptimevsra}). For more details on year fifth of DES observing, consult \cite{diehl_dark_2018}.

\subsection{Year 6}
\label{year6}

In spite of ending the five scheduled years with two good seasons, the poor weather in the first three (particularly the third) resulted in the wide survey falling short of the planned 10 tilings by a little more than one tiling. Simulations at the start of year four indicated that this would be the case, even if the weather was perfect for the remainder of the survey. In response, the project instituted a year 6 task-force to study the effect of an additional year of observing on the final DETF and other figures of merit (discussed in section~\ref{intro}). The task-force generated a suite of simulations under which the SN was dropped in year 5 (allowing more time to complete the wide survey), an additional half year was scheduled, and a full sixth year was scheduled; and estimated relative improvements to the figures of merit in each case. The simulated observing included the full range of historical weather conditions and a variety of modifications to survey strategy. The task-force concluded that an additional half year of observing (a sixth year with 52 nights) would result in a significant improvement in the final figure of merit, but a full 105 night 6th year would improve the figure of merit little beyond that. In response, the DES collaboration applied for and received an additional 52 nights in a 6th observing season.

Survey tactics for year 6 deviated from that of previous years for several reasons. First, no supernova sequences were attempted. This provided additional time for wide survey observing, but left no planned use for conditions in which the seeing was too poor for wide survey exposures to be usefully collected. Second, a handful of auxiliary observing programs were completed to improve the calibration of the the survey, explore potential improvements to photo-z measurements, and expand science programs begun with the time-domain fields. These programs were:
\begin{description}
\item[photo-sweep] While the pointing dithering scheme stabilizes the relative photometric calibration of the wide survey on small angular scales, gradual gradients over large angle are possible. To prevent these gradients, DES performed a series of calibration exposures on photometric nights with poor seeing. Each ``sweep'' consisted of a set of exposures dispersed widely across the footprint, in a time window short compared to the timescale of atmospheric transparency variation. By the nature of the program, these sweeps require long slews between exposures. To minimize the time spent waiting for dome adjustments, exposures in a photo-sweep sequence were arranged to scan the sky in azimuth. Observers completed these sweeps using exposures scripts generated by a stand-alone script that ran separately from \obstac. %
\item[deep photo-z fields] Calibration of the photometric estimates of redshifts requires an extensive training set of galaxies with both spectroscopic redshifts and deep imaging in DES filters. Sets of exposures in {\it g, r, i,} an {\it z} were collected for two fields with extensive spectroscopic samples: a field containing the cluster MACS J0416.1-2403 \citep{balestra_clash-vlt:_2016} and the Alhambra-2 field \citep{moles_alhambra_2008}. Observers completed these sequences using exposures scripts written by hand.
\item[reverberation mapping] The CTIO time allocation committee allocated the OzDES collaboration three nights of observing time to monitor the DES time-domain (``supernova'') fields to complement ongoing spectroscopic monitoring of AGN \citep{king_simulations_2015}. These three nights are included in the block of nights used for DES, in exchange for three nights worth of time during the DES survey being dedicated to these sequences on nights close to when spectra were collected. DES observers executed these sequences using hand written exposure scripts.
\item[narrow band] The survey collected a set of exposures on the time-domian survey fields using the N964 filter, an externally supplied DECam filter not normally used by DES \citep{zheng_design_2019}. Sequences of exposures in this narrow band filter were preceded and followed by exposures in {\it z} band designed to identify and eliminate variable objects. These exposures are being used to improve the photo-z redshift estimates in these fields. DES observers executed these sequences using hand written exposure scripts. 
\end{description}

Each of these programs were executed by manual insertion of the exposures into the queue using observing scripts, generated either through stand-alone programs external to \obstac\ or written directly by astronomers. 

Finally, as the survey reached completion, the wide-survey reached states that never occurred in previous years. First, there were times during which the low priority area in the south was the only part of the footprint with uncompleted exposures. If \obstac\ had continued to follow its original strategy of completing exposures by order of tiling and avoiding overlapping exposures, then there was likely that the exposures in this area would be orphaned: that there be area where there are exposures, but not enough to reach a minimum number for inclusion in the final footprint. Therefore, when working in this area:
\begin{itemize}
    \item the ordering by tiling id was removed;
    \item exposures were prioritized north to south, such that the edge of the footprint spread southward at full depth as additional exposures were added; and
    \item the requirement that exposures not overlap other overlapping exposures taken in the same night and the same filter was removed. This reduces the usefulness of this region for weak lensing, but did not effect overall depth.
\end{itemize}
Finally, when all exposures in the wide footprint are complete, \obstac\ would start redoing exposures whose $\tau$ was only slightly above the data quality cutoff. This rarely occurred, because this time was better used for the ancillary programs (listed above).

\begin{figure*}
\centering

Time usage in year 6

\includegraphics[height=0.4\textheight]{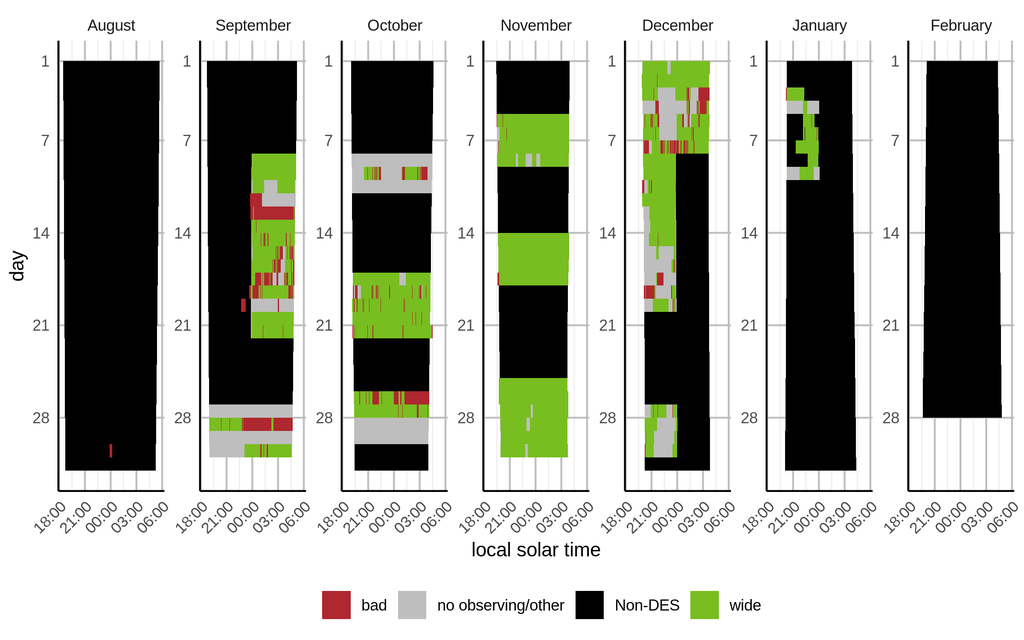}
\caption{\label{fig:desschedy6}
A graphical representation of the sixth year of DES observing. Green areas mark time during which useful wide survey exposures were taken. Red indicates time during which DES exposures were attempted, but declared bad (for example due to clouds or very poor seeing). Gray indicates time allocated to DES, but during which no wide survey exposures were taken (either due to the dome being closed due to weather, or the execution of auxiliary exposures). Black indicates time not allocated to DES.
}
\end{figure*}

Overall, the weather in year 6 was similar to that in years 4 and 5 (see figure~\ref{fig:desschedy6}), and the wide survey completed its objective of full coverage of 5000 square degrees.

For more details on year six of DES observing, consult \cite{diehl_dark_2019}.

\section{The \obstac\ application}
\label{obstac}

\SetStartEndCondition{ }{}{}%
\SetKwProg{Fn}{def}{\string:}{}
\SetKwFunction{Range}{range}%
\SetKw{KwTo}{in}\SetKwFor{For}{for}{\string:}{}%
\SetKwIF{If}{ElseIf}{Else}{if}{:}{elif}{else:}{}%
\SetKwFor{While}{while}{:}{fintq}%
\AlgoDontDisplayBlockMarkers\SetAlgoNoEnd\SetAlgoNoLine%

\subsection{Use cases}

\begin{figure*}
\centering
\includegraphics[width=.9\linewidth]{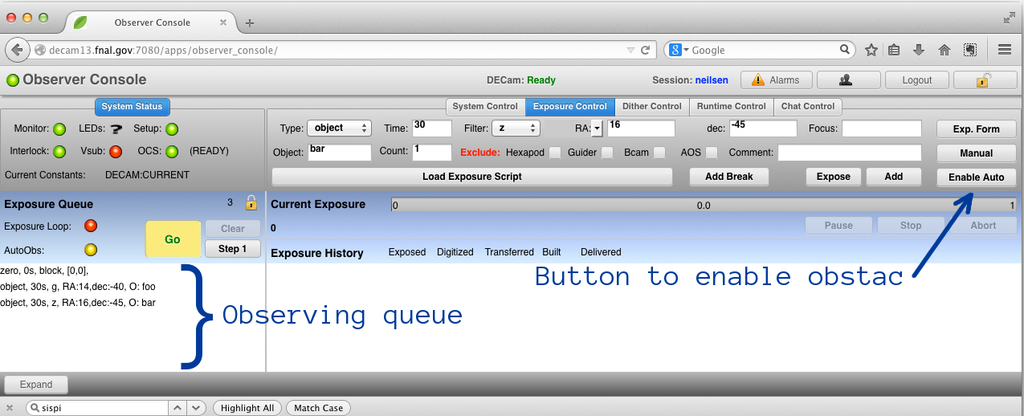}
\caption{\label{fig:expcontrol}
The SISPI exposure control console.}
\end{figure*}

The DES survey scheduler, \obstac, has three primary use cases:
\begin{description}
\item[observing] The primary job of the scheduler is to automate exposure scheduling during DES observing. In this use case, \obstac\ is a component of SISPI, the DECam data acquisition and control system software package \citep{honscheid_read-out_2008}. SISPI presents a web-based user interface to the observers, incorporating a variety of interactive web pages the let the observers monitor the health of the software system, telescope and camera telemetry, and specify and execute exposures. Among these is the exposure control console, shown in figure \ref{fig:expcontrol}. During observing without \obstac, observers set parameters for the exposures they wish to take in the panel near the center of the exposure control console hit the ``add'' button, which causes the new exposure to be added to the exposures queue, the list of exposures on the lower left of the window. When the ``go'' button (just above the observing queue) is pressed, SISPI removes the first (top) exposure from the queue, and begins executing it. When it finishes the first exposures, it moves to the next (the one now at the top of the queue). Observers may edit the queue, adding or removing exposures from the queue at any time, even during the execution of exposures.

During DES data collection, \obstac\ takes over the role of adding exposures to the queue. When the observers are ready to take science data, they hit the "enable auto" button toward the right of the exposure control window. \obstac\ then adds exposures to the queue until it reaches a length defined in the \obstac\ configuration file, choosing exposures according to DES observing strategy.

SISPI removes exposures from the queue and executes them as it does when \obstac\ is not running. Each time the contents of the queue changes (including when SISPI removes the exposure at the head of the queue when it begins to execute it), \obstac\ checks the contents of the queue and adds new exposures if necessary.

When the observers are ready to end DES science data collection, they hit the ``enable auto'' button (now relabeled ``disable auto''), and \obstac\ stops adding new exposures to the queue.

\item[survey simulation] Simulations played an essential role in designing strategy and tactics, constructing schedule requests, and setting expectations. The core \obstac\ scheduler could be run to simulate the results of a given strategy, tactics, and schedule for a given set of starting conditions, under a range of weather conditions. Simulations generated by \obstac\ created tables of all scheduled exposures, with metadata on each exposure's time, conditions, and data quality.  Supporting utilities, part of the \obstac\ product but not involving the scheduler itself, supported and partially automated the configuration, execution, and analysis of these simulations. The simulations themselves were initially run on a small cluster that served as the SISPI test stand, but were then migrated to use FermiGrid and the Open Science Grid (OSG), Fermilab's bulk computing infrastructure. An additional suite of tools in \obstac\ managed the creation and execution of \obstac\ simulation jobs on OSG, and collected and compiled results.

\item[prediction and scripting] In addition to creating full survey simulations, \obstac\ could be used to create schedules for short (up to one night) chunks of time. These simulations did not use simulated weather conditions, but rather conditions specified by the user; and instead of generating lists of exposures with full quality metadata, generated exposure scripts which could be uploaded to SISPI for execution. These scripts were generated automatically each day for the following night of observing, assuming a range of weather conditions, and the results reviewed at the ``run manager's meeting'' that preceded most nights of observing. These scripts indicated what the observers should expect of the upcoming night, alerted the operations team of any unexpected or undesirable scheduling behaviour that might occur, and the scripts themselves could have been used to observe had \obstac\ ever failed in operations (it never did).
\end{description}

In addition to these central scheduling use cases, the \obstac\ package includes utilities that support the above use cases, including tools for listing the observing conditions (sun and moon locations and distances, sky brightness, estimate $\tau$ given an atmospheric seeing, time of rising and setting, etc.) for any given DES exposure at any given time, generation of data for tables that needs to be loaded into the SISPI database for production or simulation to use, checking schedule specification files for the correct numbers of nights and their distributions, and creation of lists of wide survey exposures corresponding to a given footprint specification. 

\subsection{Core scheduling architecture}
\label{coreschedarch}

\subsubsection{Implementation of the Markov Decision Process}

\obstac\ schedules exposures following an architecture following a Markov Decision process (MDP) (see section~\ref{mdp}): the schedule selects each exposure in time order, based on the state of the system (including records of past exposures) at the time. Programmatic elements of \obstac\ (python classes and their children) map to MDP nomenclature as follows:
\begin{description}
\item[environment] \obstac\ represents the interface to the environment using an instance of the {\tt ObsCircumstance} class.  Details of how various aspects of the environment vary based on how the object is configured. For example, depending on the use case, {\tt ObsCircumstance} objects can return the seeing using instances of the {\tt AutoRegSeeingSource}, {\tt DatabaseSeeingSource}, or {\tt ConstantSeeingSource}, depending on whether the seeing is to be determined using the measurements from recent exposures, taken from a database of values used in simulation, or just return the same configured value for all calls.
\item[actor] The \obstac\ actor is an instance of a {\tt Tactician} class, or its children. The structure of the {\tt Tactician} class is slightly different from that of a traditional actor in an MDP, in that it returns a list of zero or more actions, rather than one specific action. (When it comes time to actually take the action, the first in the list is used.) This enables composition of complex {\tt Tactician} objects using combinations of other {\tt Tactician} objects; see below.
\item[actions] \obstac\ represents actions using instances of the {\tt ActionSpec} class and its subclasses, {\tt ObsSpec}, {\tt ObsSequencesSpec}, and {\tt WaitSpec}.
\end{description}

The MDP as used in \obstac\ differs somewhat from that used in dynamic programming or reinforcement learning, in that there is no reward calculated as part of the process: \obstac\ itself does not react to its past success or failure. Instead, when running simulations, the survey scientist used separate utilities to calculate and study metrics of complete simulation runs, and adjusted the algorithm used by \obstac\ by hand. Figure~\ref{fig:fundop} shows an example of using \obstac's environment, actor, and actions to select and take one action.

\begin{figure}
\begin{verbatim}
>>> from obstac import TestEnv
>>> from obstac import astroutil
>>> from obstac.tacticians.StaticTactician import StaticTactician
>>> from obstac.circumstances.ObsCircumstance import ObsCircumstance
>>> from obstac.circumstances.ConstantCloudSource import ConstantCloudSource
>>> from obstac.circumstances.ConstantSeeingSource import ConstantSeeingSource
>>> 
>>> # Instantiate a trivial test tactician 
>>> tactician = StaticTactician(1050, 2, 1, 90, 1)
>>> 
>>> # Create an ObsCircumstance object
>>> 
>>> t0 = astroutil.clock_time('2016-02-15T01:00:00Z')
>>> circ = ObsCircumstance(t0,
...                        cloud_src=ConstantCloudSource(eighths=0),
...                        seeing_src=ConstantSeeingSource(0.85))
>>> 
>>> # actually perform obstac's fundamental operation, selecting an exposure
>>> actions = tactician(circ)
>>> 
>>> # See what we get back
>>> print type(actions)
<type 'list'>
>>> print len(actions)
1
>>> 
>>> action = actions[0]
>>> print type(action)
<class 'obstac.actions.ObsSpec.ObsSpec'>
>>> print action
1 exposure on hex 1096-653, tiling 2 in g, for 90 seconds
>>>
>>> # Check the state of the environment/circumstance before the exposure
>>> print circ.coords
Coords(RA=118.67121347440916, dec=-30.165277780000004)
>>> print astroutil.utciso(circ.current_time)
2016-02-15 01:00:00Z
>>> 
>>> exposure_sequence = action.do(circ)
>>> 
>>> # Check the state of the environment/circumstance after the exposure
>>> print circ.coords
Coords(RA=107.786582844065, dec=-64.85327599999998)
>>> print astroutil.utciso(circ.current_time)
2016-02-15 01:03:31Z
\end{verbatim}
\caption{\label{fig:fundop}
A simple example of \texttt{obstac}'s fundamental operation: selecting an exposure.
}
\end{figure}

\subsubsection{{\tt ObsCircumstance}: the environment}

\begin{figure*}
\centering
\resizebox{0.9\linewidth}{!}{\input{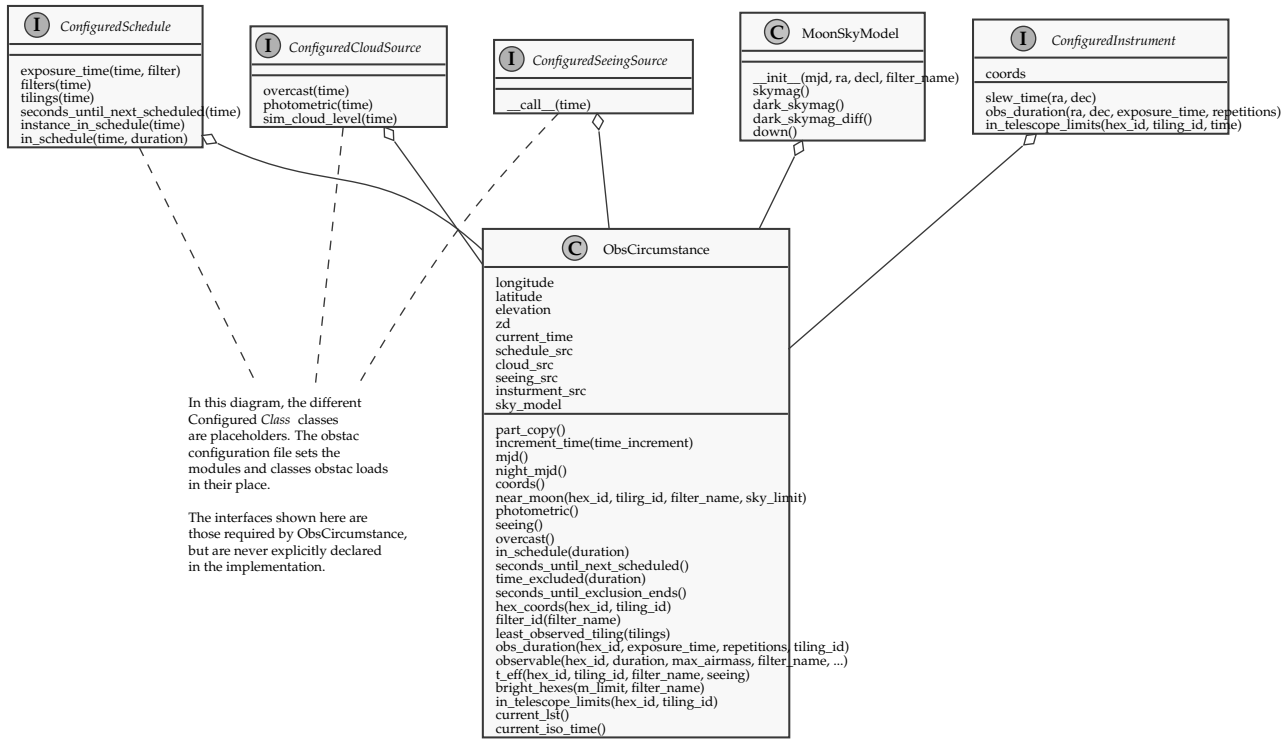}}
\caption{\label{fig:circclassdiag}
Class diagram showing interchangeable elements of \texttt{ObsCircumstance}.}
\end{figure*}

An instance of the {\tt ObsCircumstance} object provides an interface to information on the state of the system. Code used for the calculation or data access required is not generally handled in the implementation of the {\tt ObsCircumstance} class itself, but rather handled by member objects (see figure~\ref{fig:circclassdiag}) provided at the time of instantiation. These can be specified either directly as parameters during the creation of an {\tt ObsCircumstance} instance, or configured in the \obstac\ configuration file. (The configuration file can list a python module and object defined in that module to act as default values for each of these elements). Such configurable elements include:
\begin{description}
\item[\tt schedule\_src] During simulation, the {\tt ObsCircumstance} object keeps track of which time is allocated to DES, to ensure that the simulator only simulates exposures during this time. When configured for production, all time is considered DES time, which allows \obstac\ to schedule for DES even in time during which no DES exposures were expected.
\item[\tt cloud\_src] Also during simulation, {\tt ObsCircumstance} provides data on the how cloudy the conditions are. This data is not used by any of the agents ({\tt Tactician} objects), because they have no access to current data during production, but the \obstac\ simulator uses this feature of the {\tt ObsCircumstance} object to assign data quality.
\item[\tt seeing\_src] The {\tt ObsCircumstance} object provides data on the current value of the seeing. In production, the {\tt AutoRegSeeingSource} implementation uses recent exposures to estimate the seeing; in simulation, {\tt AutoRegSeeingSource} is used in the instance of the {\tt ObsCircumstance} object used by the agent, but {\tt DatabaseSeeingSource} is used in the instance used to calculate final data quality and metadata for simulated exposures. When used for short term prediction and scripting, the user specifies the seeing to be used, and the {\tt ConstantSeeingSource} always returns that value.
\item[\tt sky\_model] There are two models available to calculate the sky brightness, {\tt MoonSkyModel}, which uses the model described in appendix~\ref{skybrightness}, and {\tt KSMoonSkyModel}, which uses a model taken from \cite{krisciunas_model_1991} directly.
\item[\tt instrument\_src] The {\tt Instrument} class defines how slew and readout time are calculated, and defines the telescope limits.
\end{description}

Unlike some MDP implementations, the code in the module that defines the environment does not define how the environment changes state in response to action. Instead, actions include code that uses elements of the {\tt ObsCircumstance} object to calculate how the environment should change, and updates the passed {\tt ObsCircumstance} object; see the sample code in figure~\ref{fig:fundop}.

\subsubsection{{\tt ActionSpec}: the action}

\begin{figure*}
\centering
\resizebox{!}{0.8\textheight}{\definecolor{plantucolor0000}{RGB}{248,248,248}
\definecolor{plantucolor0001}{RGB}{56,56,56}
\definecolor{plantucolor0002}{RGB}{194,194,194}
\definecolor{plantucolor0003}{RGB}{0,0,0}
\scalebox{6.25}{
\begin{tikzpicture}[yscale=-1
,pstyle0/.style={color=plantucolor0001,fill=plantucolor0000,line width=1.5pt}
,pstyle1/.style={color=plantucolor0001,fill=plantucolor0002,line width=1.0pt}
,pstyle2/.style={color=plantucolor0001,line width=1.5pt}
,pstyle4/.style={color=plantucolor0001,line width=1.0pt}
]
\draw[pstyle0] (122.5pt,8pt) rectangle (289.6758pt,92pt);
\draw[pstyle1] (175.0857pt,22pt) ellipse (9pt and 9pt);
\node at (175.0857pt,22pt)[]{\textbf{\Large C}};
\node at (192.5857pt,15.975pt)[below right,color=black]{ActionSpec};
\draw[pstyle2] (123.5pt,36pt) -- (288.6758pt,36pt);
\node at (128.5pt,40pt)[below right,color=black]{tactician};
\draw[pstyle2] (123.5pt,54pt) -- (288.6758pt,54pt);
\node at (128.5pt,58pt)[below right,color=black]{\textit{do(circ: ObsCircumstance)}};
\node at (128.5pt,68pt)[below right,color=black]{\textit{duration(circ: ObsCircumstance)}};
\node at (128.5pt,78pt)[below right,color=black]{update\_cirtumstance(circ: ObsCircumstance)};
\draw[pstyle0] (143.5pt,212pt) rectangle (268.5074pt,296pt);
\draw[pstyle1] (179.9175pt,226pt) ellipse (9pt and 9pt);
\node at (179.9175pt,226pt)[]{\textbf{\Large C}};
\node at (197.1214pt,219.975pt)[below right,color=black]{WaitSpec};
\draw[pstyle2] (144.5pt,240pt) -- (267.5074pt,240pt);
\node at (149.5pt,244pt)[below right,color=black]{wait\_time};
\draw[pstyle2] (144.5pt,258pt) -- (267.5074pt,258pt);
\node at (149.5pt,262pt)[below right,color=black]{\_\_init\_\_(wait\_time, tactician)};
\node at (149.5pt,272pt)[below right,color=black]{do(circ: ObsCircumstance)};
\node at (149.5pt,282pt)[below right,color=black]{duration(circ: ObsCircumstance)};
\draw[pstyle0] (154pt,416pt) rectangle (326.0533pt,750pt);
\draw[pstyle1] (215.6195pt,430pt) ellipse (9pt and 9pt);
\node at (215.6195pt,430pt)[]{\textbf{\Large C}};
\node at (233.1195pt,423.975pt)[below right,color=black]{ObsSpec};
\draw[pstyle2] (155pt,444pt) -- (325.0533pt,444pt);
\node at (160pt,448pt)[below right,color=black]{hex\_id};
\node at (160pt,458pt)[below right,color=black]{tiling\_id};
\node at (160pt,468pt)[below right,color=black]{program\_id};
\node at (160pt,478pt)[below right,color=black]{filter\_id};
\node at (160pt,488pt)[below right,color=black]{exptime};
\node at (160pt,498pt)[below right,color=black]{repetitions};
\node at (160pt,508pt)[below right,color=black]{min\_airmass};
\node at (160pt,518pt)[below right,color=black]{max\_airmass};
\node at (160pt,528pt)[below right,color=black]{exptype};
\node at (160pt,538pt)[below right,color=black]{expwait};
\draw[pstyle2] (155pt,552pt) -- (325.0533pt,552pt);
\node at (160pt,556pt)[below right,color=black]{\underline{\_\_init\_\_(filter\_id, hex\_id, tiling\_id,...)}};
\node at (160pt,566pt)[below right,color=black]{\underline{fromSequence()}};
\node at (160pt,576pt)[below right,color=black]{do(circ: ObsCircumstance)};
\node at (160pt,586pt)[below right,color=black]{duration(circ: ObsCircumstance)};
\node at (160pt,596pt)[below right,color=black]{update\_circumstance(circ: ObsCircumstance)};
\node at (160pt,606pt)[below right,color=black]{to\_sequence()};
\node at (160pt,616pt)[below right,color=black]{time\_limits(circ: ObsCircumstance)};
\node at (160pt,626pt)[below right,color=black]{hex\_name()};
\node at (160pt,636pt)[below right,color=black]{filter\_name()};
\node at (160pt,646pt)[below right,color=black]{program\_name()};
\node at (160pt,656pt)[below right,color=black]{ra()};
\node at (160pt,666pt)[below right,color=black]{decl()};
\node at (160pt,676pt)[below right,color=black]{utciso\_time\_limits()};
\node at (160pt,686pt)[below right,color=black]{earliest\_start()};
\node at (160pt,696pt)[below right,color=black]{latest\_start()};
\node at (160pt,706pt)[below right,color=black]{estimated\_start()};
\node at (160pt,716pt)[below right,color=black]{estimated\_airmass()};
\node at (160pt,726pt)[below right,color=black]{queue\_dict()};
\node at (160pt,736pt)[below right,color=black]{observable(circ: ObsCircumstance, sky\_limit)};
\draw[pstyle0] (304pt,152pt) rectangle (476.0533pt,356pt);
\draw[pstyle1] (343.8544pt,166pt) ellipse (9pt and 9pt);
\node at (343.8544pt,166pt)[]{\textbf{\Large C}};
\node at (361.3544pt,159.975pt)[below right,color=black]{ObsSequenceSpec};
\draw[pstyle2] (305pt,180pt) -- (475.0533pt,180pt);
\node at (310pt,184pt)[below right,color=black]{hex\_seq};
\node at (310pt,194pt)[below right,color=black]{repetitions};
\node at (310pt,204pt)[below right,color=black]{tactician};
\node at (310pt,214pt)[below right,color=black]{max\_airmass};
\node at (310pt,224pt)[below right,color=black]{program\_id};
\node at (310pt,234pt)[below right,color=black]{hex\_id};
\node at (310pt,244pt)[below right,color=black]{seq\_id};
\node at (310pt,254pt)[below right,color=black]{ospecs};
\draw[pstyle2] (305pt,268pt) -- (475.0533pt,268pt);
\node at (310pt,272pt)[below right,color=black]{\underline{\_\_init\_\_(filter\_id, hex\_id, ...)}};
\node at (310pt,282pt)[below right,color=black]{hex\_name()};
\node at (310pt,292pt)[below right,color=black]{do(circ: ObsCircumstance)};
\node at (310pt,302pt)[below right,color=black]{duration(circ: ObsCircumstance)};
\node at (310pt,312pt)[below right,color=black]{update\_circumstance(circ: ObsCircumstance)};
\node at (310pt,322pt)[below right,color=black]{observable(circ: ObsCircumstance, sky\_limit)};
\node at (310pt,332pt)[below right,color=black]{time\_limits(circ: ObsCircumstance)};
\node at (310pt,342pt)[below right,color=black]{queue\_dicts(circ: ObsCircumstance, seqid)};
\draw[pstyle0] (6pt,496pt) rectangle (79.8667pt,670pt);
\draw[pstyle1] (31pt,510pt) ellipse (9pt and 9pt);
\node at (31pt,510pt)[]{\textbf{\Large C}};
\node at (45.6667pt,503.975pt)[below right,color=black]{Obs};
\draw[pstyle2] (7pt,524pt) -- (78.8667pt,524pt);
\node at (12pt,528pt)[below right,color=black]{ra};
\node at (12pt,538pt)[below right,color=black]{decl};
\node at (12pt,548pt)[below right,color=black]{filter};
\node at (12pt,558pt)[below right,color=black]{exptim};
\node at (12pt,568pt)[below right,color=black]{program};
\node at (12pt,578pt)[below right,color=black]{hex};
\node at (12pt,588pt)[below right,color=black]{tiling};
\node at (12pt,598pt)[below right,color=black]{eariest\_start};
\node at (12pt,608pt)[below right,color=black]{latest\_start};
\node at (12pt,618pt)[below right,color=black]{estimated\_start};
\node at (12pt,628pt)[below right,color=black]{estimated\_airmass};
\node at (12pt,638pt)[below right,color=black]{exptype};
\node at (12pt,648pt)[below right,color=black]{expwait};
\draw[pstyle2] (7pt,662pt) -- (78.8667pt,662pt);
\draw[pstyle0] (61.5pt,810pt) rectangle (172.1286pt,854pt);
\draw[pstyle1] (74.5pt,824pt) ellipse (9pt and 9pt);
\node at (74.5pt,824pt)[]{\textbf{\Large C}};
\node at (86.5pt,817.975pt)[below right,color=black]{ObsCircumstance};
\draw[pstyle2] (62.5pt,838pt) -- (171.1286pt,838pt);
\draw[pstyle2] (62.5pt,846pt) -- (171.1286pt,846pt);
\draw[fill=black,line width=1.0pt] (117pt,583pt) ellipse (2pt and 2pt);
\draw[pstyle4] (206pt,112.3782pt) ..controls (206pt,144.6039pt) and (206pt,183.0772pt) .. (206pt,211.7852pt);
\draw[pstyle4] (199.0001pt,112.269pt) -- (206pt,92.2689pt) -- (213.0001pt,112.2689pt) -- (199.0001pt,112.269pt) -- cycle;
\draw[pstyle4] (149.7934pt,108.3626pt) ..controls (140.0336pt,121.7418pt) and (131.3208pt,136.5942pt) .. (126pt,152pt) ..controls (96.4014pt,237.6993pt) and (105.0914pt,267.7771pt) .. (126pt,356pt) ..controls (132.3562pt,382.8198pt) and (142.4037pt,410.0742pt) .. (153.9613pt,435.9525pt);
\draw[pstyle4] (144.6334pt,103.6032pt) -- (162.5393pt,92.2727pt) -- (155.6074pt,112.2964pt) -- (144.6334pt,103.6032pt) -- cycle;
\draw[pstyle4] (257.8249pt,107.458pt) ..controls (272.1304pt,123.3185pt) and (288.0487pt,140.967pt) .. (303.7432pt,158.3674pt);
\draw[pstyle4] (252.3223pt,111.8087pt) -- (244.1249pt,92.2689pt) -- (262.7183pt,102.432pt) -- (252.3223pt,111.8087pt) -- cycle;
\draw[pstyle4] (337.8184pt,368.4516pt) ..controls (330.8221pt,383.7969pt) and (323.5329pt,399.7845pt) .. (316.2282pt,415.8062pt);
\draw[color=plantucolor0001,fill=plantucolor0001,line width=1.0pt] (343.3613pt,356.2942pt) -- (337.2327pt,360.0942pt) -- (338.3831pt,367.2129pt) -- (344.5118pt,363.4129pt) -- (343.3613pt,356.2942pt) -- cycle;
\draw[pstyle4] (80.2891pt,583pt) ..controls (91.8568pt,583pt) and (103.4246pt,583pt) .. (114.9924pt,583pt);
\draw[pstyle4] (119.4023pt,583pt) ..controls (130.8654pt,583pt) and (142.3285pt,583pt) .. (153.7915pt,583pt);
\draw[color=plantucolor0001,line width=1.0pt,dash pattern=on 7.0pt off 7.0pt] (117pt,585.1526pt) ..controls (117pt,604.7364pt) and (117pt,752.6416pt) .. (117pt,809.6701pt);
\end{tikzpicture}
}}
\caption{\label{fig:actionclassdiag}
Class diagram for \obstac\ actions.}
\end{figure*}
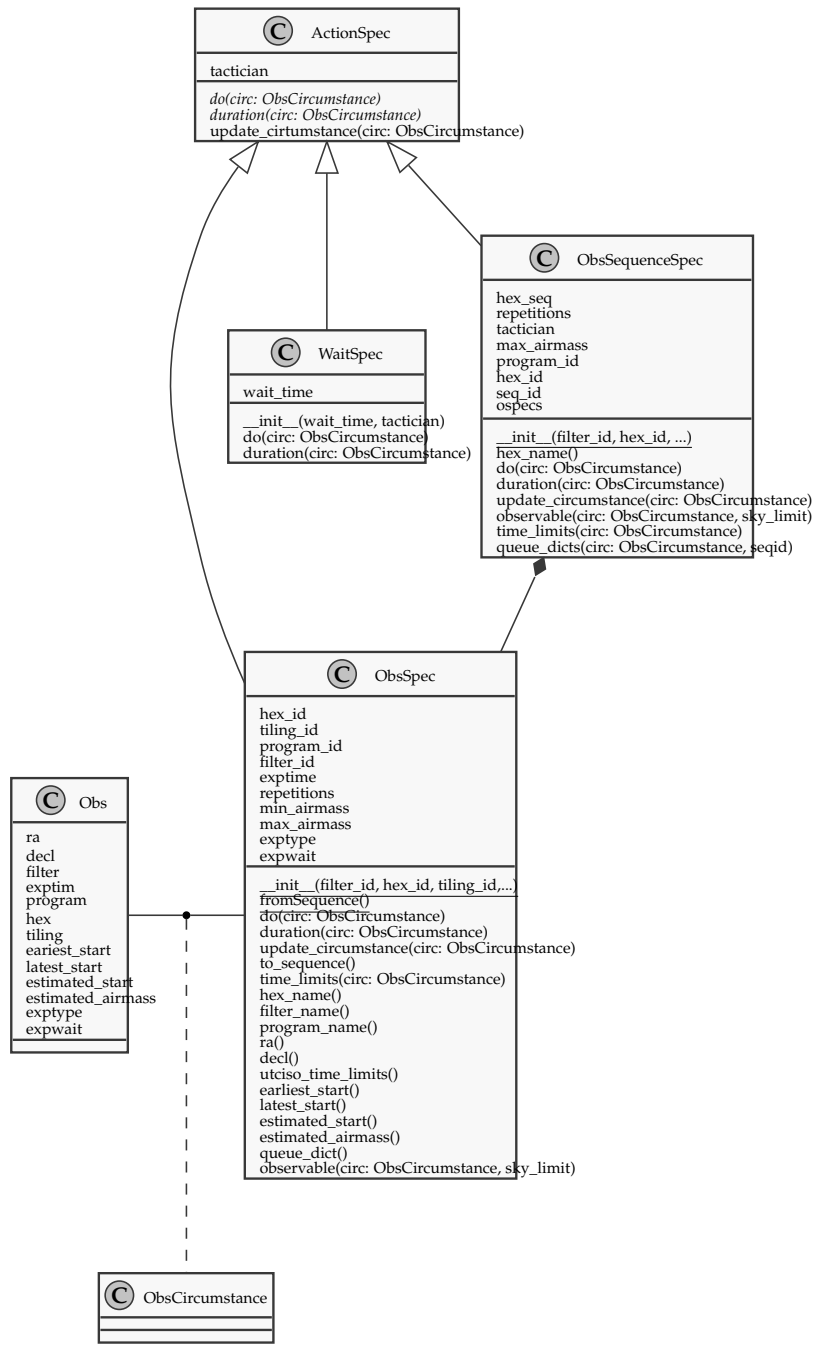

\obstac\ actions represent actions to be scheduled as a single unit. There are three defined actions: an observation ({\tt ObsSpec}), a sequence of observations ({\tt ObsSequenceSpec}), and waiting ({\tt WaitSpec}).  Action objects are also responsible for some calculations related to the actions they represent. In simulations, all action objects are responsible for being able to update a circumstance object to reflect the execution of an action. Figure~\ref{fig:actionclassdiag} shows the class diagram for action objects.

In addition to simply storing the parameters of an action, an {\tt Action} object includes a {\tt do} method, which takes an {\tt ObsCircumstance} object as a parameter, updates it in response to the action, and returns all the parameters that need to be passed to {\tt SISPI} in order to add the exposure (or sequences of exposures) to the queue. For example, a simple {\tt Action} object does not include data on the earliest or latest time at which each exposure can be started, because it depends on the time for which it will be scheduled. The {\tt namedtuple} returned by the {\tt do} method includes all of this data.

Figure~\ref{fig:actionspec} shows an example of creating an action representing a typical survey exposure.

\begin{figure*}
\begin{verbatim}
>>> from obstac import TestEnv
>>> from obstac.actions.ObsSpec import ObsSpec
>>> from obstac import astroutil
>>> from obstac.circumstances.ObsCircumstance import ObsCircumstance
>>> from obstac.circumstances.ConstantCloudSource import ConstantCloudSource
>>> from obstac.circumstances.ConstantSeeingSource import ConstantSeeingSource
>>> 
>>> # Create an ObsCircumstance object
>>> 
>>> circ = ObsCircumstance(astroutil.clock_time('2016-02-15T01:00:00Z'),
...                        cloud_src=ConstantCloudSource(eighths=0),
...                        seeing_src=ConstantSeeingSource(0.85))
>>> 
>>> action =  ObsSpec(1, 1050, 2, 2, 90)
>>> print action
1 exposure on hex 1096-653, tiling 2 in g, for 90 seconds
>>>
>>> # Check the state of the environment/circumstance before the exposure
>>> print circ.coords
Coords(RA=118.67121347440916, dec=-30.165277780000004)
>>> print astroutil.utciso(circ.current_time)
2016-02-15 01:00:00Z
>>> 
>>> exposure_sequence = action.do(circ)
>>> for exposure in exposure_sequence:
...     for key, value in zip(exposure._fields, exposure):
...         print "%
...
ra: 107.786582844
decl: -64.853276
filter: g
exptime: 90.0
program: survey
hex: 1096-653
tiling: 2
earliest_start: 2016-02-15T00:26:47Z
latest_start: 2016-02-15T09:05:05Z
estimated_start: 2016-02-15T01:00:00Z
estimated_airmass: 1.24623812163
exptype: object
expwait: False
>>> 
>>> # Check the state of the environment/circumstance after the exposure
>>> print circ.coords
Coords(RA=107.786582844065, dec=-64.85327599999998)
>>> print astroutil.utciso(circ.current_time)
2016-02-15 01:03:31Z
\end{verbatim}
\caption{\label{fig:actionspec}
Examples of \obstac\ action specifications.
}
\end{figure*}

\subsubsection{{\tt Tactician}: the agent}

An \obstac\ {\tt Tactician} object is a python function or other callable object that takes an {\tt ObsCircumstance} object as its only argument, and returns a list of zero or more actions. The true ``agent'' in the MDP is a specific {\tt Tactician} that always returns a list at least one element long, such that the higher level code is always returned at least one action to execute. {\tt Tactician} objects can be composed using other {\tt Tactician} objects, and \obstac\ includes several classes of {\tt Tactician} objects designed for such composition. For example, if we wanted a tactician that would chose an {\it i} band wide survey exposure if the seeing is better than 1.3'' and a {\it Y} band exposure otherwise, the we could define a tactician thus:
\begin{verbatim}
>>> i_tactician = SurveyTactician(seeing_limit=1.3, preferred_filters=['i'])
>>> y_tactician = SurveyTactician(preferred_filters=['Y'])
>>> tactician = CascadingTactician([g_tactician, y_tactician])
\end{verbatim}
If we wanted to ensure that \obstac\ chooses a {\tt Wait} action when the solar zenith distance was less than $100\degree$, but used the above tactician otherwise, we could do this instead line:
\begin{verbatim}
>>> i_tactician = SurveyTactician(seeing_limit=1.3, preferred_filters=['i'])
>>> y_tactician = SurveyTactician(preferred_filters=['Y'])
>>> exposure_tactician = CascadingTactician([g_tactician, y_tactician])
>>> tactician = DayTactician(zenith_distance=100, exposure_tactician)
\end{verbatim}
The top level scheduler for DES observing combines a variety of tacticians in this way to implement the tactics described in section~\ref{tactics}.

\subsection{Observing in operations}

The DECam data acquisition system, SISPI \citep{honscheid_read-out_2008}, consists of a collection of applications (the guider, readout control system, image stabilization, telescope control system interface, etc.) and infrastructure elements (a database, logger and alarm system, configuration management system) that communicate through a common communications system. The communication system supplies a mechanism for shared variables (based on a publisher/subscriber architecture) and remote procedure calls between applications (the Python Messaging Library (PML), a python implementation of the Soar Messaging Library (SML, \cite{schumacher_soar_2004}) based on the python remote objects (PYRO, \cite{uber_ai_labs_pyro_2019}) module). Two SISPI applications (``roles'') are supplied by \obstac\ itself, and these applications directly interact with several other elements of SISPI:
\begin{description}
\item[\tt OCS] The observatory control system (OCS) SISPI application coordinates camera operation in taking exposures or sequences of exposures. OCS includes a data structure that holds the specifications for upcoming exposures: on observing queue. When observing, it removes exposures from the head of the queue and coordinates other parts of the system in collecting the exposure, and then moves to the next exposure of the queue. Other parts of the system can edit (add, remove, or reorder) the queue through PML messages to OCS. The queue can be edited while observing is ongoing, such that future exposures can be planned while an exposure is taking place.
\item[\tt GUI] The primary\footnote{There is also a purely text-based interface.} user interface to SISPI is a series of web applications. One of these web applications is the exposure control console, shown in figure~\ref{fig:expcontrol}. It displays the contents of the queue (managed by OCS), and features controls that observers can use to edit the queue, and initiate or halt execution of the queue.
\item[\tt OBSTACSRVR] The \obstac\ server provides a PML interface to \obstac's scheduling functions. The primary service it offers is {\tt fill\_queue}. When another component of the SISPI system issues the {\tt fill\_queue} command to {\tt OBSTACSRVR}, it selects new exposures and adds them to the OCS queue (by issuing PML commands to OCS) until the queue reaches a (configurable) length, at which point it stops until it receives another {\tt fill\_queue} command.
\item[\tt AUTOOBS] The {\tt AutoObs} application is also supplied by \obstac\ itself, and triggers {\tt OBSTACSRVR} when necessary. {\tt AUTOOBS} publishes a shared variable that determines whether or not \obstac\ is enabled. The SISPI GUI, which subscribes to this shared variable, can be used by the observer to enable on disable it. On starting, {\tt AUTOOBS} creates a call-back to the OCS, which then alerts it whenever there is a change to the queue (for example due to OCS starting an exposure and removing the first entry from the queue). If the callback is called when {\tt AUTOOBS} is enabled, and {\tt AUTOOBS} is not already waiting for {\tt OBSTACSRVR} to add entries to the queue, then it triggers {\tt OBSTACSRVR} to fill the queue.
\end{description}

\begin{figure*}
\centering
\resizebox{!}{0.9\textheight}{\input{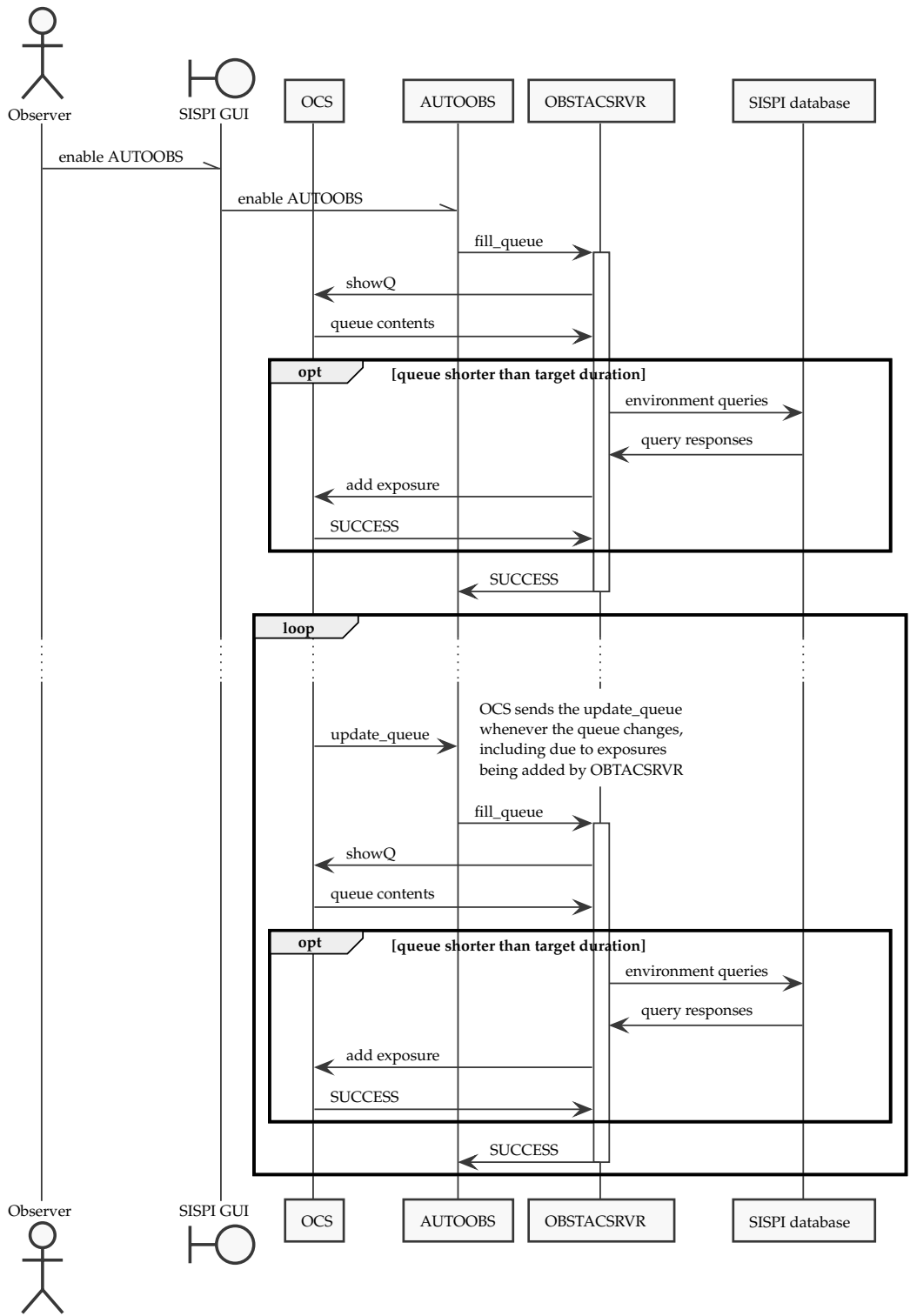}}
\caption{\label{fig:prodseq}
UML Sequence diagram showing interaction between \obstac\ roles and other elements of SISPI.}
\end{figure*}

Figure~\ref{fig:prodseq} shows the interaction between these SISPI elements during routine \obstac\ scheduled observing. 

In addition to the modules supporting the core scheduling operations (described in section~\ref{coreschedarch}), several \obstac\ submodules support interaction with SISPI and are required for operation in production. The {\tt AutoObs} module implements the {\tt AUTOOBS} role. The {\tt obstacdb} module manages interactions with the database. An {\tt ObsQueue} class provides an interface to the observing queue, and its subclass {\tt ObsQueueOCS} specializes this for interaction with the OCS queue (a separate subclass is used in simulation). 

\subsection{Simulations}

\subsubsection{Levels of simulation tools}

Simulations using \obstac\ use many of the same elements used in production. The actions and tacticians are the same. The same {\tt ObsCircumstance} class is used to manage the environment. A different instance of the same database ({\tt postgresql}) is used, using a subset of the schema used in the SISPI database: all tables and other schema elements used by \obstac\ are replicated in the databases used in simulation. Rather than use the queue implementation supplied by OCS, the simulation queue manages the contents of the queue internally, but the interface to this queue is the same as that seen in production.

The \obstac\ simulator can be viewed from three separate levels:
\begin{enumerate}
    \item Given an existing \obstac\ installation, configuration, and simulation instance of the database, the lowest level simulation driver selects exposures for a specified period of time, updates the test database, and creates a table of selected exposures (including timing and simulated data quality).
    \item Given a subversion tag for the version of \obstac\ to be used, an assortment of configuration parameters that define not only \obstac's tactics but simulation input and survey definition, the {\tt prep-sim.sh} and {\tt run-sim.sh} shell scripts check out the desired version of \obstac, create and instantiate a simulation instance of the database, and run the simulation.
    \item Given a subversion tag for the version of \obstac\ to be used and an assortment of configuration parameters that define not only \obstac's tactics but simulation input and survey definition, the {\tt obssim} collection of scripts generate input files needed for suites of simulations to be submitted to the Open Science Grid \citep{herner_advances_2016}, submit them, monitor their progress, collect the results, and produce summary statistics. These scripts manage the storage and bookkeeping of simulation specifications, metadata, results, and summary statistics.
\end{enumerate}

Although the first two levels were developed at the start of the survey and changed only modestly throughout, the third level evolved significantly over time. Initial planning was performed primarily using simulations run on a small cluster designed to be the SISPI test stand, executed mostly with simple commands and with human-updated book keeping. Over the course of the survey, simulation was migrated to more capable facilities with more complex processes for submitting and managing jobs, and the book keeping increasingly automated.

\subsubsection{Individual simulation execution}

Each level in this sequence wraps the operations of the level preceding it. The first level is the one that actually executes the simulation of survey tactics, following algorithm~\ref{alg:simdriver}. The actual implementation of the central loop of algorithm~\ref{alg:simdriver} resembles the code in figure~\ref{fig:fundop}, but in addition to iterating over the scheduled time in the survey, it assigns data quality (using simulations of seeing and replayed records of clouds) and updates the simulation database and record of exposures after every selection. 

\begin{algorithm}
\SetAlgoLined
initialization\;
\Repeat{survey over} {
 select action\;
 \If{observation}{
 assign seeing\;
 assign clouds\;
 update database\;
 update telescope pointing\;
 update exposure list\;
 }
 update time\;
}
\caption{Outline of the simulation driver}
\label{alg:simdriver}
\end{algorithm}

Without DES Data Management's data quality assessments, the \obstac\ simulations must assign data quality to complete exposures using simulated or historical data. CTIO supplied records of cloud cover data starting at the beginning of 1975, in the form of human recorded eighths cloud cover for each quarter of the night (where 0 designates a photometric quarter night, and 8 a completely overcast one), as estimated in the night logs from the observatory's telescopes. The user supplies an offset between simulated date and reference weather data as a parameter when running and \obstac\ simulation. For example, if the user sets the {\tt cloud\_offset} parameter to 30, then the \obstac\ simulator will simulate the first quarter of MJD 57672 with the recorded first quarter of MJD $57672-\mbox{\tt round}(30\times365.25)=46715$. With 38 years historical years to work with, this procedure allows for the simulation of 7 independent 5 year simulations, or 33 correlated ones. As the survey progressed and there were fewer years left to simulate, the number of independent simulations of all remaining years grew correspondingly.

\obstac\ simulations use pre-generated sets of artificial seeing data generated according to the seeing model described in appendix~\ref{seeing}. When starting a simulation, the user specified an index for the simulated seeing data to use in the {\tt seeing\_index} parameter.

\subsubsection{Automatic initialization and execution}

Execution of the first level of simulation requires an existing database. \obstac\ depends on the SISPI database not only for metadata on exposures already taken, but also tables of pointings for all desired exposures and their priorities, tables of scheduled nights, and other metadata on the nights and exposures. The second level driver for \obstac\ simulations generates this database from scratch, deriving database tables from text files that specify pointings, dithers, exposure times, time-domain sequence definitions, and nights scheduled, and querying the production SISPI database for completed exposures to serve as the starting point for the simulation.

\subsubsection{Bulk simulation on utility computing resources}

While a single simulation can give an initial idea of how well a given configuration and version of \obstac\ will perform, planning requires an understanding of how well it performs under a range of weather conditions. The third level simulation tools manage execution of simulations covering ranges of conditions, and provide the tools needed for the survey scientist to keep track of which simulations have been run with what \obstac\ versions, configurations, weather conditions, and results.

The procedure for generating a set of simulations comprised the following steps:
\begin{enumerate}
    \item Identify the {\tt subversion} tag or version number of the version of \obstac\ to use, and run {\tt obssim tar\_abstac} to check the designated \obstac\ out of {\tt subversion}, create a corresponding tarball, and store it in a standard archive directory.
    \item Create configuration files for database initialization and \obstac\ itself, and copy them into an archive directory with a unique configuration id (chosen by hand by the user, but generally numeric and increasing). Record the configuration id and a comment in a metadata file (by hand).
    \item Run {\tt obssim dag} to generate specifications for OSG jobs called DAGs.\footnote{for directed acyclic graphs, because the dependencies between jobs are specified in such a data structure.}  This command generates a collection of DAGs that simulate the specified \obstac\ version and configuration under a range of weather conditions, possibly splitting simulations into smaller sequences of dates executed successively to avoid submitting jobs that are too long to conveniently process on available computing resources.
    \item Actually submit the compute jobs using {\tt obssim submit}, which not only submits the jobs but also records metadata about the jobs so that they can be tracked, and their results retrieved.
    \item Monitor the progress of the jobs, and kill restart jobs as necessary. {\tt obssim q} lists the status of running jobs, {\tt obssim held} lists held jobs, and {\tt obssim kill} kills running jobs. {\tt obssim submit} can be used to restart a killed job.
    \item Retrieve the results with {\tt obssim results}, which copies the output files of all completed jobs into a consistent directory structure. These results include the \obstac\ log files, the table of exposures selected and their simulated data quality metrics, and a full dump of the state of the SISPI database at the end of the simulation.\footnote{If {\tt obssim dag} splitted a simulation into multiple data ranges to be submitted in turn, {\tt obssim results} will retrieve any completed intermediary results as well, which is useful for getting fast results for the initial part of the simulations}.
    \item Finally, use {\tt obssim analyze} to generate summary statistics for completed simulations, and update a web page with a table of high level statistics covering all completed simulations, and web pages with plots and lower level statistics for each simulation.
\end{enumerate}

\subsection{{\tt obscript}: prediction and scripting}

In addition to adding exposures directly to the SISPI queue and running long range survey simulations, the \obstac\ scheduler could be used to select exposures for short ranges of time (a night or less). The {\tt obscript} tool accepts a start time and date, duration, and specification of seeing conditions, and generates a schedule for a list of exposures covering that time. Unlike the full simulator, {\tt obscript} never writes to the database, and so it can use the production {\tt SISPI database} or its read-only mirror. 

This functionality can also be used to generate SISPI scripts for other purposes: modifications or completely new {\tt Tactician} classes can be written, and then called within the \obstac\ python environment to create scripts that can then be run by SISPI without the need for modifying the production version of \obstac.

\section{\obstac\ after DES}
\label{afterdes}

DES is not the only large observing program to use DECam. Other such programs also need to write scheduling software to schedule exposures. The typical mechanism for doing this has been to use stand-alone utilities that generate SISPI scripts, which are then uploaded into SISPI by the observers. After the completion of DES and the start of the DELVE survey \citep{drlica-wagner_program_2019}, the {\tt AUTOOBS} component of \obstac\ was modified so that instead of using \obstac's own scheduler, it acts as an interface between SISPI and a scheduling process running outsite of SISPI.

Like the {\tt AUTOOBS} implementation supplied by the \obstac\ version used for DES, the post-DES {\tt AUTOOBS} uses a shared variable (controllable from the SISPI GUI) to track whether or not automatic scheduling is enabled, and a callback is used in SISPI to trigger the {\tt AUTOOBS} process each time any change is made to the queue. Instead of scheduling exposures using a SISPI process, however, the post-DES {\tt AUTOOBS} does the following each time the callback is triggered:\footnote{Exact file names given here are those currently configured for use at the observatory. The files and directories are, however, configurable parameters, and can be set in the SISPI initialization file.}
\begin{enumerate}
    \item If it exists, copy \verb|~sispi/obstac/queue/current.json| to \verb|~sispi/obstac/queue/previous.json|.
    \item Query OCS for the current contents of the queue, and write them as a json file to \verb|~sispi/obstac/queue/current.json|.
    \item Query OCS for current ongoing exposures, and write them as a json file to \verb|~sispi/obstac/queue/inprogress.json|.
    \item If {\tt AUTOOBS} is enabled,
    \begin{enumerate}
        \item Write the current time (as an ISO 8601 date/time string) to the named pipe \verb|/tmp/obstac_fifo.txt|.
        \item Wait (until a timeout) for the file \verb|~sispi/obstac/inbox/queue.json| to exist, with a time newer than the time stamp written.
        \item When the file appears, move it to \verb|~sispi/obstac/loaded/queue_${DATETIME}.json|.
        \item Using the SISPI PML interface, command SISPI's OCS to load \verb|~sispi/obstac/loaded/queue_${DATETIME}.json|.
    \end{enumerate}
\end{enumerate}

This sequence allows an external scheduler to be triggered by waiting for time stamps to appear on the \verb|/tmp/obstac_fifo.txt| named pipe, and read and write simple json files in the file system to get data on what exposures are ongoing or currently on the queue, and send any created SISPI scripts to SISPI.

The post-DES \obstac\ also supplies a base class, {\tt obstac.Scheduler.Scheduler}, for creating python scripts that act as a corresponding scheduler, and an example subclass, {\tt obstac.ExampleScheduler.ExampleScheduler}, that provides a trivial complete implementation.\footnote{The post-DES \obstac\ can be found at \url{https://github.com/ehneilsen/obstac10}, and the example scheduler is in {\tt python/obstac/ExampleScheduler.py}.} DECam projects that create their own schedulers according to this pattern (either based on {\tt obstac.Scheduler.Scheduler} or entirely independent of \obstac) can write their own schedulers, and start their schedulers at the start of a night of observing. As long as this process is running, it can be used to automatically schedule exposures, and be enabled and disabled though the SISPI GUI.

\section{Summary and conclusion}
\label{conclusion}

DES is a stage III dark energy experiment, measuring cosmological equation of state parameters and improving the DETF figure of merit by a factor of 3 to 5 over stage II experiments. To do so, it used DECam on the Blanco telescope at CTIO to perform a time domain survey of 10 fields to collect supernova light curves and measure the redshift-distance relation from them; and a wide survey to collect multi-band data on 5000 square degrees in the southern Galactic cap. The survey was originally planned to take 525 nights of observing time running from August through February for five years. The third year had exceptionally poor weather, and a sixth year of 52 nights was added to enable the completion of the survey.

The time-domain fields were chosen to overlap regions of the sky in which there were already deep spectroscopic data sets, and have at least some of the fields accessible from instruments in the northern hemisphere. The area of the wide survey was chosen to be accessible to the the Blanco telescope, overlap complementing surveys (such as that performed by the SPT), avoid regions of high Galactic dust or stellar density, and cover a 5000 square degree connected region that samples both large and small angular scales well. Specific pointings and exposure times were chosen to maximize the area covered to a threshold depth, allow for tight relative calibration using, and spread nightly PSF systematic errors across many nights.

We defined a new metric for depth quantification, $\tau$ (or $t_{\mbox{eff}}$), that maps directly to limiting magnitude, but is additive with coaddition. The expected $\tau$ varies weather, airmass, and moon position, and we developed tools for predicting $\tau$ based on weather and sky models, and designed survey strategy and tactics accordingly.

Survey tactics were designed to make optimal use of the time scheduled, and reacted to environmental conditions by following a Markov decision process architecture: each specific exposure was scheduled only shortly before it was observed, based on the state of the survey and environmental at the time of observation. We developed a scheduler, \obstac, that automates the execution of these tactics, and was used to run suites of simulations that allowed development and refinement of both observing tactics and survey strategy generally. \obstac\ was developed in python, and integrated into SISPI, the DECam readout and control system. After the completion of DES, parts of \obstac\ were modified to support the automation of externally supplied schedulers with this control system, without requiring modification of SISPI itself.

The automation of scheduling was vital to the success of the survey, and the collaboration learned from our efforts. Some of the more important lessons are:
\begin{enumerate}
    \item The ability to use the same code to simulate the survey under a full range of historically informed weather realizations was an essential element of planning, but the extreme weather in year three showed that even a $\sim40$ year baseline is inadequate to show the full range of weather that might be encountered. Extreme years occur, and one can only design flexibility into the system for them, one cannot plan to them.
    \item The importance of the position of the moon, particularly when trying to maintain an observing cadence on a field near the ecliptic, is generally under-appreciated in the astronomical community, and the selection of time-domain fields near the path of the moon created special challenges for maintaining the required observing cadence. 
    \item The mismatch between the tight {\sc r.a.} distribution and the tightest possible distribution in calendar time presented significant challenges. Although there would still remain some mismatch due to seasonal variations in the weather and duration of the night, a better match would have been possible with an all-sky, all year survey (such as LSST) and thus higher data quality.
    \item Tight integration between the observing and control system (SISPI) and the automated scheduler simplified the initial development, but a better decoupling of the scheduler itself from the rest of the system would have been helpful. There were two important aspects of the coupling which could have been reduced. First, \obstac\ made extensive use of the SISPI database, to the extent that (reduced) copies of this database had to be created and an instance run during simulation. Although some interaction in production is unavoidable (the database being the source of completed exposures and telemetry data), this could have been better isolated such that a database instance was not a requirement for simulation. Second, the \obstac\ scheduler process was managed by the control system process manager (the {\tt SISPI Architect}); and run from a product installed using {\tt eups} \citep{lupton_version_2019}, the package management system used by it. This complicated process and version management by users who were not SISPI experts, for example when an observer needed to update \obstac's configuration or move to a different version.
    \item Assumptions about the DES strategy embedded in the structure of the scheduler made later adjustments more complicated, and limited the applicability of \obstac\ to other surveys. For example, the human abstraction of the set of exposures arising from combinations of tilings, offsets, and filters was mapped directly into artifacts in the code. A more general structure (continuing the example, maintaining just the simple abstraction of a desired exposure in the code) would have required a little more work initially, but the added flexibility would have been worth it.
\end{enumerate}

DES survey strategy and tactics resulted in a successful survey that made efficient scientific use of the allotted time, and scheduling required little expertise or effort on the part of observers. (For more details, see \cite{diehl_dark_2016} and \cite{diehl_dark_2018}.) Predictions set by simulations were generally accurate, the exception being one year in which the weather was significantly worse than any recorded historical year. 

\newpage
\appendix

\section{Exposure quality}
\label{dataquality}

\subsection{Conditions affecting image quality}
In addition to the properties of the instrument itself (the DECam camera and the Blanco telescope), there are three major factors that influence the quality of an exposure: 
\begin{description}
\item[{seeing ({\sc fwhm}),}] blurring of the image due to optical distortion by the atmosphere (the final image {\sc fwhm} includes atmospheric and instrumental components); 
\item[{atmospheric transmission ($\eta$),}] the fraction of light from astronomical sources that makes it through the atmosphere (and maybe clouds) to the instrument; and
\item[{sky brightness ($b$),}] light scattered or emitted by the atmosphere and other sources adds noise to the image.
\end{description}
A good survey strategy minimizes these effects as much as possible, given other constraints. 

\subsection{\texorpdfstring{$\tau$}{t} as an image quality metric}

Image quality affects the quality of the catalog of astronomical objects produced by the survey in two ways. First, when combined with blurring due to the optics of the instrument itself, the seeing sets the of size point sources (such as stars) in the images, usually measured as the full width at half maximum ({\sc fwhm}) of point sources in the image. The {\sc fwhm} in turn limits the precision with which the survey can separate nearby objects and measure the shapes of those objects (an item of particular concern for measurements of weak lensing). Second, the {\sc fwhm}, atmospheric transmission, and noise due to sky brightness all limit the brightness of objects that can be detected in an image, and the precision with which this brightness can be measured. This property is typically quantified in astronomical images as the limiting magnitude: the magnitude at which point source can be measured to a reference signal to noise ratio.

Flux from an object increases linearly with exposure time, while the noise increases as the square root of the exposure time, such that
\begin{equation}
m_{\mbox{\scriptsize lim}} = m_{\circ, \mbox{\scriptsize cond}} + 1.25 \log(\mbox{\sc exptime})
\end{equation}
where $m_{\circ, \mbox{\scriptsize cond}}$ is a function of the instrument and conditions under which the exposure was taken (the seeing, atmospheric transmission, and sky brightness). If we define a new value, $\tau$, as
\begin{equation}
\label{taudef}
\tau \equiv \eta^2 \left( \frac{0.9"}{\mbox{\sc fwhm}} \right)^2 \left( \frac{b_{\mbox{\scriptsize dark}}}{b} \right)
\end{equation}
then
\begin{equation}
m_{\mbox{\scriptsize lim}} = m_{\circ} + 1.25 \log(\tau) + 1.25 \log(\mbox{\sc exptime}) = m_{\circ} + 1.25 \log(\tau \times \mbox{\sc exptime})
\end{equation}
where $m_{\circ}$ is a function only of the instrument performance, and not the conditions \citep{neilsen_limiting_2016}. $\tau$ varies monotonically with $m_{\mbox{\scriptsize lim}}$. It is a scaling factor that determines how long an {\sc exptime} is needed in one set of conditions to match the limiting magnitude of an exposure with a different {\sc exptime} taken under a different set of conditions. Furthermore, the quantity $\tau \times \mbox{\sc exptime}$ is additive when exposures are coadded: if the limiting magnitude of a co-added exposure is
\begin{equation}
m_{\mbox{\scriptsize lim}} = m_{\circ} + 1.25 \log\left(\sum_{i} \tau_{i} \times \mbox{\sc exptime}_{i}\right)
\end{equation}
For a set of exposures with uniform exposure time, once again 
\begin{equation}
m_{\mbox{\scriptsize lim}} = m_{\circ} + 1.25 \log(\tau) + 1.25 \log(\mbox{\sc exptime})
\end{equation}
where $\tau \equiv \sum_i \tau_i$.

The accumulated $\tau$ for a given area of the sky is therefore a metric for the depth in a survey with exposures of uniform exposure time, and has the advantages (over limiting magnitude) of being a simpler function of seeing, atmospheric transmission, and sky brightness, being additive with exposures, and roughly proportional to the observing time for a given set of conditions.

\subsection{\texorpdfstring{$\tau$}{t}, airmass, and hour angle}
\label{tairmassha}

Image quality degrades with increases in airmass due to three separate factors:
\begin{itemize}
    \item Atmospheric extinction of the light from the object by the atmosphere. The magnitude of the extinction is linear with the airmass with an "extinction coefficient" generally denoted $k$ such that, in equation~\ref{taudef},
    \begin{equation}\eta = 10^{-\frac{2}{5} k (X-1)}.\end{equation}
    \item The brightness of the airglow of the sky (and therefore the photon noise in the image ``background'') increases at higher higher airmasses, because the line of sight passes through the airglow layer at an angle, increasing the effective depth of the layer. Following the van Rhijn model for skyglow (see appendix~\ref{skybrightness}),
    \begin{equation}
        \frac{b_{\mbox{\scriptsize dark}}}{b} = \sqrt{\frac{h + R\mu^{2}}{h+R}} \times 10^{\frac{2}{5}k(X-1)}
    \end{equation}
    where $h$ is the altitude of the airglow layer and R is the radius of the Earth.
    \item The atmospheric contribution to the {\sc fwhm} of the delivered PSF increases with airmass, because the the additional turbulent air causes additional refraction by variations in the atmospheric index of refraction. In equation~\ref{taudef},
    \begin{equation}
        \mbox{\sc fwhm}^{2} = \epsilon_{i}^2 + \left[\epsilon_{\mbox{\scriptsize zenith}} \mu^{-\frac{3}{5}}\right]^{2}
    \end{equation}
    where $\epsilon_{i}$ is the instrumental contribution to the {\sc fwhm}, $\epsilon_{\mbox{zenith}}$ is the atmospheric contribution an zenith, and $\mu = \cos z$. See appendix~\ref{seeing}, particularly figure~\ref{fig:seeingvsairmass}.
\end{itemize}

\begin{figure*}
\centering
\includegraphics[width=\linewidth]{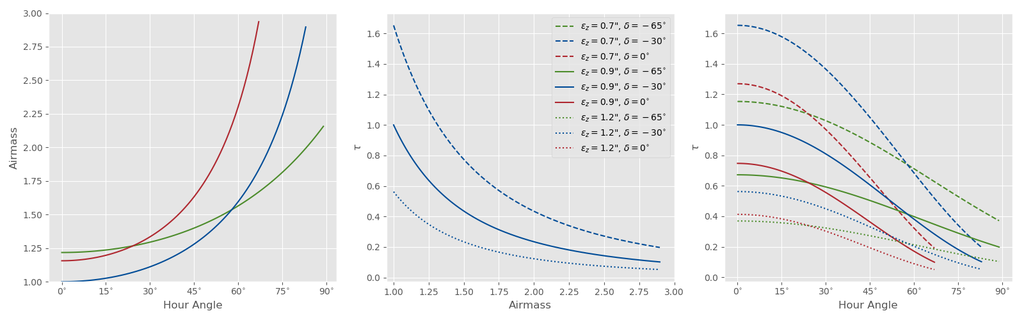}
\caption{\label{fig:tauvsha}
The left panel shows the variation of airmass with hour angle for three declinations: $\delta=-65\degree$, the southern edge of the DES footprint; $\delta=-30\degree$, which passes through the zenith at CTIO; and $\delta=0\degree$, the northern edge of the DES footprint. The center panel shows how $\tau$ varies with airmass (including the effects of the variation of seeing, sky brightness from airglow, and extinction, but excluding clouds and the scattering of moonlight) for three different values for atmospheric seeing at zenith. The right panel combines these, showing the variation of $\tau$ with hour angle for different combinations of seeing and declination.}
\end{figure*}

The central panel of figure~\ref{fig:tauvsha} shows the variation in $\tau$ with airmass, using the \cite{van_rhijn_brightness_1921} sky brightness model for the airglow with an airglow layer at 90km \citep{packer_altitudes_1961}, the Kolmogorov model for the variation of seeing with airmass \citep{lena_observational_1988}, and a characteristic atmospheric extinction of $k=0.1$. $\tau$ drops sharply with airmass. To avoid using observing time collecting exposures of low quality (low $\tau$), DES places a hard cut of 1.4 airmasses on the wide survey exposures attempted under good observing conditions. Under poor conditions, the DES scheduler dynamically modifies the cut to lower airmass. When observing supernova fields, an airmass cut of 1.5 is usually used instead, although exposures at higher airmass were sometimes attempted at times of year when fields were not accessible at lower airmass at any time of the night.

The hour angle, {\sc ha}, of a given pointing at a given time is the difference (in units of sidereal time) between the time the pointing reaches its minimum zenith distance (and therefore airmass) and the given time. \footnote{The celestial equatorial coordinates system is a spherical coordinate system in which the polar axis corresponds roughly with the Earth's axis in 2000, in which $\delta$ (the declination) is defined to be the polar distance - $90\degree$, and the azimuthal distance can be expressed in either of two different ways: the hour angle ({\sc ha}), the angle west of the local meridian, or the right ascension ($\alpha$), the angle east of an intersection between the plane of the Earth's rotation about its own axis an the plane of the Earths orbit about the Sun. As the Earth rotates, the {\sc ha} increases with time, completing one full rotation in one sidereal day. The local sidereal time, {\sc lst}, defines the offset $\alpha$ and {\sc ha}: $\mbox{\sc ha} = \mbox{\sc lst} - \alpha$. (Note that {\sc ha} and $\alpha$ increase in opposite directions.) The change in the {\sc ha} over a given time interval is the change in {\sc lst}, so the current {\sc lst} uniquely defines the positions (and therefore zenith distance and airmass) of celestial coordinates relative to the observer.} The zenith distance of a pointing can be calculated as
\begin{equation} \label{eq:mudeltaphiha}
    \mu = \cos \delta \cos \phi \cos \mbox{\sc ha} + \sin \delta \sin \phi\
\end{equation}
where $\mu \equiv \cos(z)$.\footnote{using a direct application of the cosine formula for spherical geometry, described, for example, in \cite{smart_textbook_1977}, p. 7.}
Therefore, although pointings with the same declination vary with hour angle in the same way, objects with different declinations do not: at a given observatory with latitude $\phi$, the minimum zenith distance varies as $\sin \delta$, while the change in zenith distance with {\sc ha} (and therefore {\sc lst}) varies as $\cos \delta$. The left panel of figure~\ref{fig:tauvsha} shows this variation for CTIO and the northern and southern edges of the DES footprint, as well as a pointing that passes through the zenith.
One can combine the relations between hour angle and airmass and between airmass and $\tau$ to arrive at the relation between hour angle and $\tau$ for pointings at different declinations under different seeing conditions, shown in the right panel of figure~\ref{fig:tauvsha}.

To map civil date and time to areas of the footprint visible with a given airmass (and therefore $\tau$, except for weather and the moon), we begin by calculating the {\sc lst}\footnote{following \cite{meeus_astronomical_1998} ch. 12, but ignoring higher precision terms that have no effect on strategy or tactics. Note that $280.46062\degree$ was the Greenwich Sidereal Time at MJD=515445.5 (2000-01-01 00:00:00Z), the offset by $\lambda$ converts that to the Local Sidereal Time at the observatory, and $\frac{366.242}{365.242}\times360\degree = 360.98565\degree$, where 366.242 is the number of sidereal days per tropical (equinox to equinox) year and 365.242 is the number of solar days per tropical year, so $360.98565\degree$ is the number of sidereal degrees per solar day. The sidereal day is shorter than a solar day (such that there is one additional day per year) because, while the sidereal day varies only with the rotation of the Earth around its own axis (and the precession of the equinoxes, which is negligible for this purpose), the solar day varies with the rotation of the Earth around the sun as well, and one complete rotation takes one (tropical) year, by definition.} (from which we can derive the {\sc ha}, the airmass $X$, and finally $\tau$):
\begin{eqnarray}
\mbox{\sc lst} & = & 280.46062\degree + \lambda + (\mbox{\sc mjd}-51544.5 \mbox{ days}) \times 360.98565\degree/\mbox{day} \\
\mbox{\sc ha} & = & \mbox{\sc lst} - \alpha \\
\mu & = & \cos \delta \cos \phi \cos \mbox{\sc ha} + \sin \delta \sin \phi \\
X & = & \sqrt{ a^{2}\mu^{2} + 2 a + 1 } - a \mu \\
\tau & = & 10^{-\frac{2}{5} k (X-1)} 
       \times \frac{\epsilon_{\circ}^{2}}{\epsilon_{i}^2 + \left[\epsilon_{\mbox{\scriptsize zenith}} \mu^{-\frac{3}{5}}\right]^{2}}
       \times \sqrt{\frac{h + R \mu^2}{h + R}}\\
\end{eqnarray}
At a reasonable zenith distance for observing, $X \simeq \frac{1}{\mu}$ and $\sqrt{\frac{h + R \mu^2}{h + R}} \simeq \mu = \frac{1}{X}$, so
\begin{equation}
\label{eq:tauandx}
\tau  \simeq  \frac{10^{-\frac{2}{5} k (X-1)}}{X} \times \frac{\epsilon_{\circ}^{2}}{\epsilon_{i}^2 + X^{\frac{6}{5}}\epsilon_{\mbox{\scriptsize zenith}}^2}
\end{equation}

\subsection{\texorpdfstring{$\tau$}{t}, the sun, and the moon}

\begin{figure*}
\centering
\includegraphics[height=.4\textheight]{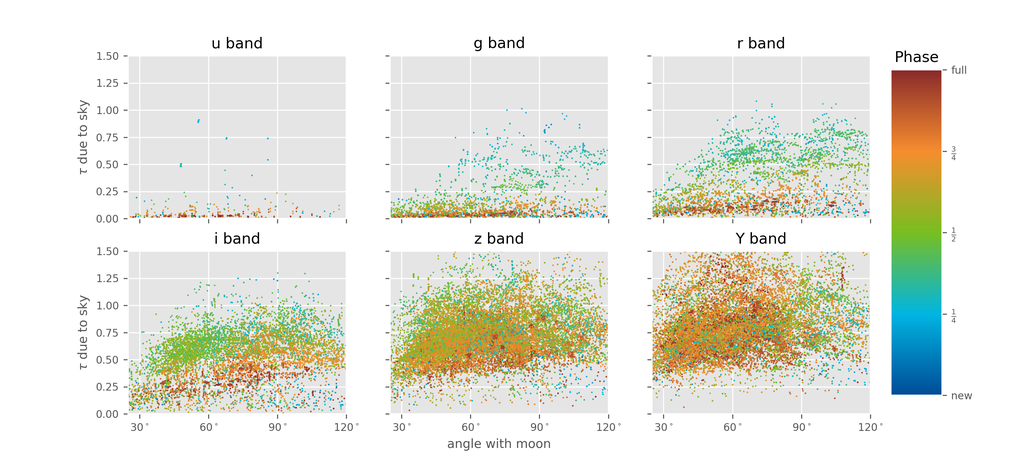}
\caption{\label{fig:skytvsmoonangle}
These figure show the relation between the sky brightness factor in $\tau$ (and therefore $t_{\mbox{eff}}$) for exposures processed by DES-DM, as a function of the angle between the pointing an the moon. Only exposures where the cloud extinction factor in $\tau$ is near one and the moon has a zenith distance of less than $80\degree$ are included. Points are color coded by the phase of the moon. In bluer bands ($u$, $g$, and $r$), Rayleigh scattering of moonlight is important, so except when the moon is near new it strongly affects the sky brightness even when the moon is far from the pointing. In redder bands ($z$ and $Y$), Rayleigh scattering is less important, and scattered moonlight is only prominent when Mie scattering is significant, when the moon is near the pointing.
}
\end{figure*}

\begin{figure*}
\centering
\includegraphics[height=.4\textheight]{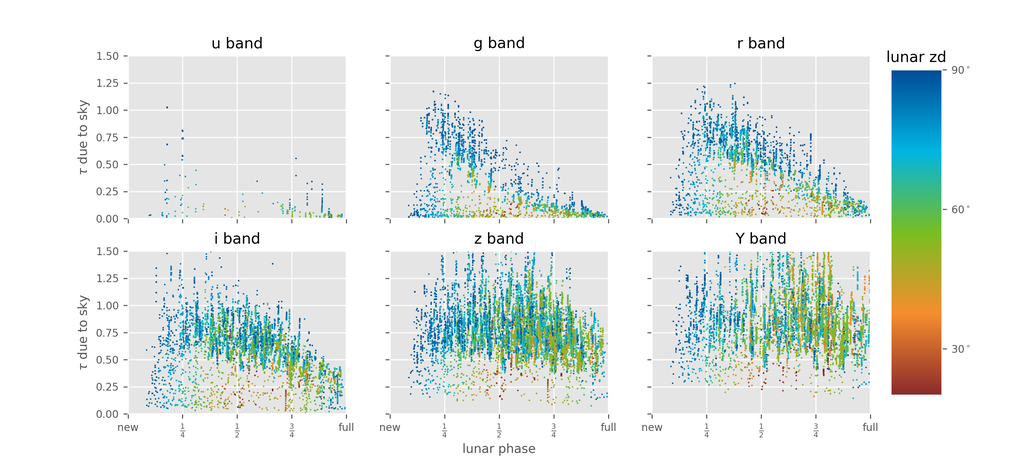}
\caption{\label{fig:skytvsmoonphase}
These figure show the relation between the sky brightness factor in $\tau$ (and therefore $t_{\mbox{eff}}$) for exposures processed by DES-DM, as a function of lunar phase. Exposures with significant extinction due to clouds, where the moon is below the horizon, or where the pointing is within $60\degree$ of the moon (so Mie scattering maybe important) are excluded. In bluer bands ($u$, $g$, and $r$), where Rayleigh scattering is prominent, $\tau$ is near zero even when the moon is faint: such bands can only be effectively observed either when the moon is down entirely, or when the moon is both faint (near new) and close to the horizon (blue in the plot). In redder bands ($z$, $Y$), the moon has little effect at any zenith distance or phase, unless the phase is very near full. 
}
\end{figure*}

\begin{figure*}
\centering
\includegraphics[height=.4\textheight]{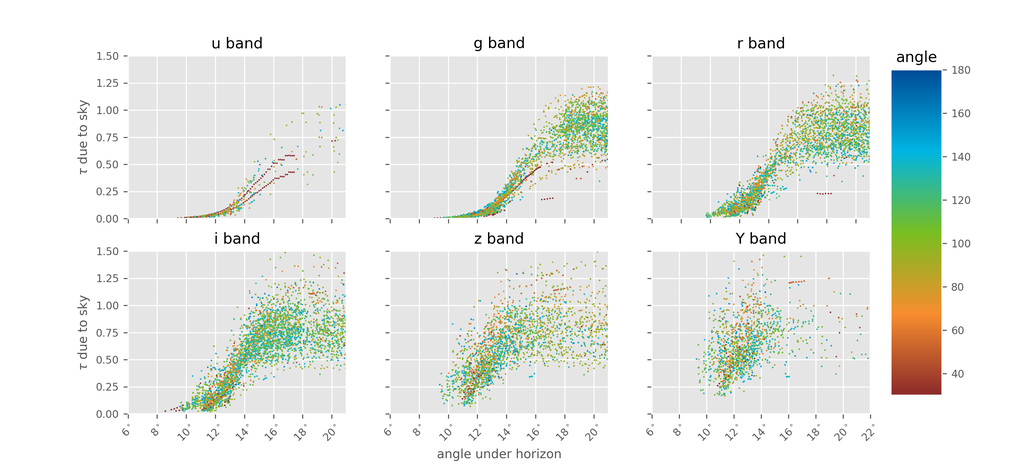}
\caption{\label{fig:skytvssunzd}
This figure shows the relation between the sky brightness factor in $\tau$ and the altitude of the sun, for exposures taken twilight and without significant moon or cloud extinction. Point colors represent the angle between the sun and the pointing of the exposure. Conventional wisdom that twilight has little effect on sky brightness at astronomical twilight (greater than $18\degree$ below the horizon) appears to describe the data well. The transition between futile and optimal happens at smaller angles for redder filters, with exposures in $z$ and $Y$ effective at angles down to $14\degree$.
}
\end{figure*}

Another major factor in data quality is sky brightness. \obstac\ included a sky brightness model, detailed in appendix~\ref{skybrightness}. The observed relationship between sky brightness related parameters and $\tau$ are shown in figures~\ref{fig:skytvsmoonangle},~\ref{fig:skytvsmoonphase}, and~\ref{fig:skytvssunzd}. In figure~\ref{fig:skytvsmoonangle}, the separation between the near-full exposures (red points) and near-new exposures (blue points) in {\it g}, {\it r}, and {\it i} show the strong dependence of sky brightness on moon phase in these filters. The lack of points in the upper left (high $\tau$, low angular separation) in these subplots indicates the strong dependence of sky brightness on angular separation from the moon for these filters (where Rayleigh scattering dominates). In contrast, $\tau$ shows a weaker dependence on angular separation in redder filters (where Mie scattering dominates).

Figure~\ref{fig:skytvssunzd} shows the effect of twilight on data quality. In each band, data collection appears futile when solar altitudes are greater than $-10\degree$. As the altitude falls, data quality rises and then flattens. Exposures in visible light ({\it g} and {\it r} band) follow conventional wisdom that the sky is fully dark at the end of astronomical twilight, when the solar altitude is less than $-18\degree$, and exposures further into the infrared attain ``full'' data quality at progressively higher altitudes.

\section{Airmass}
\label{airmass}

Several factors in image quality depend critically on the airmass, the density of the atmosphere integrated along the line of sight, relative to the density of the atmosphere integrated to zenith (see section~\ref{tairmassha}). At zenith angles typically used for astronomical observing, the airmass $X$ is often approximated by an atmosphere modeled as a flat plane, such that $X=\sec z$. Because the height of the Earth's atmosphere is small compared to the radius of the Earth, this approximation is usually sufficient for zeniths angles accessible to the Blanco telescope.

\begin{figure*}
\centering
\includegraphics[width=\linewidth]{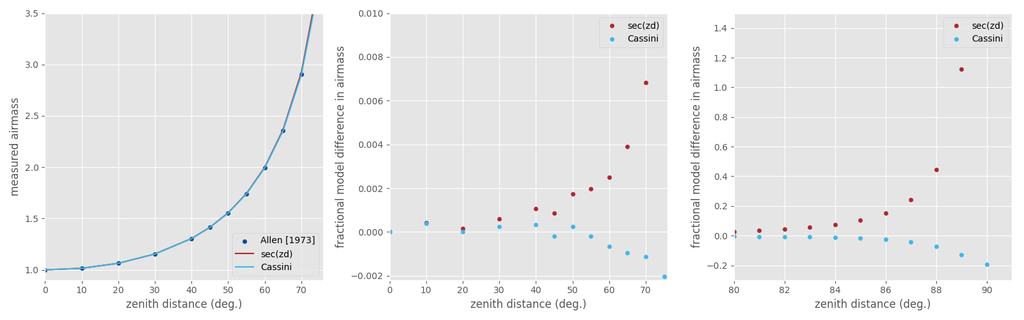}
\caption{\label{fig:airmassvszd}
The left panel shows the variation in airmass with zenith distance. The dark blue points show empirical measurements compiled in \cite{allen_astrophysical_1973}. The light blue and red (mostly obscured by the light blue) show the Cassini (uniform spherical shell) and $\sec(z)$ (flat slab) approximations. Note that they agree well for most zenith distances. The central and right panels show the fractional difference between the two models and the empirical measurements for the two models, in two ranges of zenith. Both models match the empirical measurements to within 1\% for zenith distances accessible to the Blanco telescope. At higher airmasses, however, the $\sec(z)$ diverges, approaching infinity at $z=90\degree$. This can become relevant for sky brightness estimates, because sky brightness from scattered moonlight depends not only on the airmass of the telescope pointing, but also on the airmass of the moon, which may be near the horizon during observing. The coefficient of extinction is of order 0.1, so although neglecting the extinction of the moon entirely can result in an overestimate of the sky brightness from scattered moonlight by several magnitudes, estimating the airmass to within 20\% will result only in a 2\% error in the sky brightness. 
}
\end{figure*}

This approximation breaks down near the horizon, where $\sec z$ goes to infinity. Here, the observer can see out of the atmosphere due to the curvature of the Earth, such that the maximum $X \simeq 38$. When estimating sky brightness due to scattered moonlight, extinction of the moonlight by the Earth's atmosphere is proportional to the airmass {\em of the moon} when it is near the horizon: if a flat plane approximation of the airmass is used, sky brightness from the moon will be significantly underestimated near the horizon. The red points in figure~\ref{fig:airmassvszd} compare the flat plane approximation to measured values.

\begin{figure*}
\centering
\includegraphics[width=0.7\linewidth]{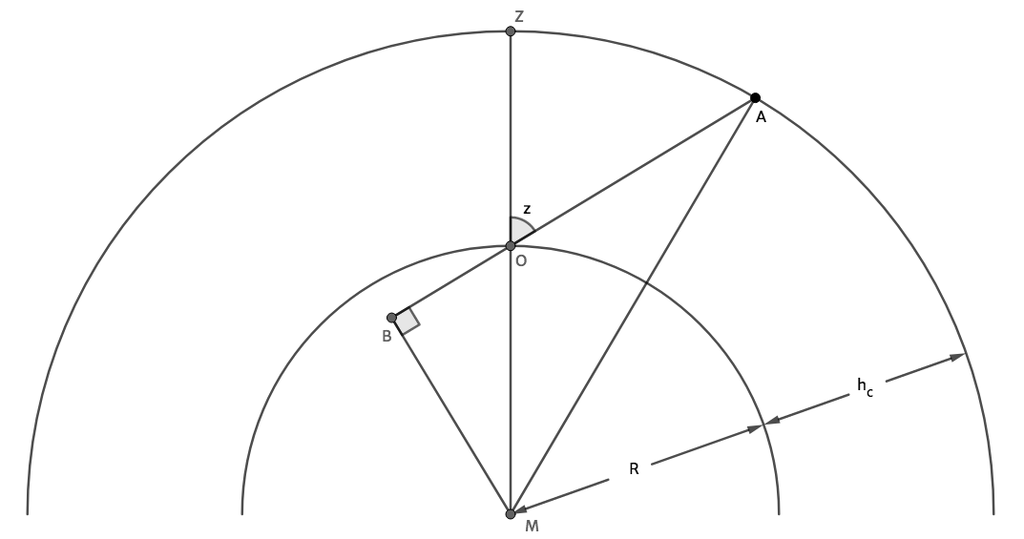}
\caption{\label{fig:cassini}
The Cassini model for the atmosphere, in which it is approximated by a spherical shell of uniform density and finite thickness $h_{C}$. In this diagram, $M$ marks the center of the Earth, $R$ is the radius of the Earth, $O$ is the location of the observer, $Z$ is directly over the observer, who observes along line of sight $OA$ at a zenith angle $z$. 
}
\end{figure*}

One way of improving on the traditional "flat uniform slab" model of the atmosphere is to model the atmosphere as a uniform spherical shell of finite depth. This model is attributed to Cassini by \cite{kristensen_astronomical_1998}, who provides a derivation of the airmass from such a model, repeated here.

The airmass \(X\) is the ratio of the path length through to
atmosphere to the path length through zenith:
\[ X = \frac{OA}{h_{C}} \]
We begin with the application of the Pythagorean theorem on triangle
\(ABM\):
\[ MA^2 = MB^2 + BA^2 \]
We then apply the Pythagorean theorem on triangle \(OBM\), substitute
known distances, and solve for \(MB^2\):
\begin{eqnarray*}
MO^2 & = & MB^2 + BO^2 \\
R^2 & = & MB^2 + R^2 \cos^2 z \\
MB^2 & = & R^2 (1 - \cos^2 z)
\end{eqnarray*}
Introducing the shorthand
\[\mu = \cos z \] and substituting the above value for \(MB^2\) into
our earlier equation for \(MA^2\), we arrive at:
\[ MA^2 = R^2 (1 - \mu^2) + BA^2\]
We can now solve for \(BA^2\) and substitute in the known value for
\(MA\), \(R+h_{C}\):
\begin{eqnarray*}
BA^2 & = & (R + h_{C})^2 + (\mu^2 - 1) R^2 \\
     & = & \mu^2 R^2 + 2 R h_{C} + h_{C}^2
\end{eqnarray*}
We substitute our values into the airmass equation:
\begin{eqnarray*}
X & = & \frac{AO}{h_{C}} \\
 & = & \frac{BA - BO}{h_{C}}\\
 & = & \frac{BA - R \cos z}{h_{C}}\\
 & = & \frac{\sqrt{\mu^2 R^2 + 2 R h_{C} + h_{C}^2} - R \mu}{h_{C}}\\
 & = & \sqrt{\mu^2 \frac{R}{h_{C}}^2 + 2 \frac{R}{h_{C}} + 1} - \frac{R}{h_{C}} \mu 
\end{eqnarray*}
We can let
\begin{equation}
a = \frac{R}{h_{C}}
\end{equation}
to get
\begin{equation}
X = \sqrt{a^2 \mu^2 + 2 a + 1} - a \mu
\end{equation}

The blue points in figure~\ref{fig:airmassvszd} compare measured values to a Cassini model with $a=470$, which estimates the airmass to within $\sim 1$\% up to a zenith distance of $70\degree$, and never exceeds 20\%, even at the horizon.

\section{Seeing}
\label{seeing}

A model for the seeing and its effect on the delivered point spread function (PSF) full width at half maximum (FWHM) plays two important roles in strategy development and execution: first, a model is required to generate simulated seeing data so that realistic survey simulations can be made; and second, \obstac\ needs to estimate the PSF FWHM of candidate exposures during the scheduling process.

\begin{figure*}
\centering
\includegraphics[width=\linewidth]{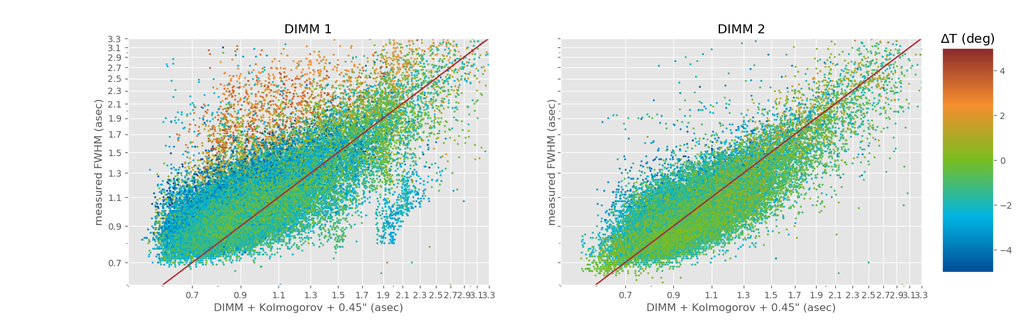}
\caption{\label{fig:fwhmvsdimmkolmogorov}
An estimate for the delivered {\sc fwhm} of an image can be made using a DIMM measurement, an airmass, a band-pass, and an instrumental contribution to the seeing. These plots plot such an estimate for all DES exposures with corresponding DIMM measurements. Each plot represents a different DIMM; DIMM 1 operated during the early part of the survey, DIMM 2 at the end. The assumed instrumental contribution of 0.45" assumes thermal equilibrium between the temperature of the telescope mirror and surrounding air. The color scale shows the difference in temperature, such that colors other than green indicate an instrumental contribution of 0.45" is expected to be an underestimate.
}
\end{figure*}

To generate such a model, we used data from the CTIO differential image motion monitor (DIMM) covering dates from April 4, 2001 through March 16, 2010. A DIMM measures the atmospheric turbulence by measuring the relative positions of two images of the same star as seen through two apertures on the same telescope, in this case two 8cm diameter apertures separated by 17cm \citep{kornilov_combined_2007, els_four_2009}, and so measures the atmospheric turbulence at these spatial wavelengths. CTIO then followed standard practice, and used a Kolomogorov turbulence model to estimate the atmospheric contribution to the PSF FWHM on an exposure at zenith, in 500nm light. DES survey data shows that the DIMM measurements at a derived FWHM for DECam images are well correlated, but that there is significant scatter; see figure~\ref{fig:fwhmvsdimmkolmogorov}.

Two important relations derived from the Kolmogorov model, required to relate a seeing that represents the turbulence of the atmosphere as a whole to the PSF FWHM for a specific image, are those between the atmospheric contribution to the PSF at one airmass to that an another,
\begin{equation}
    \mbox{\sc fwhm} \propto X^{\frac{3}{5}}
\end{equation}
and the contribution to the PSF at in one filter (frequency band) to another,
\begin{equation}
    \mbox{\sc fwhm} \propto \lambda^{-\frac{1}{5}}.
\end{equation}

\begin{figure*}
\centering
\includegraphics[width=\linewidth]{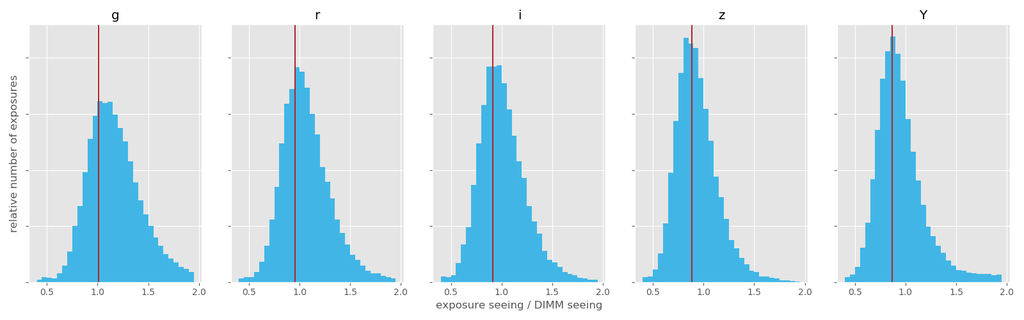}
\caption{\label{fig:exposuredimmratiohists}
These histograms show the distributions of the ratio of the measured {\sc FWHM} of each image (corrected for an instrumental contribution of 0.45'' and to an airmass of one using the Kolgomorov model) to the seeing value derived from the CTIO DIMM, which uses the Kolmogorov model and and effective wavelength of 500nm. Offsets from a ratio of one should be due to differences between this nominal DIMM wavelength and the true band-passes of the DES filters. The vertical red lines show the offsets predicted by the Kolmogorov model.
}
\end{figure*}

Figure~\ref{fig:exposuredimmratiohists} shows the distribution of the ratio of measured PSF FWHM values to DIMM values for DES data around in different bands (blue) compared to the value predicted by a Kolmogorov model (red). In all cases, the value indicated by the Kolomogorov model matches the peak of the distribution well (compared to the overall width of the distribution), but there is a significant scatter, and there is a long tail in which the delivered seeing is poor compared to what the DIMM data would indicate. 

\begin{figure*}
\centering
\includegraphics[width=0.5\linewidth]{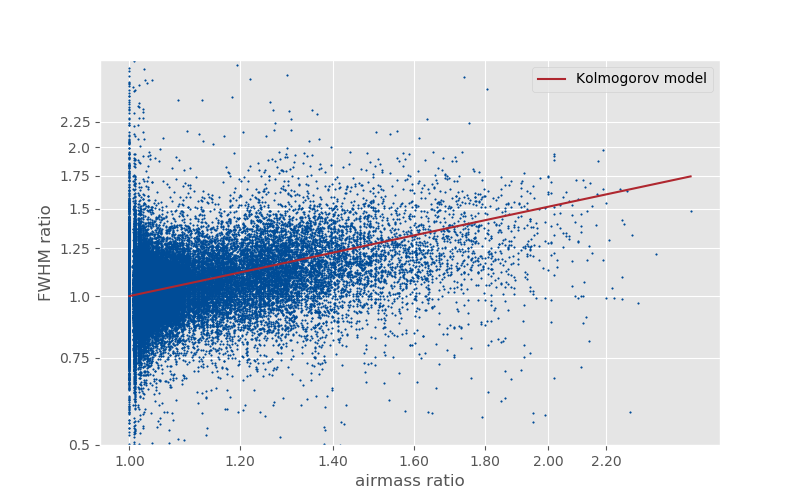}
\caption{\label{fig:seeingvsairmass}
Each point represents a pair of DES exposures taken in the same band in succession with at most a 10 minute gap. The horizontal axis shows the ratios of the airmass of the exposures with the larger airmass to the airmass of the one with the lower, and the vertical shows the corresponding ratio of the measured FWHM of the PSF for the exposures. According to the Kolmogorov turbulence model, $\mbox{\sc fwhm} \propto X^{\frac{3}{5}}$, so we expect that
$\frac{\mbox{\sc fwhm}_{1}}{\mbox{\sc fwhm}_{2}} \propto \left(\frac{X_{1}}{X_{2}}\right)^{\frac{3}{5}}$. This plot is in a logarithmic scale, so we expect this exponential relationship to appear linear. The red line show shows the line with the slope expected from the Kolmogorov model.
}
\end{figure*}

\begin{figure*}
\centering
\includegraphics[width=\linewidth]{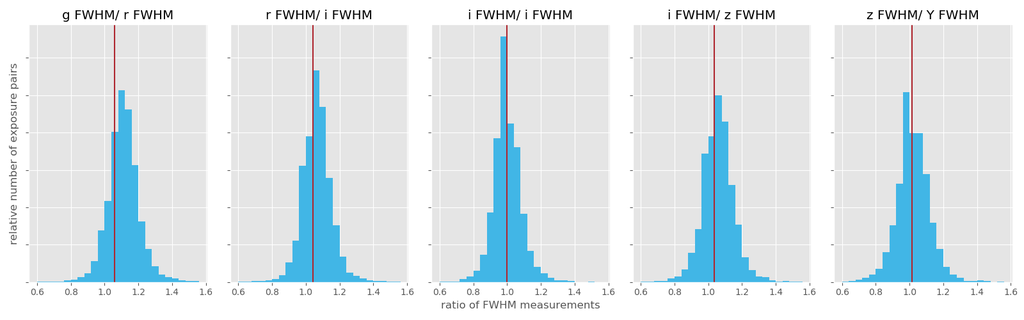}
\caption{\label{fig:filterpairfwhmratiohists}
These histograms show the distributions of {\sc fwhm} ratios of pairs of DES exposures taken in a given pair of bands in succession with at most a 10 minute gap and at similar airmass. According to the Kolmogorov model, $\mbox{\sc fwhm} \propto \lambda^{-\frac{1}{5}}$. The red lines mark the ratios derived from this relation and the centers of the DES filter band-passes.
}
\end{figure*}

Figures~\ref{fig:seeingvsairmass} and~\ref{fig:filterpairfwhmratiohists} examine the effects of airmass and wavelength separately. In Figure~\ref{fig:seeingvsairmass} FWHM ratios of neighboring exposures in the same filter are compared as a function of the ratio of their airmasses, a red line indicating the Kolmogorov model. Figure~\ref{fig:filterpairfwhmratiohists} shows histograms of ratios of FWHM measurements of neighboring exposures in different bands, but similar airmasses, with a red line indicating the prediction of the Kolmogorov model. In all cases there is significant scatter, even in the case where the two exposures are in the same filter, while the peak roughly matches the Kolmogorov prediction, indicating that most of the scatter is due to changes in the atmospheric turbulence on short time scales. The largest deviation between the prediction and measurement is in {\it g} band, where the peak in the measured ratio is slightly worse than that estimated by the model. So, while there are indications that the Kolomgorov model is not perfect, it appears that such deviation is small compared to short timescale variations in the atmospheric seeing.

To model the time variation the seeing, we analyzed the CTIO DIMM data as an auto-regressive (AR) time series using the {\tt xts} package in {\tt R}. In an AR(p) model, the value at timestep $t$ is a weighted mean of the $p$ most recent values, some global average, plus a random step from this weighted mean \citep{cryer_time_2008}:
\begin{equation}
    Y_{t} = \phi_{1}Y_{t-1} + \phi_{2}Y_{t-2} + ... + \phi_{p}Y_{t-p} + e_{t}
\end{equation}
where $Y$ are values in the time series, $\phi_{n}$ are weights, and $e_{t}$ are random steps. If $e_{t}$ are normally distributed and the set of $\phi$ values meet a set of stationary conditions, $Y$ will follow a normal distribution centered on zero. For an AR(1) model (where $p=1$), the AR model reduces to a damped random walk, and the stationary condition is that $|\phi|<1$. For an AR(2) model, the conditions are that $\phi_{1} + \phi_{2} < 1$, $\phi_{2} - \phi_{1} < 1$, and $|\phi_{2}|<1$. See \cite{cryer_time_2008} pp. 66-77 for a more in depth discussion.

We began by transforming the DIMM values to zero centered, normal distribution of $Y$ values. We then re-sampled the data into even 5 minute intervals, and fit model parameters in an AR(2) (that is, $p=2$) model for each calendar month. These fit models were used directly to generate artificial $Y$ values, transformed back into DIMM {\sc fwhm} values, and used the results for \obstac\ simulations. 

For prediction of seeing while scheduling, we used a slightly simpler model. The distribution of $\log \mbox{\sc fwhm}$ values in any given calendar month is well fit by the ``skew-normal'' distribution of \cite{ohagan_bayes_1976}. We therefore fit the $\log \mbox{\sc fwhm}$ of each month of DIMM data to the location ($\xi$), scale ($\omega$), and shape ($\alpha$) parameters of this distribution, and transform between $Y$ and measured {\sc fwhm} values accordingly. For the actual AR(2) model, we used a set of global values across all months:
\begin{equation}
    Y_{t} = 0.8 \times Y_{t-1} + 0.14 \times Y_{t-2}.
\end{equation}

\section{Sky brightness}
\label{skybrightness}

\subsection{Contributions to sky brightness}

The sky brightness model used by \obstac\ follows the general approach used by \cite{krisciunas_model_1991}, estimating the overall sky brightness by adding flux from three major contributors:
\begin{itemize}
    \item airglow, or emission from ions in the Earth's atmosphere;
    \item Rayleigh scattering of moonlight by atoms in the Earth's atmosphere; and
    \item Mie scattering of moonlight by aerosols in the Earth's atmosphere.
\end{itemize}
Extinction due to airmass is included in all three cases. In addition to the scattering of moonlight considered by \cite{krisciunas_model_1991}, \obstac's model also includes scattering of sunlight, but only in very rough approximation. (Improved modeling of twilight would require modeling the shadow of the Earth on the Earth's atmosphere, which makes the problem significantly more complex, and proved unnecessary for \obstac's purposes).

Several significant factors were not considered in either the model presented by \cite{krisciunas_model_1991} or that used by \obstac. In particular, zodiacal light was ignored, light pollution by neighboring town and cities (La Serena and Vicu\~na) is completely absent from these models, and no attempt was made to model the decline in airglow over the course of the night. 

The {\tt skybright} \citep{neilsen_skybright_2019} python package and command line application provides an implementation of the model described here.

\subsection{van Rhijn Airglow}

\begin{figure*}
\centering
\includegraphics[width=0.7\linewidth]{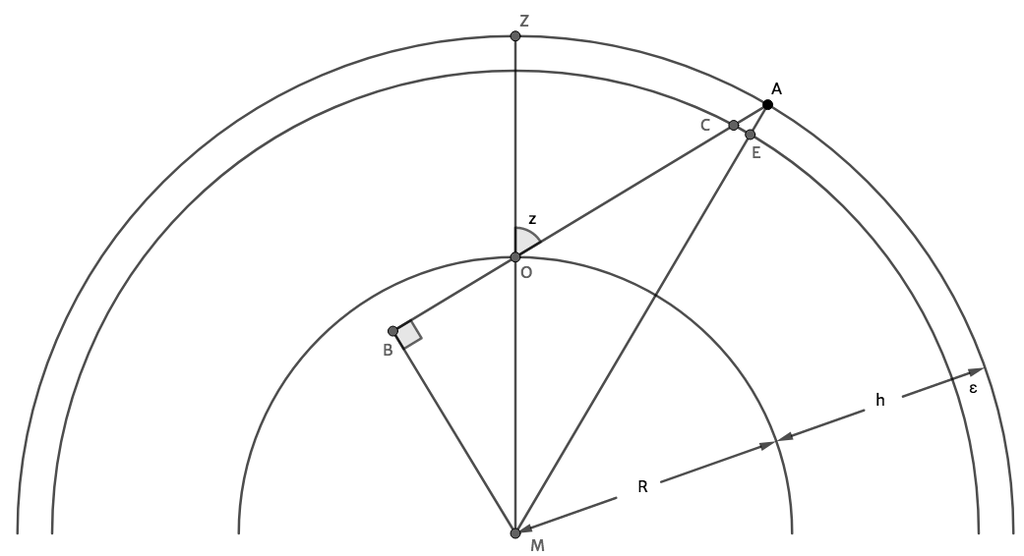}
\caption{\label{fig:vanrhijn}
This diagram shows the various components of the \cite{van_rhijn_brightness_1921} model for airglow, in which light is emitted from a thin shell of width $\epsilon$ at height $h$ above the surface of the Earth (with radius $R$). $C$ and $A$ mark the points at which the line of sight intersect with the near and far edge of this thin shell, and $z$ is the angle between the line of sight and the direction of the zenith, $Z$, above the observer.
}
\end{figure*}

\obstac\ follows \cite{krisciunas_model_1991}, and uses the \cite{van_rhijn_brightness_1921} model
for airglow. In this model, the all light from airglow comes from a thin spherical shell,
shown in figure \ref{fig:vanrhijn}. The light received from a thin
emitting shell is proportional to the length of the line of sight
through that shell. So, for an observer standing at \(O\) and looking
at a zenith angle \(z\), the light seen will be proportional to the
length \(CA\).

Following figure \ref{fig:vanrhijn},
\begin{eqnarray}
OA & = & BA - BO \\
OC & = & BC - BO \\
OA - OC & = & BA - BC \\
& = & BA \left[ 1 - \frac{BC}{BA}\right]
\end{eqnarray}
From the Pythagorean theorem on \(\bigtriangleup BAM\),
\begin{equation} BA = \sqrt{MA^2 - MB^2} \end{equation}
and similarly on \(\bigtriangleup OBM\),
\begin{equation} MB^2 = R^2 - BO^2 \end{equation}
so
\begin{equation} BA = \sqrt{MA^2 - R^2 + BO^2}. \end{equation}
Noting that
\begin{equation} BO = R \cos z \end{equation}
and
\begin{equation} MA = R + h \end{equation}
we arrive at
\begin{equation} BA = \sqrt{(R+h)^2 - R^2 + R^2 \cos^2 z} .\end{equation}

Applying the Pythagorean theorem to \(\bigtriangleup BCM\),
\begin{eqnarray}
MC^2 & = & BC^2 + BM^2 \\
& = & BC^2 + R^2 - BO^2 \\
BC^2 & = & MC^2 - R^2 + BO^2 \\
& = & (MA - \epsilon)^2 - R^2 + BO^2 \\
& = & MA^2 - R^2 + BO^2 - 2 \epsilon MA + \epsilon^2 \\
& = & BA^2 - 2 \epsilon MA + \epsilon^2 \\
BC & = & \sqrt{BA^2 - 2 \epsilon MA + \epsilon^2}
\end{eqnarray}

Finally, solve for \(CA\):
\begin{eqnarray}
CA & = & OA - OC  \\
   & = & BA \left[ 1 - \frac{BC}{BA} \right] \\
   & = & BA \left[ 1 - \sqrt{\frac{BA^2 - 2 \epsilon MA + \epsilon^2}{BA^2}} \right] \\
   & = & BA \left[ 1 - \sqrt{1 - \frac{2 \epsilon MA - \epsilon^2}{BA^2}} \right]
\end{eqnarray}

Apply the Taylor series of \(\sqrt{1-x}\) and drop \(\epsilon^2\) and
higher order terms. Noting that \(MA=R+h\), arrive at:
\begin{eqnarray}
CA & = & \frac{\epsilon MA}{BA} \\
   & = & \frac{\epsilon MA}{\sqrt{MA^2 - R^2 + BO^2}} \\
   & = & \frac{\epsilon (R+h)}{\sqrt{(R+h)^2 - R^2 + BO^2}} \\
   & = & \frac{\epsilon (R+h)}{\sqrt{(R+h)^2 - R^2 + R^2 \cos^2 z}} \\
   & = & \frac{\epsilon}{\sqrt{1 - \frac{R}{R+h} \sin^2 z}} \\
\end{eqnarray}

At zenith, the path \(z=0\), so \(CA = \epsilon\). If we take the
surface brigtness of the sky at zenith to be \(\mathrm{m_{zenith}}\),
then:
\begin{eqnarray}
\mathrm{m} & = & \mathrm{m_{zenith}} - 2.5 \log_{10} ( [ 1 - \frac{R}{R+h} \sin^2 z ]^{-\frac{1}{2}} ) \\
& = & \mathrm{m_{zenith}} + 1.25 \log_{10} (1 - \frac{R}{R+h} \sin^2 z )
\end{eqnarray}

Using the Cassini model of the atmosphere (a spherical shell of
uniform density), the height of the shell is roughly 20km, well below
the 80 to 300 km altitude expected for the emitting layer. Therefore,
the standard applications for the extinction as a function of airmass
should apply, so
\begin{eqnarray}
\mathrm{m} & = & \mathrm{m_{zenith}} - k + 1.25 \log_{10} (1 - \frac{R}{R+h} \sin^2 z ) + k X \\
 & = & \mathrm{m_{zenith}} + 1.25 \log_{10} (1 - \frac{R}{R+h} \sin^2 z ) + k (X - 1)
\end{eqnarray}

\begin{figure*}
\centering
\includegraphics[width=\linewidth]{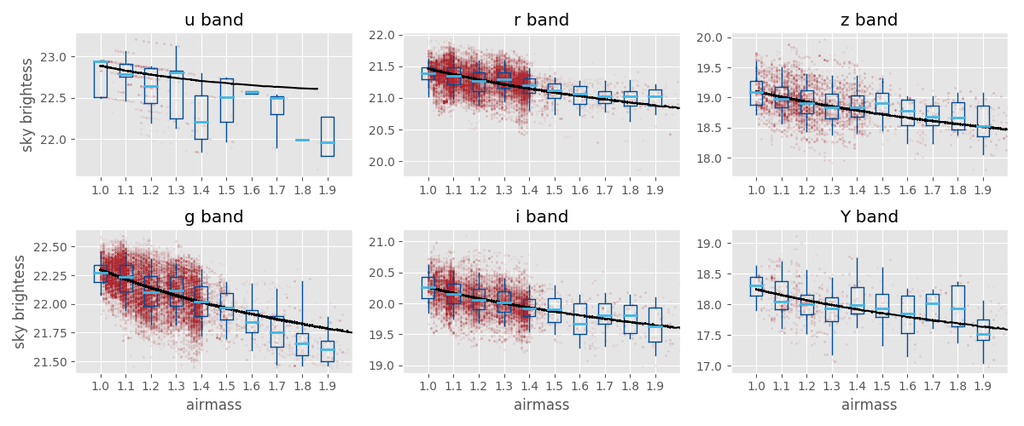}
\caption{\label{fig:darkskyvsairmass}
This plot shows the sky brightness of the fully dark (no twilight, moon, or clouds) sky as a function of airmass. Red points show sky brightness measurements from DES exposures. The black line shows the van~Rhijn model, with $m_{zenith}$ fit to the data. The blue markers indicate the distribution of exposures in different airmass bins: the whiskers extend to the 5th and 95th percentiles, the boxes mark the first and third quartiles, and the light blue bar shows the median.
}
\end{figure*}

Figure~\ref{fig:darkskyvsairmass} shows the van~Rhijn model plotted over the measured values in DES exposures. The behaviour seems generally correct, but the variance from factors other than airmass is similar to or greater than the variation with airmass alone.

\subsection{Scattered moonlight}
\label{scatteredmoonlight}

\begin{figure*}
\centering
\includegraphics[width=0.7\linewidth]{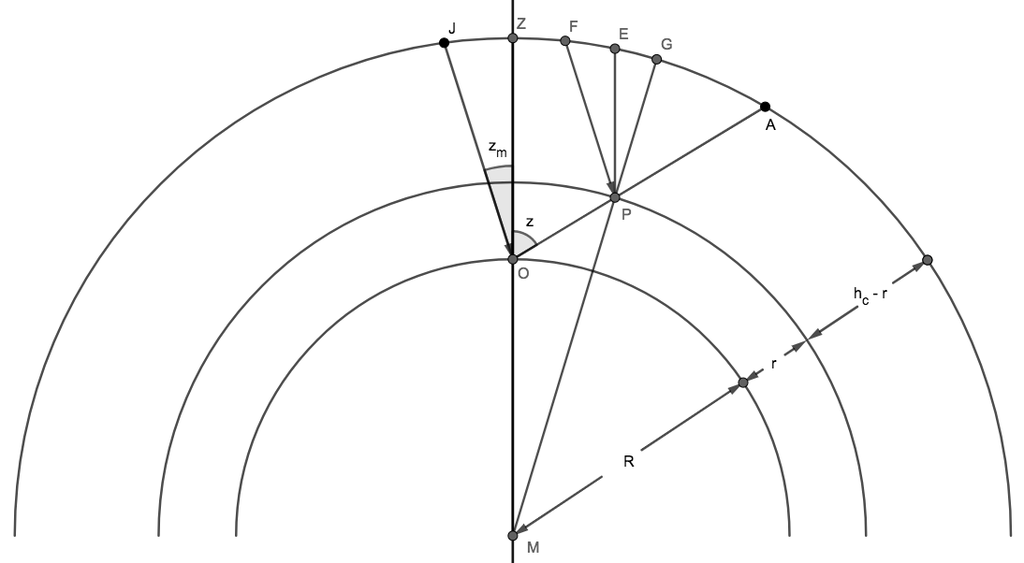}
\caption{\label{fig:moonk}
This diagram shows the geometry of the scattering of moonlight off of the atmosphere along a line of sight from an observer at $O$. $A$ mark the points at which the line of sight intersects with an idealized outer edge of the atmosphere (at altitude $h_C$), and $z$ is the angle between the line of sight and the direction of the zenith, $Z$, above the observer. Moonlight hitting the observer directly through $J$ has a zenith angle $z_m$ at the observer. Because the distance to the moon is much greater than the height of the atmosphere, moonlight hitting a scattering particle in the atmosphere at $P$ at altitude $r$ through point $F$ is effectively parallel to the moonlight hitting the observer.
}
\end{figure*}

Consider a scattering particle at distance \(r\) from the observer. If
light scatters off of this particle from the moon with incident
intensity \(I(r)\) through an angle \(\rho\), we have
\begin{equation} L = \beta f(\rho) I(r) \end{equation}
where \(f(\rho)\) is the scattering function. At a distance \(r\) from
this scattering particle, the intensity will drop according to the
square of the distance and the extinction between the observer and
the particle:
\begin{equation} B = \frac{\beta}{4 \pi r^2} f(\rho) I(r) 10^{-0.4 k X(r)}\end{equation}
If the scatterers have a density \(D\) such that the number of
scatterers is
\begin{equation} dN = D dV, \end{equation}
then the total light from all scatterers at distance \(r\) in solid
angle \(\Omega\) is
\begin{eqnarray}
dB & = & \frac{\beta}{4 \pi r^2} f(\rho) \, I(r) \, D(r) \, 10^{-0.4 k X(r)} dV \\
   & = & \frac{\beta}{4 \pi r^2} f(\rho) \, I(r) \, D(r) \, 4 \pi r^2 \, \Omega \, 10^{-0.4 k X(r)} dr \\
   & = & \beta D(r) \, f(\rho) \, I(r) \, \Omega \, 10^{-0.4 k X(r)} dr
\end{eqnarray}
So, the surface brightness is:
\begin{equation} \frac{B}{\Omega} = \beta \, f(\rho) \int_0^{h_C} D(r) \, I(r) \, 10^{-0.4 k X(r)} dr \end{equation}

\(I(r)\) is the intensity of moonlight on the optical path at \(r\),
and is equal to the intensity of moonlight outside the atmosphere,
\(I^*\), reduced by the extinction between the outside of the
atmosphere and the scattering particle. Note that the airmass
appropriate for this extinction is the airmass of the moon, less the airmass of the moon to the height of the
scattering particle, resulting in an intensity of moonlight
\begin{equation} I(r) = I^* 10^{-0.4\,k\,[X_m(h_C)-X_m(r)]} \end{equation}
so
\begin{equation} \frac{B}{\Omega} = \beta \, f(\rho)\,  I^* \, 10^{-0.4\,k\,X_m(h_C)} \int_0^{h_C} D(r) \, 10^{-0.4\,k \,[X(r) - X_m(r)]} dr \end{equation}

Now, introduce \(x\), the airmass of a particle at height \(r\)
from an observer looking at zenith. Using a plane parallel
approximation of the atmosphere such that the zenith angle of the moon from P equals the zenith angle of the moon from O\footnote{
The true zenith angle of the moon from P, $\angle BPF$, is $z_{m} + \angle BPE$. $\angle BPE= \angle PMO$, so by the law of sines, $\frac{OP}{\sin \angle BPE} = \frac{PM}{\sin 180\degree - z}$. $\angle PMO$ reaches its maximum when $P=A$ and $z$ is near the telescope limit, $z = 70\degree$, at which point (using the Cassini model) $OP= X h$ and $MP = R + h$. When $z=70\degree$, $X=2.9$ and $\sin 180\degree - z = 0.94$. So, $\sin \angle PMO = 0.94 \times 2.9 \times \frac{h}{R+h} = \frac{2.7}{a+1}$ where $a \simeq 470$ (see appendix~\ref{airmass}), at which point $\angle PMO \simeq 0.33\degree$, which is negligible for our purposes here.
},
\begin{equation} x = \frac{X(r)}{X(h_C)} = \frac{X_m(r)}{X_m(h_C)} \end{equation}
so
\begin{eqnarray}
X(r) & = & x X(h_C) \\
X_m(r) & = & x X_m(h_C) \\
\end{eqnarray}
By the definition of airmass,
\begin{eqnarray}
dx & \propto & D(r) dr \\
D(r) dr & = & c_{0} dx
\end{eqnarray}
substituting,
\begin{eqnarray}
\frac{B}{\Omega} & = & \beta \, f(\rho)\,  I^* \, 10^{-0.4\,k\,X_m(h_C)} \int_0^1 D(r) \, 10^{-0.4\,k \,[X(h_C) - X_m(h_C)]x} dx \\
& = & \beta \, f(\rho)\,  I^* \, 10^{-0.4\,k\,X_m(h_C)}  \frac{c_{0} \, 10^{-0.4\,k \,[X(h_C) - X_m(h_C)]}-1}{-0.4\,k\,[X(h_C)-X_m(h_C)]} \\
& = & \beta \, f(\rho)\,  I_0 p \, 10^{-0.4\,k\,X_m(h_C)}  \frac{c_{0} \, 10^{-0.4\,k \,[X(h_C) - X_m(h_C)]}-1}{-0.4\,k\,[X(h_C)-X_m(h_C)]} 
\end{eqnarray}
in which \(I^{*} = I_0 p\), where \(I_0\) is the illuminance from the full moon and \(p\) is the lunar phase function.
Combining constants and setting \(X=X(h_C)\) and \(X_m=X_m(h_C)\)
leaves us with:
\begin{eqnarray}
\frac{B}{\Omega} & = & c f(\rho)\,  p \, 10^{-0.4\,k\,X_m} \frac{10^{-0.4\,k \,[X - X_m]} - 1}{-0.4\,k\,[X-X_m]} \\
& = & c f(\rho)\,  p  \frac{10^{-0.4\,k\,X} - 10^{-0.4\,k\,X_m}}{-0.4\,k\,[X-X_m]} \label{eq:bomega}
\end{eqnarray}
where $I_0$, $\beta$ and $c_{0}$ are combined into a single multiplicative constant $c$.

\subsection{The scattering function}

The scattering function of moonlight has two components: Rayleigh scattering off of gas in the atmosphere, and Mie scattering off of aerosols. The factor $c f(\rho)$ in equation~\ref{eq:bomega} can therefore be expressed as:
\begin{equation} \label{eq:frho}
    c f(\rho) = c_{R} f_{R}(\rho) + c_{M} f_{M}(\rho) 
\end{equation}

Rayleigh scattering can be approximated thus:
\begin{equation} \label{eq:raylfrho} f_{R}(\rho) = \frac{3}{4} (1 + \cos^2\rho )\end{equation}
The derivation is presented in many textbooks, for example \cite{petty_first_2006} sections 12.2.1 and 12.2.2.

The exact solution for the Mie scattering function, $f_{M}(\rho)$, is inconvenient, both
in form and because it relies on knowledge of the distribution of the
shape and size of the scattering particles\footnote{\cite{petty_first_2006} plots the full scattering function on p.~366, and notes the advantage considering the log of the scattering function on the bottom of page~368.
Petty defines the asymmetry parameter $g$ on p.~329, and presents the Henyey-Greenstein phase function on p.~331.}. \cite{henyey_diffuse_1941} (equation 2) provide a widely adopted approximation:
\begin{equation}f_{M}(\rho) = \frac{\gamma(1-g^2)}{4\pi}\frac{1}{(1+g^2-2 g \cos\rho)^\frac{3}{2}} \end{equation}
where \(g\) is an asymmetry parameter between 0 and 1. \cite{cornette_physically_1992} offer a refinement of the Henyey-Greenstein function:
\begin{equation} \label{eq:cornette}
f_{M}(\rho) = \frac{3}{2} \frac{1-g^2}{2+g^2}
            \frac{1 + \cos^2 \rho}{(1 + g^2 - 2 g \cos\rho)^{\frac{3}{2}}}
\end{equation}
Combining equation~\ref{eq:bomega} with equations \ref{eq:frho}, \ref{eq:raylfrho}, and \ref{eq:cornette}, we get:
\begin{equation} \label{eq:moonscatter}
    \frac{B}{\Omega} = \left[ c_{R} \times \left( \frac{3}{4} (1 + \cos^2\rho ) \right)
                            + c_{M} \times \left( \frac{3}{2} \frac{1-g^2}{2+g^2} \frac{1 + \cos^2 \rho}{(1 + g^2 - 2 g \cos\rho)^{\frac{3}{2}}}\right)
                            \right] \times p \times \left[ \frac{10^{-0.4\,k\,X} - 10^{-0.4\,k\,X_m}}{-0.4\,k\,[X-X_m]} \right]
\end{equation}
in which $c_{R}$, $c_{M}$, $g$, and $k$ can be fit to historical data, and other parameters can be calculated from the time and pointing.

\subsection{Twilight}
When the sun and moon fall beneath the horizon, the Earth casts a shadow on the atmosphere along the line of sight, and as the shadow travels up the column of air along the line of sight, the sky brightness in that direction will fall. The \obstac\ sky model approximates this by continuing to use the scattering model from equation~\ref{eq:moonscatter} for the relative brightness of different locations on the sky, but decreasing the $\log(\mbox{flux})$ by a quadratic fit in altitude of the body in twilight (sun or moon). 

\subsection{Assessment}

\begin{figure*}
\centering
\includegraphics[width=\linewidth]{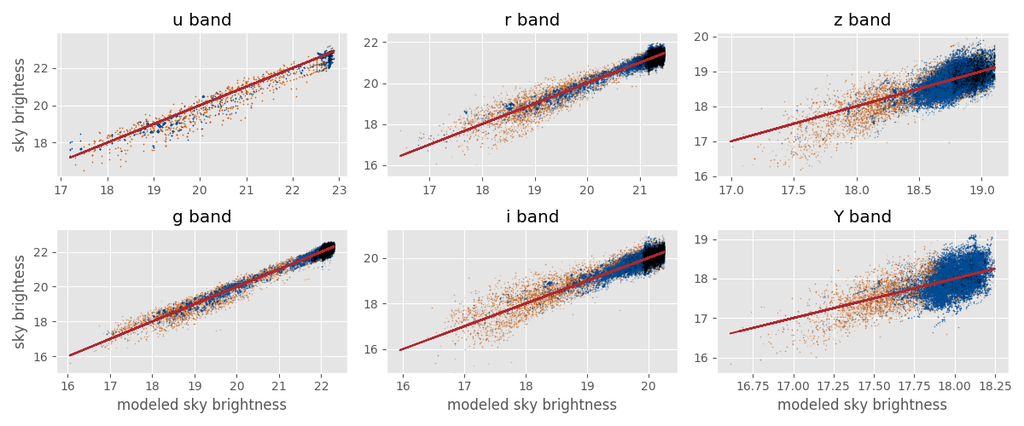}
\caption{\label{fig:skyvsmodel}
This plot compares sky brightness measured from DES images to sky brightness estimated from the model, for exposures with no clouds. The red line shows the perfect fit to the model. The orange points show exposures with twilight but no moon, blue with moon but no twilight, and black with neither twilight nor moon.
}
\end{figure*}

Figure~\ref{fig:skyvsmodel} compares the combined model, including airglow, scattered moonlight, and twilight, with the data collected by DES. The standard deviation of the residuals ranges from 0.2 to 0.3 magnitudes for dark and moony sky, with bluer bands having smaller residuals. Residuals for twilight were higher, between 0.3 and 0.4 magnitudes, which is unsurprising given the simplicity of the twilight model.

\section{Notation}
\label{notation}

\begin{tabular}{ll}
\hline
$\alpha$ & Right Ascension (equatorial celestial coordinate) \\
$\delta$ & declination (equatorial celestial coordinate) \\
$\epsilon_{i}$ & instrumental contribution to the delivered FWHM \\
$\epsilon_{\circ}$ & reference seeing for $\tau=0$ (typically $\epsilon_{\circ}=0.9''$)\\
$\epsilon_{\mbox{zenith}}$ & contribution of the atmosphere to the FWHM when pointed at zenith \\ 
$\eta$ & atmospheric transparency \\
$\tau$ & exposure time scaling factor \citep{neilsen_limiting_2016} \\
$\lambda$ & longitude of the telescope (east is positive)\\
$\mu$ & the cosine of the zenith distance \\
$\phi$ & latitude of the telescope \\
\hline
$a$ & ratio of the radius of the Earth to the height of its atmosphere, idealized as a uniform shell \\
$b$ & flux per unit solid angle from the sky \\
$b_{\mbox{dark}}$ & flux per unit solid angle from the sky, in moonless conditions at zenith, in reference conditions \\
{\sc exptime} & exposure time (in seconds) \\
$g$ & symmetry paramater for Mie scattering \\
{\sc gst} & Greenwich sidereal time, the hour angle of the vernal equinox at Greenwich \citep[pp.~48-49]{seidelmann_explanatory_1992}\\
{h} & height of the airglow layer of the atmosphere \\
$h_{C}$ & height of the of the atmosphere in the Cassini (spherical shell of finite thickness) model of the atmosphere \\
{\sc ha} & the Hour Angle of a pointing, $\mbox{\sc ha} = \mbox{\sc lst} - \alpha$ \\
{\sc lst} & local sidereal time, $\mbox{\sc lst} = \mbox{\sc gst} + \lambda$ \citep[pp.~48-49]{seidelmann_explanatory_1992}\\
$k$ & the coefficient of atmospheric extinction \\
$m_{\mbox{lim}}$ & limiting magnitude for an exposure \\
$m_{\circ, \mbox{circ}}$ & zero point for the limiting magnitude, for a given instrument and set of atmospheric conditions \\
$m_{\circ}$ & zero point for the limiting magnitude, for a given instrument \\
{\sc mjd} & modified Julian date, the (dynamical) time since 1858-11-17 00:00:00.0 UTC in days \citep[pp.~55-56]{seidelmann_explanatory_1992}\\
$p$ & lunar phase \\
$R$ & radius of the Earth \\
$X$ & airmass of the telescope pointing \\
$X_{m}$ & airmass of the moon \\
$z$ & zenith distance
\end{tabular}

\bibliographystyle{aasjournal_ehnmod}
\bibliography{zotero-refs}

\section*{Acknowledgements}
\addcontentsline{toc}{section}{Acknowledgements}

This manuscript has been authored by Fermi Research Alliance, LLC under Contract No. DE-AC02-07CH11359 with the U.S. Department of Energy, Office of Science, Office of High Energy Physics. The United States Government retains and the publisher, by accepting the article for publication, acknowledges that the United States Government retains a non-exclusive, paid-up, irrevocable, world-wide license to publish or reproduce the published form of this manuscript, or allow others to do so, for United States Government purposes.

Funding for the DES Projects has been provided by the U.S. Department of Energy, the U.S. National Science Foundation, the Ministry of Science and Education of Spain, 
the Science and Technology Facilities Council of the United Kingdom, the Higher Education Funding Council for England, the National Center for Supercomputing 
Applications at the University of Illinois at Urbana-Champaign, the Kavli Institute of Cosmological Physics at the University of Chicago, 
the Center for Cosmology and Astro-Particle Physics at the Ohio State University,
the Mitchell Institute for Fundamental Physics and Astronomy at Texas A\&M University, Financiadora de Estudos e Projetos, 
Funda{\c c}{\~a}o Carlos Chagas Filho de Amparo {\`a} Pesquisa do Estado do Rio de Janeiro, Conselho Nacional de Desenvolvimento Cient{\'i}fico e Tecnol{\'o}gico and 
the Minist{\'e}rio da Ci{\^e}ncia, Tecnologia e Inova{\c c}{\~a}o, the Deutsche Forschungsgemeinschaft and the Collaborating Institutions in the Dark Energy Survey. 

The Collaborating Institutions are Argonne National Laboratory, the University of California at Santa Cruz, the University of Cambridge, Centro de Investigaciones Energ{\'e}ticas, 
Medioambientales y Tecnol{\'o}gicas-Madrid, the University of Chicago, University College London, the DES-Brazil Consortium, the University of Edinburgh, 
the Eidgen{\"o}ssische Technische Hochschule (ETH) Z{\"u}rich, 
Fermi National Accelerator Laboratory, the University of Illinois at Urbana-Champaign, the Institut de Ci{\`e}ncies de l'Espai (IEEC/CSIC), 
the Institut de F{\'i}sica d'Altes Energies, Lawrence Berkeley National Laboratory, the Ludwig-Maximilians Universit{\"a}t M{\"u}nchen and the associated Excellence Cluster Universe, 
the University of Michigan, the National Optical Astronomy Observatory, the University of Nottingham, The Ohio State University, the University of Pennsylvania, the University of Portsmouth, 
SLAC National Accelerator Laboratory, Stanford University, the University of Sussex, Texas A\&M University, and the OzDES Membership Consortium.

Based in part on observations at Cerro Tololo Inter-American Observatory, National Optical Astronomy Observatory, which is operated by the Association of 
Universities for Research in Astronomy (AURA) under a cooperative agreement with the National Science Foundation.

The DES data management system is supported by the National Science Foundation under Grant Numbers AST-1138766 and AST-1536171.
The DES participants from Spanish institutions are partially supported by MINECO under grants AYA2015-71825, ESP2015-66861, FPA2015-68048, SEV-2016-0588, SEV-2016-0597, and MDM-2015-0509, 
some of which include ERDF funds from the European Union. IFAE is partially funded by the CERCA program of the Generalitat de Catalunya.
Research leading to these results has received funding from the European Research
Council under the European Union's Seventh Framework Program (FP7/2007-2013) including ERC grant agreements 240672, 291329, and 306478.
We  acknowledge support from the Brazilian Instituto Nacional de Ci\^encia
e Tecnologia (INCT) e-Universe (CNPq grant 465376/2014-2).

This manuscript has been authored by Fermi Research Alliance, LLC under Contract No. DE-AC02-07CH11359 with the U.S. Department of Energy, Office of Science, Office of High Energy Physics.

This publication makes use of data products from the Two Micron All
Sky Survey, which is a joint project of the University of
Massachusetts and the Infrared Processing and Analysis
Center/California Institute of Technology, funded by the National
Aeronautics and Space Administration and the National Science
Foundation.

This research has made use of NASA's Astrophysics Data System.

This research made use of SciPy \citep{jones_scipy.org_2019}.

This research made use of NumPy \citep{walt_numpy_2011}.

This research made use of matplotlib, a Python library for publication quality graphics \citep{hunter_matplotlib:_2007}.

This research made use of Astropy, a community-developed core Python package for Astronomy \citep{astropy_collaboration_astropy_2018}.

This research made use of Cartopy \citep{met_office_cartopy:_2019}.

This research made use of PAL: A Positional Astronomy Library \citep{jenness_pal:_2013}.

This research made use of SLALIB \citep{wallace_slalib_1994}.

This research made use of pandas \citep{mckinney_pandas:_2011}.

\end{document}